\documentclass[12pt]{article}

\usepackage[a4paper,left=1.5cm,right=1.5cm,top=1.5cm,bottom=1.5cm]{geometry}
\usepackage{graphicx}
\usepackage{amsmath}        
\usepackage{etoolbox}    

\usepackage{subcaption}     
\usepackage{placeins}
\usepackage{appendix}
\usepackage{url}
\usepackage{authblk}

\usepackage{setspace}
\usepackage[mathlines]{lineno}

\usepackage[authoryear]{natbib}

\usepackage{fancyhdr}
\AtBeginDocument{%
  \setlength\abovedisplayskip{6pt plus 2pt minus 2pt}%
  \setlength\belowdisplayskip{6pt plus 2pt minus 2pt}%
  \setlength\abovedisplayshortskip{4pt plus 2pt minus 2pt}%
  \setlength\belowdisplayshortskip{4pt plus 2pt minus 2pt}%
  \setlength\jot{3pt}%
}

\BeforeBeginEnvironment{equation}{\begingroup\scriptsize}
\AfterEndEnvironment{equation}{\endgroup}

\BeforeBeginEnvironment{equation*}{\begingroup\scriptsize}
\AfterEndEnvironment{equation*}{\endgroup}

\BeforeBeginEnvironment{align}{\begingroup\scriptsize}
\AfterEndEnvironment{align}{\endgroup}

\BeforeBeginEnvironment{align*}{\begingroup\scriptsize}
\AfterEndEnvironment{align*}{\endgroup}

\BeforeBeginEnvironment{eqnarray}{\begingroup\scriptsize}
\AfterEndEnvironment{eqnarray}{\endgroup}
\BeforeBeginEnvironment{eqnarray*}{\begingroup\scriptsize}
\AfterEndEnvironment{eqnarray*}{\endgroup}

\title{Detective quantum efficiency based comparison of HRTEM and ptychography phase imaging}

\author[1]{Felix Bennemann}
\author[1,2]{Angus I. Kirkland}
\author[3]{David A. Muller}
\author[1]{Peter Nellist\thanks{Email: peter.nellist@materials.ox.ac.uk}}

\affil[1]{University of Oxford, Department of Materials, Parks Road, Oxford OX1 3PH, UK}
\affil[2]{The Rosalind Franklin Institute, Building R113,
Rutherford Appleton Laboratory, Harwell Campus, Didcot, Oxfordshire OX11 0QX, UK}
\affil[3]{School of Applied and Engineering Physics, Cornell University, Ithaca, New York 14853, United States}

\date{}

\begin{document}

\maketitle

\abstract{High-resolution transmission electron microscopy (HRTEM) is an important method for imaging beam sensitive materials often under cryo conditions. Electron ptychography in the scanning transmission electron microscope (STEM) has been shown to reconstruct  low-noise phase data at a reduced fluence for such materials. This raises the question of whether ptychography or HRTEM provides a more fluence-efficient imaging technique.
Even though the transfer function is a common metric for evaluating the performance of an imaging method, it only describes the signal transfer with respect to spatial frequency, irrespective of the noise transfer. It can also not be well defined for methods, such as ptychography, that use an algorithm to form the final image. Here we apply the concept of detective quantum efficiency (DQE) to electron microscopy as a fluence independent and sample independent measure of technique performance. We find that, for a weak-phase object, ptychography can never reach the efficiency of a perfect Zernike phase imaging microscope but that ptychography is more robust to partial coherence.}
\section{Introduction}
When an electron wave interacts with a specimen its amplitude and phase can be modified. Particularly for weakly scattering objects, such as biological samples or 2D materials, the modification of the phase is the dominant effect. For a microscope to image these materials, it is therefore crucial to be sensitive to this introduced phase change. The most common technique to image this phase change in microscopy is Zernike phase contrast microscopy. It works by introducing a phase shift of $\pi/2$ for all scattered waves except for the direct beam. In light optics, this is achieved by use of a phase plate.  For imaging beam-sensitive specimens using the conventional transmission electron microscope (CTEM), the required phase shift is mostly commonly introduced by using aberrations such as defocus.  To enable sufficiently low-frequency phase transfer, defocus values into the micron range are used. With the help of cryogenic electron microscopy (cryo-EM) techniques in the CTEM it is possible to image biological specimens close to their native state\citep{HENDERSON1990899, PENCZEK199233, Dubochet_Adrian_Chang_Homo_Lepault_McDowall_Schultz_1988}. Through single particle analysis, complex biological structures have been reconstructed at atomic resolution\citep{Nakane2020, Yip2020}.

In scanning transmission electron microscopy (STEM), electron ptychography has been proven to achieve similar low-noise images at low fluence\citep{Zhou2020}. This poses the question as to which technique is most fluence efficient. While metrics like resolution and signal to noise ratio (SNR) provide useful insights into image quality, they are dependent on the fluence and, in the case of SNR, also on the sample. Electron ptychography has so far relied on the use of phase contrast transfer functions (PCTFs) to assess the performance of reconstruction techniques. The actual transfer delivered by common ptychography techniques such as the Wigner distribution deconvolution (WDD) or ePIE iterative method varies depending on the algorithm. Furthermore, PCTFs only provide information about the modulation of the signal at a given frequency, and do not describe the transfer of the noise. This is crucial, though, because while the noise in Zernike phase contrast microscopy and annular dark field (ADF) electron microscopy is Poisson noise, ptychography imposes significant frequency-dependent noise filtering in the 4D dataset. Noise-normalised transfer functions have been calculated for 4D STEM methods, but are fluence-dependent\citep{SEKI2018118}.

A first approach at comparing the fluence efficiency of electron ptychography with Zernike phase contrast microscopy came from \cite{CDwyer2024}. Using information theory they concluded that four dimensional scanning transmission electron microscopy (4D-STEM) has a maximum quantum efficiency of half that of Zernike phase contrast microscopy for high spatial frequencies. Despite that, they concluded that 4D-STEM is the preferred choice for atomic-resolution imaging due to its superior robustness against aberrations impacting high spatial frequencies.

In this work, we use random processes to model noise within both the Zernike phase contrast microscope and 4D-STEM imaging. This allows us to determine their detective quantum efficiencies (DQEs). An initial attempt at this has been made by \cite{tcBFMuller}. Here we rigorously derive the analytical expressions for the two techniques under a variety of conditions. We also suggest possible optimisations of ptychography.

Comparing optical systems based on their DQEs allows for a sample and fluence independent comparison while retaining straightforward practical application. In the field of electron microscopy DQEs have historically been used to quantify the performance of detectors. However, in other fields such as medical imaging, it is common to define a DQE for the entire imaging process including any \textit{in silico} processing\citep{JPBissonnetteCunnigham, ConebeamCT, SiewardsenConebeam3D}.

For a linear optical system the DQE can be defined as
\begin{equation}\label{DQE}
    DQE(\mathbf{Q_p})=\frac{SNR_\text{out}^2(\mathbf{Q_p})}{SNR_\text{in}^2(\mathbf{Q_p})}
\end{equation}
where \(SNR_\text{out}\) is the output signal to noise ratio, \(SNR_\text{in}\) is the input signal to noise ratio or the ideal signal to noise ratio constrained solely by Poisson noise and $\mathbf{Q_p}$ is the spatial frequency \citep{jones1959advances}. The signal to noise ratio (SNR) is defined as
\begin{equation}\label{snr_from_power}
    SNR(\mathbf{Q_p})^2=\frac{\vert P_\text{signal}(\mathbf{Q_p})\vert^2}{\vert P_\text{noise}(\mathbf{Q_p})\vert^2}
\end{equation}
where $P_\text{signal}(\mathbf{Q_p})$ is the signal power and $P_\text{noise}(\mathbf{Q_p})$ is the noise power. In the field of electron microscopy the number of electrons detected is commonly defined as the signal power. To an approximation, the DQE represents the percentage of electrons contributing to the image to their maximal extent.

\section{DQE framework}
In order to find an analytical expression for the DQE of an imaging technique, an expression for the square of the signal to noise ratio of the ideal microscope needs to be constructed as shown in \eqref{DQE}. For a weak phase object, Zernike phase contrast microscopy was chosen as the reference, similar to \citet{CDwyer2024}, because it provides a theoretically ideal method for converting phase changes into amplitude modulation and thereby image contrast. In the following section an expression for the SNR a CTEM using aberrations to form the phase contrast is found. In section \ref{SSB_ptych_analytical_sec}, an expression for the signal to noise ratio of single sideband ptychography (SSB) is then found and used to find the DQE of SSB ptychography together with the signal to noise ratio of Zernike phase contrast microscopy. The SSB approach is used here as a ptychography method that extracts all the available information in the case of the WPOA.

\subsection{Zernike phase contrast imaging}\label{Zernike_foundation}
This subsection focuses on obtaining an expression for the DQE of Zernike phase contrast microscopy. Given that Zernike phase contrast microscopy is regarded here as the ideal phase contrast method, we might expect it to have a DQE of 1. In practice, experimental effects such as partial coherence and the reliance on lens aberrations to give the required Zernike phase shift will have an impact on the DQE. In segment \ref{Zernike_sig_pow_sect} the signal power of Zernike phase contrast microscopy is derived and in segment \ref{Zernike_noise_pow_sect} the noise power of Zernike phase contrast microscopy is derived. In \ref{Zernike_snr_sq_sect} the two previous segments are combined to derive the signal to noise ratio of Zernike phase contrast micrsocopy. Finally, the detective quantum efficiency is derived in \ref{Zernike_dqe_sect}.
\subsubsection{Zernike phase contrast signal power}\label{Zernike_sig_pow_sect}
Considering a weak phase object, the specimen transfer function is defined as
\begin{equation}\label{wpoa_approx}
    \psi_0(\mathbf{R_p})=1-i\phi(\mathbf{R_p})
\end{equation}
, where $\phi(\mathbf{R_p})$ is the phase shift and is proportional to the projected atomic potential\citep{Cowley1981}.
In the spatial frequency domain this results in
\begin{equation}
    \psi_0(\mathbf{Q_p})=\mathcal{F}\big\{\psi_0(\mathbf{R_p})\big\}=\delta(\mathbf{Q_p})-\mathcal{F}\big[i\phi(\mathbf{R_p})\big]
\end{equation}
An ideal Zernike phase plate introduces a phase shift of \(\pi/2\) for all spatial frequencies except for the direct beam\citep{ZERNIKE1942686}.
\begin{equation}
    \psi_i(\mathbf{Q_p})=\delta(\mathbf{Q_p})-e^{i\pi/2}\mathcal{F}\big[i\phi(\mathbf{R_p})\big]
\end{equation}
Finally the electron wave intensity is recorded in real space
\begin{equation}\label{general_phase_intensity}
    I(\mathbf{R_p})=\vert\psi_i(\mathbf{R_p})\vert^2=\vert 1-e^{i\pi/2}\big[i\phi(\mathbf{R_p})\big]\vert^2
\end{equation}
When invoking the WPOA, this can be simplified to
\begin{equation}\label{Zernike_intensity_ewave_real_space}
    I(\mathbf{R_p})=\vert\psi_i(\mathbf{R_p})\vert^2=\vert 1-i\big[i\phi\big]\vert^2=\vert 1+\phi\vert^2=1+2\phi+\phi^2
\end{equation}
Ignoring the term \(\phi^2\) due to the small angle of \(\phi\), the factor 2 becomes apparent. When combining this result with the envelope function \(E(\mathbf{Q_p})\) and the aberration function \(\chi(\mathbf{Q_p})\), this yields the well known result for the intensity transmission function. The effect of the aperture function on the DQE of Zernike phase contrast microscopy lies beyond the scope of this paper. The function \(\chi(\mathbf{Q_p})\) describes the phase shift introduced of the electron wave introduced by the microscope. For a perfect Zernike phase plate, \(\chi(\mathbf{Q_p})=\pi/2\) except for the direct beam at $\mathbf{Q_p}=0$. In a real microscope lens imperfections and defocus lead to additional phase shifts. The combination of defocus and aberrations can also be used to approximate a phase plate. For a given aberration function, Scherzer defocus provides the closest approximation to the ideal phase plate \citep{ScherzerDefocus}. $E(\mathbf{Q_p})$ represents the product of all envelope functions that can be accounted for. This includes the envelopes for partial temporal coherence, partial spatial coherence, specimen drift, specimen vibration and the detector performance. The intensity transmission function is also called the phase contrast transfer function of Zernike microscopy. We will continue to refer to it as the intensity transmission function as it highlights clearly that the detector measures electron intensity which contains information about the phase. The detector does not measure the phase of the electron wave directly.
\begin{equation}\label{zernike_int_transmission}
T_Z(\mathbf{Q_p})=2E(\mathbf{Q_p})\sin\chi(\mathbf{Q_p})
\end{equation}
The intensity transmission function in Eq. \eqref{zernike_int_transmission} describes the intensity transfer in the case where only the direct beam remains unmodified \citep{Kirkland2020}. In the case of real phase plates, however, there is a range of unshifted spatial frequencies from 0 to $k_z$. To take this into account, a normalisation factor has been proposed but we will not consider it here \citep{ibanez2024}.

To find the signal power, \eqref{zernike_int_transmission} needs to be scaled by the number of electrons per pixel \(N_e\) in the image and the discrete Fourier transform (DFT) normalisation constants \(N_x\) and \(N_y\) representing the dimensions of the 2D discrete signal along the x- and y-axes, respectively. Here, we follow the convention of normalizing the inverse transform while leaving the forward transform unnormalised.

\begin{align}
X_{k_x, k_y} &= \sum_{n_x=0}^{N_x-1} \sum_{n_y=0}^{N_y-1} 
    x_{n_x, n_y} \, 
    e^{-2\pi i \left( \frac{k_x n_x}{N_x} + \frac{k_y n_y}{N_y} \right)} \\
x_{n_x, n_y} &= \frac{1}{N_x N_y} 
    \sum_{k_x=0}^{N_x-1} \sum_{k_y=0}^{N_y-1} 
    X_{k_x, k_y} \, 
    e^{2\pi i \left( \frac{k_x n_x}{N_x} + \frac{k_y n_y}{N_y} \right)}
\end{align}

\(N_e\) can be calculated from the sampling and the fluence \(D\) given in \(e/\)\AA$^2$ and the real space pixel size \(d_{x,y}\) given in \AA.
\begin{align}
    P_\text{sig, Z}(\mathbf{Q_p})&=N_xN_yN_e2E(\mathbf{Q_p})\sin\chi(\mathbf{Q_p})\nonumber\\
    &=N_xN_yDd_{x,y}^22E(\mathbf{Q_p})\sin\chi(\mathbf{Q_p})
\end{align}
Since a DFT was taken the expression describes the power of each frequency bin with finite width $1/N_{x,y}d_{x,y}$ to the total signal. The factor of 2 should be noted as it has important implications for the signal to noise ratio calculated later in this section.
\subsubsection{Zernike phase contrast noise power}\label{Zernike_noise_pow_sect}
The noise on the detector, which is the noise in the signal intensity can be modelled as a random Poisson process \(X_{Z,I}(\mathbf{R_p})\). The subscript $I$ indicates that the Poisson process refers to the intensity recorded on the detector within a Zernike phase contrast microscope.
\begin{align}
    E\big[X_{Z, I}(\mathbf{R_p})\big]&=Dd_{x,y}^2\vert\psi_i(\mathbf{R_p})\vert^2\\
    \text{Var}(X_{Z, I}(\mathbf{R_p}))&=Dd_{x,y}^2\vert\psi_i(\mathbf{R_p})\vert^2
\end{align}
where \(D\) is the electron fluence given in \(e/\)\AA$^2$ and \(d_{x,y}\) is the real space pixel size given in \AA. $\vert\psi_i(\mathbf{R_p})\vert^2$ is the intensity of the electron wave on the detector as given in \eqref{general_phase_intensity} and \eqref{Zernike_intensity_ewave_real_space}. Because the noise is random, it is uncorrelated between different probe positions. That means the autocorrelation function $R_{X_{Z, I}X_{Z, I}}(\mathbf{R_p})$ is nonzero only at zero lag as indicated by the delta function below.
\begin{equation}
    R_{X_{Z, I}X_{Z, I}}(\mathbf{R_p})=\delta(\mathbf{R_p})Dd_{x,y}^2\overline{\vert\psi_i(\mathbf{R_p})\vert^2}
\end{equation}
$\overline{\vert\psi_i(\mathbf{R_p})\vert^2}$ is the mean signal across the different probe positions. Considering that $\phi(\mathbf{R_p})<<1$, $\overline{\vert\psi_i(\mathbf{R_p})\vert^2}\approx1$. The noise power spectrum is the Fourier transform of the autocorrelation function as stated by the Wiener–Khinchin theorem \citep{wiener1964time,Khintchine1934}.
\begin{align}
    NPS_{Z, I}(\mathbf{Q_p})&=\mathcal{F}\big\{\delta(\mathbf{R_p})Dd_{x,y}^2\big\}\\
    &=N_xN_yDd_{x,y}^2
\end{align}
As expected for Poisson noise, the noise power spectrum is not dependent on spatial frequency. It should also be noted that this is the noise power of the intensity of the signal. That means it is the square of the noise power of the signal itself. Therefore
\begin{equation}\label{zernike_noise_sq}
    P_\text{noise, Z}^2(\mathbf{Q_p})=NPS_{Z, I}=N_xN_yDd_{x,y}^2
\end{equation}
\subsubsection{Zernike phase contrast $\text{SNR}^2$}\label{Zernike_snr_sq_sect}
Combining \eqref{snr_from_power} and \eqref{zernike_noise_sq} an expression for the signal to noise ratio squared of Zernike phase contrast microscopy can now be constructed.
\begin{align}\label{snr_sq_zernike}
    \text{SNR}_{Z}^2(\mathbf{Q_p})&=\frac{P_\text{sig, Z}^2(\mathbf{Q_p})}{P_\text{noise, Z}^2(\mathbf{Q_p})}\nonumber\\\notag&=\frac{N_x^2N_y^2D^2d_{x,y}^44E(\mathbf{Q_p})^2\sin\chi(\mathbf{Q_p})^2}{N_xN_yDd_{x,y}^2}\notag\\&=N_xN_yDd_{x,y}^24E(\mathbf{Q_p})^2\sin\chi(\mathbf{Q_p})^2
\end{align}
This shows the effect of the factor of 2 in the \eqref{zernike_int_transmission}. The factor of 2 in the intensity transmission function leads to a factor of 4 in the square of the signal to noise ratio, as shown in \eqref{snr_sq_zernike}.
\subsubsection{Zernike phase contrast DQE}\label{Zernike_dqe_sect}
To calculate the DQE of Zernike phase contrast imaging, the ideal microscope is defined as the ideal Zernike phase contrast microscope. Such a microscope does not contain any apertures, partial incoherence or aberrations. Therefore, $E(\mathbf{Q_p})$ and $\sin\chi(\mathbf{Q_p})$ are equal one in this case.
\begin{equation}\label{snr_z_ideal_sq}
    \text{SNR}_{Z, \text{ideal}}^2(\mathbf{Q_p})=4N_xN_yDd_{x,y}^2
\end{equation}
Finally, \eqref{snr_z_ideal_sq} can be combined with \eqref{snr_sq_zernike} using \eqref{DQE} to find the expression for the DQE of Zernike phase contrast microscopy.
\begin{align}\label{Zernike_DQE}
    &DQE_Z(\mathbf{Q_p})=\frac{SNR_Z(\mathbf{Q_p})^2}{SNR_{Z, \text{ideal}}(\mathbf{Q_p})^2}\nonumber\\=&\frac{4\vert E(\mathbf{Q_p})\vert^2\vert\sin\chi(\mathbf{Q_p})\vert^2N_xN_yDd_{x,y}^2}{4N_xN_yDd_{x,y}^2}\notag\\
    =&\vert E(\mathbf{Q_p})\vert^2\vert\sin\chi(\mathbf{Q_p})\vert^2
\end{align}
\subsection{SSB ptychography}\label{SSB_ptych_analytical_sec}
This subsection focuses on obtaining an expression for the DQE of SSB ptychography. In segment \ref{SSB_ptych_sig_pow_sec} the signal power of SSB ptychography is derived and in segment \ref{SSB_ptych_noise_pow_sec} the noise power of Zernike phase contrast microscopy is derived. In \ref{SSB_snr_sq_sect} the two previous segments are combined to derive the signal to noise ratio of SSB ptychography. Finally, the detective quantum efficiency is derived in \ref{SSB_ideal_dqe_sect}.
\subsubsection{SSB ptychography signal power}\label{SSB_ptych_sig_pow_sec}
As in the case of Zernike phase contrast microscopy, the weak phase object approximation (WPO) is invoked to describe the specimen transmission function. A 4D STEM dataset $\vert M(\mathbf{K_f}, \mathbf{R_p}))\vert^2$ can then be described as the modulus-squared of the convolution of \(\Psi(\mathbf{K_f})\) with the aperture function \(A(\mathbf{K_f})\).
\begin{equation}
    \vert M(\mathbf{K_f}, \mathbf{R_p}))\vert^2=\frac{Dd_{x,y}^2}{N_\alpha}\int\int A(\mathbf{K_f'})\Psi(\mathbf{K_f}-\mathbf{K_f'})A^\ast(\mathbf{K_f''})\Psi^\ast(\mathbf{K_f}-\mathbf{K_f''})e^{i2\pi\mathbf{R_p}\cdot(\mathbf{K_f'}-\mathbf{K_f''})}d\mathbf{K_f'}d\mathbf{K_f''}
\end{equation}
where $D$ is the electron fluence given in $e/\text{\AA}^2$, $d_{x,y}$ is the probe step size given in \AA, and $N_\alpha$ is the number of detector pixels in the bright field disc in the detector plane. For a weak phase object the entire incoming fluence is distributed across the brightfield disc. \(\Psi(\mathbf{K_f})\) is the Fourier transform of \(\psi(\mathbf{R_p})\). \(\mathbf{K_f'}\) and \(\mathbf{K_f''}\) were introduced as dummy variables for the convolution. \(G(\mathbf{K_f},\mathbf{Q_p})\) can then be calculated by taking the Fourier transform. In \cite{RODENBURG1993304} it has been shown that if the specimen can be modelled as a WPO, \(G(\mathbf{K_f},\mathbf{Q_p})\) can be written as
\begin{equation}\label{G_wpo}
    G(\mathbf{K_f},\mathbf{Q_p}) = \frac{N_xN_yDd_{x,y}^2}{N_\alpha}\big(\vert A(\mathbf{K_f})\vert^2\delta(\mathbf{Q_p})+A(\mathbf{K_f})A^*(\mathbf{K_f}+\mathbf{Q_p})\Psi_s^*(-\mathbf{Q_p})+A^*(\mathbf{K_f})A(\mathbf{K_f}-\mathbf{Q_p})\Psi_s(\mathbf{Q_p})\big)
\end{equation}
where $\Psi(\mathbf{K_f})$ is split into the direct beam and the scattered beam $\Psi_s(\mathbf{K_f})$.
\begin{eqnarray}\label{psi_kf_scatter}
    \Psi(\mathbf{K_f})&=&\delta(\mathbf{K_f})+\Psi_s(\mathbf{K_f})\nonumber\\
    &=&\delta(\mathbf{K_f})-i\mathcal{F}\{\phi(\mathbf{r})\}
\end{eqnarray}

\begin{figure}[h]
  \centering
  \begin{subfigure}[b]{0.75\textwidth}
    \includegraphics[width=\textwidth]{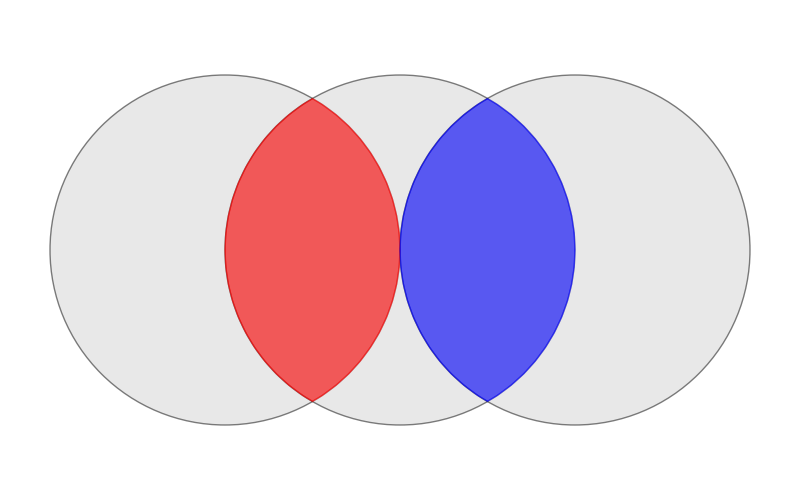}
    \caption{$Q_p=\alpha$}
    \label{fig:ssb_overlap_alpha}
  \end{subfigure}

  \vfill

  \begin{subfigure}[b]{0.4\textwidth}
    \includegraphics[width=\textwidth]{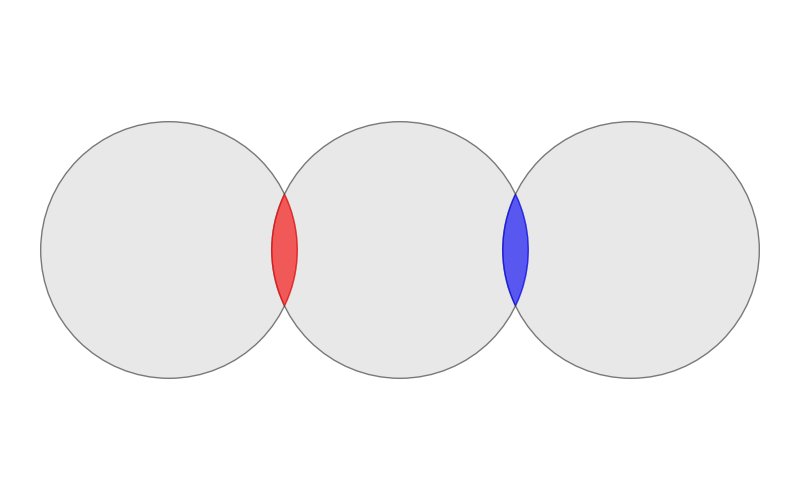}
    \caption{$Q_p=1.8\alpha$}
    \label{fig:ssb_overlap_1_8_alpha}
  \end{subfigure}
  \hfill
  \begin{subfigure}[b]{0.4\textwidth}
    \includegraphics[width=\textwidth]{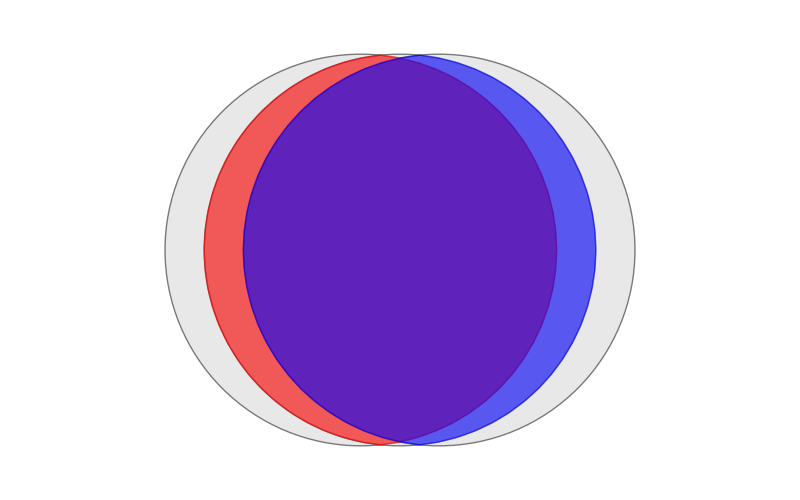}
    \caption{$Q_p=0.2\alpha$}
    \label{fig:ssb_overlap_0_2_alpha}
  \end{subfigure}

  \caption{Illustration of the double overlap and triple overlap region in $G(\mathbf{K_f},\mathbf{Q_p})$ at different spatial frequencies. The double overlap regions are shown in blue on the right and red on the left. The triple overlap region is shaded in purple.}
  \label{fig:ssb_do_to_overlap_illustration}
\end{figure}

For a weak phase object, it is possible to exclusively sum over the regions of $G(\mathbf{K_f},\mathbf{Q_p})$ that contain information about $\Psi_s(\mathbf{K_f})$. This technique is known as single sideband ptychography (SSB)\citep{PENNYCOOK2015160, YANG2015232}. The proportion of the area containing the signal, known as double overlap region, to the area of the bright field disc is known as the phase contrast transfer function (PCTF) of SSB ptychography\citep{YANG2015232}. The overlapping discs in $G(\mathbf{K_f},\mathbf{Q_p})$ and the double overlap regions are visualised in Figure \ref{fig:ssb_do_to_overlap_illustration}.
\begin{eqnarray}
    G_{SSB}(\mathbf{Q_p})&=&\frac{N_xN_yDd_{x,y}^2}{N_\alpha}\sum_{SSB,\mathbf{K}_{\mathbf{f},i}}\vert A(\mathbf{K}_{\mathbf{f},i})\vert^2\delta(\mathbf{Q_p})+A(\mathbf{K}_{\mathbf{f},i})A^*(\mathbf{K}_{\mathbf{f},i}+\mathbf{Q_p})\Psi_s^*(-\mathbf{Q_p})+A^*(\mathbf{K}_{\mathbf{f},i})A(\mathbf{K}_{\mathbf{f},i}-\mathbf{Q_p})\Psi_s(\mathbf{Q_p})\nonumber\\
    &=&N_xN_yDd_{x,y}^2 PCTF(\mathbf{Q_p})+N_xN_yDd_{x,y}^2 PCTF(\mathbf{Q_p})\Psi_s^*(-\mathbf{Q_p})
\end{eqnarray}
Even though $G_{SSB}(\mathbf{Q_p})$ is a representation of the exit wave, the normalisation factor $Dd_{x,y}^2$ suggests that it is in units of electrons. As units of electrons represent the signal power rather than the signal, an expression for the signal power of SSB electron ptychography can be constructed.
\begin{equation}\label{ssb_sig_pw}
    P_\text{sig, SSB}(\mathbf{Q_p})=N_xN_yDd_{x,y}^2 PCTF(\mathbf{Q_p})
\end{equation}
\subsubsection{SSB ptychography noise power}\label{SSB_ptych_noise_pow_sec}
The intensity of a recorded electron diffraction pattern at probe position $\mathbf{R_p}$ in a 4D dataset can be modelled as a random process \(X_{M, I}(\mathbf{K_f},\mathbf{R_p})\) comprising of a signal component \(E\big[X_{M, I}(\mathbf{K_f},\mathbf{R_p})\big]\) and a noise component $\eta_{M, I}(\mathbf{K_f},\mathbf{R_p})$. The subscript $I$ indicates that the Poisson distribution refers to the recorded intensity on the detector.
\begin{equation}\label{4D_stem_random_process}
    X_{M, I}(\mathbf{K_f},\mathbf{R_p})=E\big[X_{M, I}(\mathbf{K_f},\mathbf{R_p})\big]+\eta_{M, I}(\mathbf{K_f},\mathbf{R_p})
\end{equation}
When an electron diffraction pattern is recorded with a fluence of $D$ \(e/\)\AA$^2$, the expected value of electrons recorded is given by,
\begin{equation}\label{4DSTEM_M_expectation}
    E\big[X_{M, I}(\mathbf{K_f},\mathbf{R_p})\big]=\vert M(\mathbf{K_f}, \mathbf{R_p}))\vert^2
\end{equation}
If the noise is purely governed by a Poisson process 
\begin{equation}\label{4DSTEM_Mnoise}
    \text{Var}(X_{M, I}(\mathbf{K_f},\mathbf{R_p}))=\text{Var}(\eta_{M, I}(\mathbf{K_f},\mathbf{R_p}))=\vert M(\mathbf{K_f}, \mathbf{R_p}))\vert^2
\end{equation}

The next step in the SSB algorithm is a 2D Fourier transform with respect to the probe position $\mathbf{R_p}$. Expressions for the real and imaginary part of the result can be obtained as derived in Appendix \ref{Formal_noise_G}. As before in \ref{Zernike_noise_pow_sect} we have assumed minor variation within the CBED patterns. Therefore, the pixel intensity within the brightfield discs of the 4D-STEM dataset is approximated as the average fluence per pixel $\overline{\vert M(\mathbf{K_f}, \mathbf{R_p}))\vert^2}\approx\frac{Dd_{x,y}^2}{N_\alpha}$.
\begin{eqnarray}\label{Variance_G_real}
    \text{Var}(Re\{X_{G,I}(\mathbf{K_f},\mathbf{Q_p})\})&=&E\big[(Re\{\eta_{G,I}(\mathbf{K_f},\mathbf{Q_p})\})^2\big]\nonumber\\&=&\frac{Dd_{x,y}^2}{2N_\alpha}N_xN_y
\end{eqnarray}
\begin{eqnarray}\label{Variance_G_imag}
    \text{Var}(Im\{X_{G,I}(\mathbf{K_f},\mathbf{Q_p})\})&=&E\big[(Im\{\eta_{G,I}(\mathbf{K_f},\mathbf{Q_p})\})^2\big]\nonumber\\&=&\frac{Dd_{x,y}^2}{2N_\alpha}N_xN_y
\end{eqnarray}
The factor of $\frac{1}{2}$ that is introduced in \eqref{Variance_G_real} and \eqref{Variance_G_imag} can be interpreted as the noise power being split equally into the real and imaginary domains.

To find an expression for the variance of the resulting SSB summation over the detector plane, the noise of the sum over all $\mathbf{K}_{\mathbf{f}}$ values inside the double overlap region needs to be considered.
\begin{eqnarray}\label{Cumulative_SSB_variance_imag1}
    \text{Var}(Im\{X_{G:SSB, I}(\mathbf{Q_p})\})&=&E\Bigg[\Big(\sum_{SSB,\mathbf{K}_{\mathbf{f},i}}Im\{\eta_{G,I}(\mathbf{K}_{\mathbf{f},i},\mathbf{Q_p})\}\Big)^2\Bigg]\nonumber\\
    &=&\begin{aligned}[t]E\Bigg[\sum_{SSB,\mathbf{K}_{\mathbf{f},i}}\Big(&Im\{\eta_{G,I}(\mathbf{K}_{\mathbf{f},i},\mathbf{Q_p})\}\Big)^2\\&+2\sum_{\substack{SSB,\ \mathbf{K}_{\mathbf{f}, i},\ \mathbf{K}_{\mathbf{f}, j} \\[1pt] i < j}}Im\{\eta_{G,I}(\mathbf{K}_{\mathbf{f},i},\mathbf{Q_p})\}Im\{\eta_{G,I}(\mathbf{K}_{\mathbf{f},j},\mathbf{Q_p})\}\Bigg]
    \end{aligned}
\end{eqnarray}
The first term is simply the sum of all the individual variances of $Im\{\eta_{G,I}(\mathbf{K}_{\mathbf{f},i},\mathbf{Q_p})\}$ at the various $\mathbf{K}_{\mathbf{f},i}$s. The subscript $i$ indicates the different pixels corresponding to different $\mathbf{K}_{\mathbf{f}}$ values on the detector. The second term contains the cross terms. Considering that the noise at different $\mathbf{K}_{\mathbf{f},i}$s is independent from each other and that $E[Im\{\eta_{G,I}(\mathbf{K}_{\mathbf{f},i},\mathbf{Q_p})\}]=0$, the cross terms vanish and therefore the entire second term. Using Eq. \eqref{Variance_G_imag}, the variance of the imaginary part of the SSB summation can now be finalised.
\begin{eqnarray}\label{Cumulative_SSB_variance_imag_final}
    \text{Var}(Im\{X_{G:SSB, I}(\mathbf{Q_p})\})&=&\sum_{SSB,\mathbf{K}_{\mathbf{f},i}}E\Bigg[\Big(Im\{\eta_{G,I}(\mathbf{K}_{\mathbf{f},i},\mathbf{Q_p})\}\Big)^2\Bigg]\nonumber\\
    &=&N_{DO,i}(\mathbf{Q_p})E\Bigg[\Big(Im\{\eta_{G,I}(\mathbf{K}_{\mathbf{f}},\mathbf{Q_p})\}\Big)^2\Bigg]\nonumber\\
    &=&\frac{Dd_{x,y}^2}{2N_\alpha}N_xN_yN_{DO,i}(\mathbf{Q_p})\nonumber\\
    &=&\frac{Dd_{x,y}^2}{2}N_xN_yPCTF(\mathbf{Q_p})
\end{eqnarray}

where $PCTF(\mathbf{Q_p})$ is defined as the number of pixels in one double overlap region, $N_{DO,i}$, relative to the number of pixels in the bright field disc, $N_\alpha$, $\frac{N_{DO,i}(\mathbf{Q_p})}{N_\alpha}$.

The same analysis can be performed for the real part.
\begin{eqnarray}\label{Cumulative_SSB_variance_real}
    \text{Var}(\text{Re}\{X_{G:SSB, I}&(&\mathbf{Q_p})\})=\frac{Dd_{x,y}^2}{2N_\alpha}N_xN_yN_{DO,i}(\mathbf{Q_p})\nonumber\\&=&\frac{Dd_{x,y}^2}{2}N_xN_yPCTF(\mathbf{Q_p})
\end{eqnarray}

\eqref{wpoa_approx} shows that for a weak phase object, all the information about the potential is contained in the imaginary domain in real space. This is the reason why the final reconstruction is only formed from the imaginary component of $\mathcal{F}^{-1}\{G_{SSB}(\mathbf{Q_p})\}$. However, this does not necessarily correspond to the imaginary component of $G_{SSB}(\mathbf{Q_p})$. To isolate the component in frequency space that transforms to the purely imaginary part, we can use the property of any complex number $z$, that $\text{Im}(z)=\frac{z-z*}{2i}$.
\begin{equation}
    \text{Im}\{X_{G:SSB, I}(\mathbf{R_p})\}=\frac{1}{2i}(X_{G:SSB, I}(\mathbf{R_p})-X^*_{G:SSB, I}(\mathbf{R_p}))
\end{equation}
Using the definition $\mathcal{F}\{X_{G:SSB, I}(\mathbf{R_p})\}=X_{G:SSB, I}(\mathbf{Q_p})$ and Friedel's law $\mathcal{F}\{X_{G:SSB, I}(\mathbf{R_p})\}=X_{G:SSB, I}^*(-\mathbf{Q_p})$,
\begin{equation}
    \mathcal{F}\{\text{Im}\{X_{G:SSB, I}(\mathbf{R_p})\}\}=\frac{1}{2i}(X_{G:SSB, I}(\mathbf{Q_p})-X^*_{G:SSB, I}(-\mathbf{Q_p}))
\end{equation}
The noise power of the imaginary component in real space can now be determined.
\begin{align}
    \text{Var}(\mathcal{F}\{\text{Im}\{X_{G:SSB, I}(\mathbf{R_p})\}\})=E[\vert\frac{1}{2i}(X_{G:SSB, I}(\mathbf{Q_p})-X^*_{G:SSB, I}(-\mathbf{Q_p}))\vert^2]\nonumber\\
    =\frac{1}{4}\Big(E[\vert X_{G:SSB, I}(\mathbf{Q_p})\vert^2]+E[\vert X^*_{G:SSB, I}(\mathbf{-Q_p})\vert^2]-2\text{Re}\{E[X_{G:SSB, I}(\mathbf{Q_p})X_{G:SSB, I}(\mathbf{-Q_p})]\}\Big)
\end{align}
$X_{G:SSB, I}(\mathbf{Q_p})$ and $X_{G:SSB, I}(\mathbf{-Q_p})$ are independent since the SSB summation is performed over different regions in $\mathbf{K_f}$ space. Therefore, $E[X_{G:SSB, I}(\mathbf{Q_p})X_{G:SSB, I}(\mathbf{-Q_p})]=0$.
\begin{align}
    \text{Var}(\mathcal{F}\{\text{Im}\{X_{G:SSB, I}(\mathbf{R_p})\}\})=\frac{1}{4}\Big(E[\vert X_{G:SSB, I}(\mathbf{Q_p})\vert^2]+E[\vert X^*_{G:SSB, I}(\mathbf{-Q_p})\vert^2]\Big)\nonumber\\
    =\frac{1}{4}\Big(\text{Var}(\text{Re}\{X_{G:SSB, I}(\mathbf{Q_p})\})+\text{Var}(\text{Im}\{X_{G:SSB, I}(\mathbf{Q_p})\})+\text{Var}(\text{Re}\{X^*_{G:SSB, I}(\mathbf{-Q_p})\})+\text{Var}(\text{Im}\{X^*_{G:SSB, I}(\mathbf{-Q_p})\})\Big)
\end{align}
By using \eqref{Cumulative_SSB_variance_imag_final} and \eqref{Cumulative_SSB_variance_real}, a final expression for the variance in reciprocal space stemming from the noise in the imaginary part in real space can be constructed.
\begin{align}
    \text{Var}(\mathcal{F}\{\text{Im}\{X_{G:SSB, I}(\mathbf{R_p})\}\})=\frac{1}{4}\Big(2Dd_{x,y}^2N_xN_yPCTF(\mathbf{Q_p})\Big)=\frac{Dd_{x,y}^2}{2}N_xN_yPCTF(\mathbf{Q_p})\nonumber\\
    =\text{Var}(Im\{X_{G:SSB, I}(\mathbf{Q_p})\})
\end{align}
Even though, $G(\mathbf{K_f},\mathbf{Q_p}))$ is a complex quantity, it is the Fourier transform of $\vert M(\mathbf{K_f}, \mathbf{R_p}))\vert^2$, an intensity measured in units of electrons. Thus $G(\mathbf{K_f},\mathbf{Q_p}))$ is also in units of intensity. Therefore, \eqref{Cumulative_SSB_variance_imag_final} describes the noise power spectrum of the intensity. The noise power spectrum of the intensity is the equivalent to the square of the noise power spectrum of the signal.
\begin{eqnarray}\label{ssb_noise_sq}
    P_\text{noise, SSB}^2(\mathbf{Q_p})&=&\text{Var}(Im\{X_{G:SSB, I}(\mathbf{Q_p})\})\nonumber\\
    &=&\frac{Dd_{x,y}^2}{2}N_xN_yPCTF(\mathbf{Q_p})
\end{eqnarray}
\subsubsection{SSB ptychography $SNR^2$}\label{SSB_snr_sq_sect}
\eqref{ssb_sig_pw} and \eqref{ssb_noise_sq} can now be combined to find an expression for the square of the signal to noise ratio of SSB ptychography.
\begin{eqnarray}\label{ssb_snr_sq}
    \text{SNR}_{SSB}^2(\mathbf{Q_p})&=&\frac{P_\text{sig, SSB}^2(\mathbf{Q_p})}{P_\text{noise, SSB}^2(\mathbf{Q_p})}\nonumber\\&=&\frac{N_x^2N_y^2D^2d_{x,y}^4 PCTF(\mathbf{Q_p})^2}{\frac{Dd_{x,y}^2}{2}N_xN_yPCTF(\mathbf{Q_p})}\nonumber\\&=&2Dd_{x,y}^2N_xN_yPCTF(\mathbf{Q_p})
\end{eqnarray}
The prefactor of 2 in the square of the signal to noise ratio of SSB ptychography is due to split of the noise power into the real and imaginary domain. Since, the information about the potential lies only in the imaginary domain for the case of a weak phase object, the noise power in the real domain doesn't affect the reconstruction. This results in an effective doubling of the square of the signal to noise ratio of SSB ptychography.

\subsubsection{Ideal SSB ptychography DQE}\label{SSB_ideal_dqe_sect}
\eqref{snr_z_ideal_sq} can be combined with \eqref{ssb_snr_sq} using \eqref{DQE} to find the expression for the DQE of SSB ptychography in the absence of partial coherence and aberrations.
\begin{eqnarray}\label{SSB_DQE_ideal}
    DQE_{SSB, \text{ideal}}(\mathbf{Q_p})&=&\frac{SNR_{SSB, \text{ideal}}(\mathbf{Q_p})^2}{SNR_{Z, \text{ideal}}(\mathbf{Q_p})^2}\nonumber\\&=&\frac{2Dd_{x,y}^2N_xN_yPCTF(\mathbf{Q_p})}{4N_xN_yDd_{x,y}^2}\notag\\
    &=&\frac{PCTF(\mathbf{Q_p})}{2}
\end{eqnarray}

\begin{figure}[h]
\includegraphics[width=0.65\textwidth]{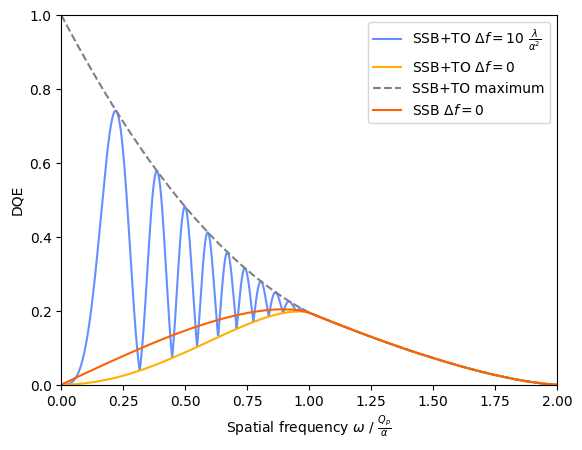}
\caption{Plot showing the DQE of SSB ptychography in orange compared to the DQE of SSB ptychography with additional information from the triple overlap region in blue. The dashed grey line indicates the maximum DQE for a weak phase object attainable for any given spatial frequency if the $-\mathbf{Q_p}$ and $+\mathbf{Q_p}$ beams display perfect constructive interference across the entire triple overlap region.\label{fig:SSB_TO_DQE}}
\end{figure}

This result shows that for in focus 4D-STEM imaging, the maximum detective quantum efficiency is reached at around 20.5\% at a spatial frequency of $\alpha$, where the $PCTF$ of SSB ptychography has its maximum. The plot of the DQE for ideal in-focus SSB is shown by the orange graph in Figure \ref{fig:SSB_TO_DQE}. For simplicity, spatial frequency $Q_p$ is expressed in units of angle (mrad) and normalised by the probe convergence semi-angle $\alpha$. For in focus ptychography the signal is purely located in the double overlap regions. When observing Figure \ref{fig:ssb_overlap_0_2_alpha} one can see that at low spatial frequencies the double overlap regions are small compared to the size of the brightfield disk due to a large triple overlap region. As spatial frequency increases, the double-overlap area grows and reaches its maximum close to $\alpha$. Exactly at $\alpha$, the triple-overlap region disappears (see Fig. \ref{fig:ssb_overlap_alpha}). Beyond that point, further increases in spatial frequency push the disks farther apart, so the double-overlap region begins to shrink again (see \ref{fig:ssb_overlap_1_8_alpha}), ultimately producing the characteristic SSB $PCTF$.

Out of focus, the signal is located in the triple overlap region as well as the double overlap region which leads to the oscillatory nature of the DQE below $\alpha$. This is discussed in more detail in section \ref{SSB_TO_DQE_sect}. The defocus parameter in Fig. \ref{fig:SSB_TO_DQE} is normalised, since the phase shift for a given defocus depends on the spatial frequency, normalisable by $\alpha$, and on the electron wavelength $\lambda$. Therefore, for a given normalised spatial frequency $\omega$, the phase shift due to defocus expressed as a multiple of $\lambda/\alpha^2$ stays constant under variation of both acceleration voltage and convergence angle.

A tempting way to improve the single sideband method would be to include the other sideband. For each $\mathbf{Q_p}$ value, the single sideband method uses either the outer or the inner sideband to perform the summation. The term inner sideband refers to the sideband that the vector $-\mathbf{Q_p}$ points towards from the origin in the $\mathbf{K_f}$ plane, $G((0,0),\mathbf{Q_p})$. The term outer sideband refers to the sideband located in the direction $\mathbf{Q_p}$ relative to the origin in the $\mathbf{K_f}$ plane, $G((0,0),\mathbf{Q_p})$.

One idea is to combine the outer sideband at $\mathbf{Q_p}$ with the inner sideband at $-\mathbf{Q_p}$ with an appropriate phase shift since both should contain the same signal information due to symmetry of the Fourier transform of the scattered component of the electron wave \citep{RODENBURG1993304}.
\begin{equation}\label{scattered_wave_symmetry}
    \Psi_s(\mathbf{Q_p})=-\Psi_s^*(-\mathbf{Q_p})
\end{equation}
However, this effect comes from the intrinsic property of the Fourier transform of any real function stated by Friedel's law. Therefore, when transforming the original dataset, not only the signal but also the detail of the noise in the two double overlap regions being summed is perfectly correlated. This results in simple doubling of the signal and the noise leaving the signal to noise ratio unaffected.

Alternatively, the summation could be performed by combining the two sidebands at any given $\mathbf{Q_p}$ by changing the phase of one sideband by $\pi$ before adding it to the other. When performing the sum this way, the pixels in the summation are not correlated. However, now the summed noise at $\mathbf{Q_p}$ and $-\mathbf{Q_p}$ is correlated resulting in the same noise power as described in the alternative summation. Both possible summation methods lead to the same signal to noise ratio as SSB ptychography and therefore no gains in the DQE can be achieved by including the second sideband. A mathematical analysis of the DQE of double sideband ptychography can be found in Appendix \ref{SSB_is_all_u_need}.

Even though improving the single sideband method by using the second double overlap region yields no benefit, including information from the triple overlap region can be beneficial under certain circumstances which will be explored in \ref{SSB_TO_DQE_sect}.

\subsection{SSB vs iCOM vs WDD comparison}
To compare the analytical results of the DQE of SSB ptychography to integrated centre of mass (iCOM) imaging\citep{LAZIC2016265} and wigner distribution deconvolution (WDD)\citep{rodenburg1992theory}, an empirical approach was taken. A set of 500 noise realisations of the 4DSTEM dataset were performed and reconstructed with both SSB, iCOM and WDD. The signal and noise power of the resulting phase images was then calculated.
\begin{eqnarray}
    \vert P_\text{signal}(\mathbf{Q_p})\vert^2&=&\langle\vert \mathcal{F}\{I(\mathbf{R_p})\}\vert^2\rangle\\
    \vert P_\text{noise}(\mathbf{Q_p})\vert^2&=&\langle\vert \mathcal{F}\{I(\mathbf{R_p})-\overline{I(\mathbf{R_p})}\}\vert^2\rangle
\end{eqnarray}
, where \(\mathcal{F}\{I(\mathbf{R_p})\}\) is the Fourier transform of a reconstructed 2D image, with \(\mathbf{R_p}\) being the two dimensional real space vector. The symbol \(\langle\rangle\) denotes the ensemble average over all noise realisations and $\overline{I(\mathbf{R_p})}$ denotes the mean reconstructed image in real space. In the limit of infinite noise realisations, this mean image represents the infinite dose ground truth image.
\begin{figure}[h]
\includegraphics[width=0.65\textwidth]{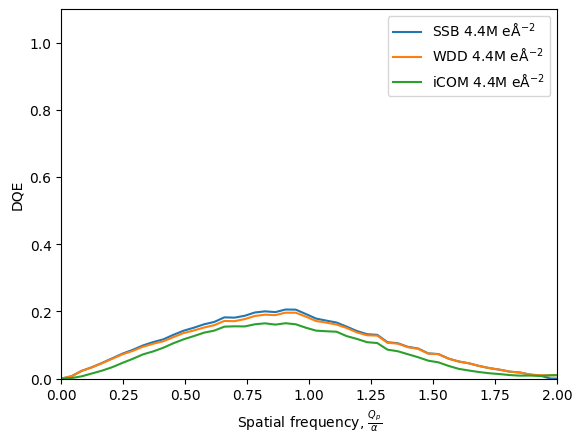}
\caption{Comparison of the DQE of SSB ptychography, WDD ptychography and iCOM based on 500 noise realisations for each technique. A single hydrogen atom was used as the sample. A convergence angle of 22 mrad was assumed with a step size of 0.15 \AA. A dose of 4.4M e\AA$^{-2}$ was used for each simulation.}\label{empirical_dqe_comp}
\end{figure}
Using equations \eqref{snr_from_power} and \eqref{DQE}, the frequency dependent DQEs can be obtained for the different techniques which are shown in Fig. \ref{empirical_dqe_comp}. As expected, there is almost no difference between the DQEs of WDD and SSB for a weak phase object. The absolute difference between the iCOM and the other algorithms is small, as might be expected since they are access the same information in the 4D data set. However, particularly at low spatial frequencies and close to the 2$\alpha$ limit, SSB and WDD show a significant percentage performance improvement compared to iCOM. Near these limits the double overlap region is small, as shown in Fig. \ref{fig:ssb_overlap_0_2_alpha} and Fig. \ref{fig:ssb_overlap_1_8_alpha}. SSB and WDD isolate precisely these small regions avoiding excess noise across the remainder of the bright field disk. iCOM on the other hand does not perform such noise filtering.
\subsection{DQE of defocused SSB, TO ptychography}\label{SSB_TO_DQE_sect}
Eq. \eqref{G_wpo} provides an expression for the expectation value of $G(\mathbf{K_f},\mathbf{Q_p})$. While we have so far only considered the case where SSB ptychography is used to reconstruct a 4D STEM dataset recorded in focus, it can also be adapted for defocus ptychography.
\begin{align}
    G(\mathbf{K_f},\mathbf{Q_p}) &= \frac{N_xN_yDd_{x,y}^2}{N_\alpha}\big(\vert A(\mathbf{K_f})\vert^2\delta(\mathbf{Q_p})+A(\mathbf{K_f})A^*(\mathbf{K_f}+\mathbf{Q_p})\Psi_s^*(-\mathbf{Q_p})+A^*(\mathbf{K_f})A(\mathbf{K_f}-\mathbf{Q_p})\Psi_s(\mathbf{Q_p})\big)\nonumber\\
    &= \frac{N_xN_yDd_{x,y}^2}{N_\alpha}\big(\vert A(\mathbf{K_f})\vert^2\delta(\mathbf{Q_p})-A(\mathbf{K_f})A^*(\mathbf{K_f}+\mathbf{Q_p})\Psi_s(\mathbf{Q_p})+A^*(\mathbf{K_f})A(\mathbf{K_f}-\mathbf{Q_p})\Psi_s(\mathbf{Q_p})\big)\nonumber\\
    &= \frac{N_xN_yDd_{x,y}^2}{N_\alpha}\bigg(\vert A(\mathbf{K_f})\vert^2\delta(\mathbf{Q_p})+\Psi_s(\mathbf{Q_p})\Big(A^*(\mathbf{K_f})A(\mathbf{K_f}-\mathbf{Q_p})-A(\mathbf{K_f})A^*(\mathbf{K_f}+\mathbf{Q_p})\Big)\bigg)
\end{align}
In the case of nonzero defocus, $A(\mathbf{K_f}, \mathbf{Q_p})$ is a complex quantity. Therefore, the expectation value can be nonzero in the triple overlap region. Therefore, the triple overlap region should be included in the summation. $A^*(\mathbf{K_f})A(\mathbf{K_f}-\mathbf{Q_p})-A(\mathbf{K_f})A^*(\mathbf{K_f}+\mathbf{Q_p})$ is a complex quantity. The scattered wave function $\Psi_s(\mathbf{Q_p})$ is an imaginary quantity. For the phase image to solely represent the scattered wave function, the modulus of $A^*(\mathbf{K_f})A(\mathbf{K_f}-\mathbf{Q_p})-A(\mathbf{K_f})A^*(\mathbf{K_f}+\mathbf{Q_p})$ needs to be taken by multiplying with $e^{-i\text{ arg(}A^*(\mathbf{K_f})A(\mathbf{K_f}-\mathbf{Q_p})-A(\mathbf{K_f})A^*(\mathbf{K_f}+\mathbf{Q_p})\text{)}}$. The notation $\sum_{SSB, TO}$ indicates that the sum in $\mathbf{K_f}$ space is taken over one of the double overlap regions as well as the entire triple overlap region.
\begin{align}
\begin{aligned}
    G_{SSB,TO}(\mathbf{Q_p})=\frac{N_xN_yDd_{x,y}^2}{N_\alpha}\sum_{SSB,TO}&\Bigg[e^{-i\text{ arg(}A^*(\mathbf{K_f})A(\mathbf{K_f}-\mathbf{Q_p})-A(\mathbf{K_f})A^*(\mathbf{K_f}+\mathbf{Q_p})\text{)}}\bigg( \vert A(\mathbf{K_f})\vert^2\delta(\mathbf{Q_p})\\&+\Psi_s(\mathbf{Q_p})\Big( A^*(\mathbf{K_f})A(\mathbf{K_f}-\mathbf{Q_p})-A(\mathbf{K_f})A^*(\mathbf{K_f}+\mathbf{Q_p})\Big)\bigg)\Bigg]
\end{aligned}\nonumber\\
\begin{aligned}
    G_{SSB,TO}(\mathbf{Q_p})=\frac{N_xN_yDd_{x,y}^2}{N_\alpha}\sum_{SSB,TO}\bigg(\vert A(\mathbf{K_f})\vert^2\delta(\mathbf{Q_p})+\Psi_s(\mathbf{Q_p})\Big\vert A^*(\mathbf{K_f})A(\mathbf{K_f}-\mathbf{Q_p})-A(\mathbf{K_f})A^*(\mathbf{K_f}+\mathbf{Q_p})\Big\vert\bigg)\label{defocus_ssb_step_by_step}
\end{aligned}
\end{align}
\subsubsection{Signal power of defocused SSB, TO ptychography}
The signal power of defocused SSB electron ptychography can now be expressed as
\begin{equation}\label{ssb_TO_sig_sq}
    P_\text{sig, SSB, TO}(\mathbf{Q_p})=\frac{N_xN_yDd_{x,y}^2}{N_\alpha}\sum_{SSB,TO}\Big\vert A^*(\mathbf{K_f})A(\mathbf{K_f}-\mathbf{Q_p})-A(\mathbf{K_f})A^*(\mathbf{K_f}+\mathbf{Q_p})\Big\vert
\end{equation}
\subsubsection{Noise power of defocused SSB, TO ptychography}
In Eq. \eqref{random_process_noise_G} it was shown that through modelling the 4D STEM acquisition as a random process, its Fourier transform can be described as,
\begin{equation}
    X_{G,I}(\mathbf{K_f},\mathbf{Q_p})=E\big[X_{G, I}(\mathbf{K_f},\mathbf{Q_p})\big]+\eta_{G, I}(\mathbf{K_f},\mathbf{Q_p})\nonumber
\end{equation}
Applying the necessary modification to the 4D STEM dataset shown above,
\begin{align}
    X_{G,TO\text{ sum prep},I}(\mathbf{K_f},\mathbf{Q_p})=e^{-i\text{ arg(}A^*(\mathbf{K_f})A(\mathbf{K_f}-\mathbf{Q_p})-A(\mathbf{K_f})A^*(\mathbf{K_f}+\mathbf{Q_p})\text{)}} \Big(E\big[X_{G, I}(\mathbf{K_f},\mathbf{Q_p})\big]+\eta_{G, I}(\mathbf{K_f},\mathbf{Q_p})\Big)
\end{align}
This can be simplified using Eq. \eqref{defocus_ssb_step_by_step}
\begin{align}
    X_{G,TO\text{ sum prep},I}(\mathbf{K_f},\mathbf{Q_p})=\frac{N_xN_yDd_{x,y}^2}{N_\alpha}\bigg(\vert A(\mathbf{K_f})\vert^2\delta(\mathbf{Q_p})+\Psi_s(\mathbf{Q_p})\Big\vert A^*(\mathbf{K_f})A(\mathbf{K_f}-\mathbf{Q_p})-A(\mathbf{K_f})A^*(\mathbf{K_f}+\mathbf{Q_p})\Big\vert\bigg)\nonumber\\+e^{-i\text{ arg(}A^*(\mathbf{K_f})A(\mathbf{K_f}-\mathbf{Q_p})-A(\mathbf{K_f})A^*(\mathbf{K_f}+\mathbf{Q_p})\text{)}}\eta_{G, I}(\mathbf{K_f},\mathbf{Q_p})\label{4D_STEM_TO_sumprep_random_proc}
\end{align}
Given the complex nature of Eq. \eqref{4D_STEM_TO_sumprep_random_proc}, variance of the real and imaginary components needs to be considered separately. Since 
\begin{eqnarray}
    \text{Var}\bigg(Re\Big\{\frac{N_xN_yDd_{x,y}^2}{N_\alpha}\Psi_s(\mathbf{Q_p})\Big\vert A^*(\mathbf{K_f})A(\mathbf{K_f}-\mathbf{Q_p})-A(\mathbf{K_f})A^*(\mathbf{K_f}+\mathbf{Q_p})\Big\vert\Big\}\bigg)\nonumber\\=\text{Var}\bigg(Im\Big\{\frac{N_xN_yDd_{x,y}^2}{N_\alpha}\Psi_s(\mathbf{Q_p})\Big\vert A^*(\mathbf{K_f})A(\mathbf{K_f}-\mathbf{Q_p})-A(\mathbf{K_f})A^*(\mathbf{K_f}+\mathbf{Q_p})\Big\vert\Big\}\bigg)=0
\end{eqnarray}
\begin{eqnarray}
    \text{Var}(Re\{X_{G,TO\text{ sum prep},I}(\mathbf{K_f},\mathbf{Q_p})\})=E\big[(Re\{e^{-i\text{ arg(}A^*(\mathbf{K_f})A(\mathbf{K_f}-\mathbf{Q_p})-A(\mathbf{K_f})A^*(\mathbf{K_f}+\mathbf{Q_p})\text{)}}\eta_{G,I}(\mathbf{K_f},\mathbf{Q_p})\})^2\big]\label{Variance_G_real_TO_sumprep}\\
    \text{Var}(Im\{X_{G,TO\text{ sum prep},I}(\mathbf{K_f},\mathbf{Q_p})\})=E\big[(Im\{e^{-i\text{ arg(}A^*(\mathbf{K_f})A(\mathbf{K_f}-\mathbf{Q_p})-A(\mathbf{K_f})A^*(\mathbf{K_f}+\mathbf{Q_p})\text{)}}\eta_{G,I}(\mathbf{K_f},\mathbf{Q_p})\})^2\big]\label{Variance_G_imag_TO_sumprep}
\end{eqnarray}
For two arbitrary complex numbers $z$ and $w$, $(\text{Re}\{zw\})^2=(\text{Re}(z)\text{Re}(w)-\text{Im}(z)\text{Im}(w))^2$. Similarly, $(\text{Im}\{zw\})^2=(\text{Re}(z)\text{Im}(w)+\text{Im}(z)\text{Re}(w))^2$.

Using these expressions the variance of the real and imaginary components of the phase corrected $X_{G,TO\text{ sum prep},I}(\mathbf{K_f},\mathbf{Q_p})$ can now be determined as shown in Appendix \ref{TO_phase_corr_noise}.
\begin{align}
    \text{Var}(\text{Re}\{X_{G,TO\text{ sum prep},I}(\mathbf{K_f},\mathbf{Q_p})\})=\frac{Dd_{x,y}^2}{2N_\alpha}N_xN_y\\
    \text{Var}(\text{Im}\{X_{G,TO\text{ sum prep},I}(\mathbf{K_f},\mathbf{Q_p})\})=\frac{Dd_{x,y}^2}{2N_\alpha}N_xN_y
\end{align}

To determine the variance of the summation, the sum of the variances of the double and triple overlap regions needs to be taken. Since through the corrections, the information content lies purely in the imaginary domain, the remaining section will focus on the imaginary noise power. This is equivalent to ignoring the noise power in the DC component as no signal is situated there in the case of a weak phase object.

As in the case of double sideband ptychography described above, there are also two methods of summing over the triple overlap region. One can either sum over the inner half of the triple overlap region at $\mathbf{Q_p}$ and combine it with a summation over the outer half at $-\mathbf{Q_p}$ or vice versa (option I). Alternatively one can sum both the inner and outer section of the triple overlap region at any given $\mathbf{Q_p}$ (option II). In Appendix \ref{app_symm_xgto_sumprep} it is shown that the two ways of summing the triple overlap region produce the exact same image. Therefore, in the following discussion we choose to sum as described in option I in Appendix \ref{app_symm_xgto_sumprep}.
\begin{align}\label{SSB_TO_def_sum_I}
    \text{Var}(\text{Im}\{X_{G:SSB,TO,I}(\mathbf{Q_p})\})=E\Bigg[\bigg(\sum_{SSB,TO}\text{Im}\{X_{G,TO\text{ sum prep},I}(\mathbf{K_f},\mathbf{Q_p})\}\bigg)^2\Bigg]\nonumber\\
    =E\Bigg[\bigg(\sum_{SSB}\text{Im}\{X_{G,TO\text{ sum prep},I}(\mathbf{K_f},\mathbf{Q_p})\}+\sum_{TO,i}\text{Im}\{X_{G,TO\text{ sum prep},I}(\mathbf{K_f},\mathbf{Q_p})\}+\sum_{TO,o}\text{Im}\{-X_{G,TO\text{ sum prep},I}^*(\mathbf{K_f},-\mathbf{Q_p})\}\bigg)^2\Bigg]
\end{align}
Using the symmetry properties of $X_{G,TO\text{ sum prep},I}(\mathbf{K_f},\mathbf{Q_p})$ established in \ref{app_symm_xgto_sumprep}, and the fact that for any complex number $z$, $\text{Im}(z)=\text{Im}(-z^*)$,
\begin{align}\label{SSB_TO_def_sum_II}
    \text{Var}(\text{Im}\{X_{G:SSB,TO,I}(\mathbf{Q_p})\})=E\Bigg[\bigg(\sum_{SSB}\text{Im}\{X_{G,TO\text{ sum prep},I}(\mathbf{K_f},\mathbf{Q_p})\}+2\sum_{TO,i}\text{Im}\{X_{G,TO\text{ sum prep},I}(\mathbf{K_f},\mathbf{Q_p})\}\bigg)^2\Bigg]
\end{align}
Since the sum over the inner half of the triple overlap region at $\mathbf{Q_p}$ is not correlated with either double overlap region,
\begin{align}\label{SSB_TO_def_sum_III}
    \text{Var}(\text{Im}\{X_{G:SSB,TO,I}(\mathbf{Q_p})\})=E\Bigg[\bigg(\sum_{SSB}\text{Im}\{X_{G,TO\text{ sum prep},I}(\mathbf{K_f},\mathbf{Q_p})\}\bigg)^2+4\bigg(\sum_{TO,i}\text{Im}\{X_{G,TO\text{ sum prep},I}(\mathbf{K_f},\mathbf{Q_p})\}\bigg)^2\Bigg]
\end{align}
The pixels within a sideband are independent of each other. The pixels within the inner half of the triple overlap region are also not correlated with each other. Using the linearity of the expectation operator,
\begin{align}\label{SSB_TO_def_sum_IV}
    \text{Var}(\text{Im}\{X_{G:SSB,TO,I}(\mathbf{Q_p})\})=E\Bigg[\sum_{SSB}\Big(\text{Im}\{X_{G,TO\text{ sum prep},I}(\mathbf{K_f},\mathbf{Q_p})\}\Big)^2+4\sum_{TO,i}\Big(\text{Im}\{X_{G,TO\text{ sum prep},I}(\mathbf{K_f},\mathbf{Q_p})\}\Big)^2\Bigg]\nonumber\\
    =\sum_{SSB}E\Bigg[\Big(\text{Im}\{X_{G,TO\text{ sum prep},I}(\mathbf{K_f},\mathbf{Q_p})\}\Big)^2\Bigg]+4\sum_{TO,i}E\Bigg[\Big(\text{Im}\{X_{G,TO\text{ sum prep},I}(\mathbf{K_f},\mathbf{Q_p})\}\Big)^2\Bigg]\nonumber\\
    =\sum_{SSB}\text{Var}(\text{Im}\{X_{G,TO\text{ sum prep},I}(\mathbf{K_f},\mathbf{Q_p})\})+4\sum_{TO,i}\text{Var}(\text{Im}\{X_{G,TO\text{ sum prep},I}(\mathbf{K_f},\mathbf{Q_p})\})\nonumber\\
    =\frac{Dd_{x,y}^2}{2N_\alpha}N_xN_y\big(N_{DO,i}(\mathbf{Q_p})+4N_{TO,i}(\mathbf{Q_p})\big)
\end{align}
\begin{eqnarray}\label{ssb_TO_noise_sq}
    P_\text{noise, SSB, TO}^2(\mathbf{Q_p})&=&\text{Var}(\text{Im}\{X_{G:SSB,TO,I}(\mathbf{Q_p})\})\nonumber\\
    &=&\frac{Dd_{x,y}^2}{2N_\alpha}N_xN_y\big(N_{DO,i}(\mathbf{Q_p})+4N_{TO,i}(\mathbf{Q_p})\big)
\end{eqnarray}
\subsubsection{\(SNR^2\) of defocused SSB, TO ptychography}
Through combining the expressions \eqref{ssb_TO_sig_sq} and \eqref{ssb_TO_noise_sq}, the square of the signal to noise ratio can be derived.
\begin{align}\label{SSB_TO_SNR_sq}
    SNR_{SSB, TO}^2(\mathbf{Q_p})&=\frac{\bigg(\frac{N_xN_yDd_{x,y}^2}{N_\alpha}\sum_{SSB,TO}\Big\vert A^*(\mathbf{K_f})A(\mathbf{K_f}-\mathbf{Q_p})-A(\mathbf{K_f})A^*(\mathbf{K_f}+\mathbf{Q_p})\Big\vert\bigg)^2}{\frac{Dd_{x,y}^2}{2N_\alpha}N_xN_y\big(N_{DO,i}(\mathbf{Q_p})+4N_{TO,i}(\mathbf{Q_p})\big)}\nonumber\\
    &=\frac{2N_xN_yDd_{x,y}^2\bigg(\sum_{SSB,TO}\Big\vert A^*(\mathbf{K_f})A(\mathbf{K_f}-\mathbf{Q_p})-A(\mathbf{K_f})A^*(\mathbf{K_f}+\mathbf{Q_p})\Big\vert\bigg)^2}{N_\alpha(N_{DO,i}(\mathbf{Q_p})+4N_{TO,i}(\mathbf{Q_p}))}
\end{align}
\subsubsection{DQE of defocused SSB, TO ptychography}
Finally, the square of the signal to noise ratio of SSB ptychography combined with the triple overlap, \eqref{SSB_TO_SNR_sq}, can be combined with that of Zernike phase contrast microscopy, \eqref{snr_sq_zernike}, to yield an expression for the DQE of SSB and triple overlap ptychography.
\begin{align}\label{SSB_TO_DQE}
    DQE_{SSB, TO}(\mathbf{Q_p})&=\frac{\frac{\frac{2N_xN_yDd_{x,y}^2}{N_\alpha}\bigg(\sum_{SSB,TO}\Big\vert A^*(\mathbf{K_f})A(\mathbf{K_f}-\mathbf{Q_p})-A(\mathbf{K_f})A^*(\mathbf{K_f}+\mathbf{Q_p})\Big\vert\bigg)^2}{N_{DO,i}(\mathbf{Q_p})+4N_{TO,i}(\mathbf{Q_p})}}{4N_xN_yDd_{x,y}^2}\nonumber\\
    &=\frac{\bigg(\sum_{SSB,TO}\Big\vert A^*(\mathbf{K_f})A(\mathbf{K_f}-\mathbf{Q_p})-A(\mathbf{K_f})A^*(\mathbf{K_f}+\mathbf{Q_p})\Big\vert\bigg)^2}{2N_\alpha\big(N_{DO,i}(\mathbf{Q_p})+4N_{TO,i}(\mathbf{Q_p})\big)}
\end{align}
This can be rewritten for the sum to take place over both sidebands and both sides of the triple overlap region, $N_{DO,TO}(\mathbf{Q_p})$.
\begin{align}\label{SSB_TO_DQE_2}
    DQE_{SSB, TO}(\mathbf{Q_p})&=\frac{\bigg(\sum_{DSB,TO}\Big\vert A^*(\mathbf{K_f})A(\mathbf{K_f}-\mathbf{Q_p})-A(\mathbf{K_f})A^*(\mathbf{K_f}+\mathbf{Q_p})\Big\vert\bigg)^2}{2N_\alpha\big(4N_{DO,i}(\mathbf{Q_p})+4N_{TO,i}(\mathbf{Q_p})\big)}\nonumber\\&=\frac{\bigg(\sum_{DSB,TO}\Big\vert A^*(\mathbf{K_f})A(\mathbf{K_f}-\mathbf{Q_p})-A(\mathbf{K_f})A^*(\mathbf{K_f}+\mathbf{Q_p})\Big\vert\bigg)^2}{4N_\alpha N_{DO,TO}(\mathbf{Q_p})}
\end{align}
The impact of including the triple overlap region under defocus when compared to standard SSB ptychography is shown in Fig. \ref{fig:SSB_TO_DQE}. One can see that at spatial frequencies above 1$\alpha$ there is no change in the DQE when including the triple overlap region. At lower spatial frequencies the DQE oscillates under defocus due to the constructive and destructive interference of the $-\mathbf{Q_p}$ and $+\mathbf{Q_p}$ beams. This can lead to an improvement of the DQE at many spatial frequencies. However, at the spatial frequencies where destructive interference eliminates the signal in the triple overlap, including it in the summation simply increases the noise and therefore leads to a lower DQE value. This is particularly visible in the extreme case of zero defocus where the triple overlap is dominated by destructive interference of the $-\mathbf{Q_p}$ and $+\mathbf{Q_p}$ beams at all spatial frequencies leading to a decreased DQE at all spatial frequencies from 0 to 1$\alpha$ when including the triple overlap region. The figure also shows the theoretical maximum envelope which represents perfect constructive interference across the entire triple overlap region. As spatial frequency tends to zero, the triple overlap region approaches the size of the entire bright field disk. In that case all electrons maximally contribute to the final image and the DQE tends to 1.
\subsection{Effect of partial coherence on SSB DQE}
For the case of Zernike phase contrast the previously stated envelope function $E(\mathbf{Q_p})$ includes the envelopes for temporal partial incoherence and spatial partial incoherence. In the case of SSB ptychography, the envelopes are 4 dimensional and need to be applied to $G(\mathbf{Q_p},\mathbf{K_f})$ \citep{NELLIST199461}.

The chromatic envelope in particular is both $\mathbf{K_f}$  and $\mathbf{Q_p}$ dependent.
\begin{equation}\label{eq_chr_envelope}
    E_\text{chr}^G(\mathbf{K_f}, \mathbf{Q_p})=e^{-\frac{1}{2}\pi^2\lambda^2\delta^2\big[2\mathbf{K_f}\cdot\mathbf{Q_p}+\vert\mathbf{Q_p}\vert^2\big]^2}
\end{equation}
\subsubsection{SSB signal power under partial coherence}
Since the noise remains Poisson noise, the expectation value of \(G(\mathbf{K_f}, \mathbf{Q_p})\) needs to be scaled by the fluence and the constants corresponding to the DFT as stated previously.
\begin{eqnarray}
    G(\mathbf{K_f}, \mathbf{Q_p})_\text{PC}=\frac{N_xN_yDd_{x,y}^2}{N_\alpha}E_\text{sou}^G(\mathbf{Q_p})&\big(&\vert A(\mathbf{K_f})\vert^2\delta(\mathbf{Q_p})\nonumber\\
    &+&E_\text{chr}^G(\mathbf{K_f}, -\mathbf{Q_p})A(\mathbf{K_f})A^*(\mathbf{K_f}+\mathbf{Q_p})\Psi_s^*(-\mathbf{Q_p})\nonumber\\
    &+&E_\text{chr}^G(\mathbf{K_f}, \mathbf{Q_p})A^*(\mathbf{K_f})A(\mathbf{K_f}-\mathbf{Q_p})\Psi_s(\mathbf{Q_p})\big)
\end{eqnarray}
Considering that $E_\text{sou}^G(\mathbf{Q_p})$ does not depend on $\mathbf{K_f}$ it can be moved outside of the summation over $\mathbf{K_f}$. Inside the double overlap region, $\vert A(\mathbf{K_f})\vert^2+A(\mathbf{K_f})A^*(\mathbf{K_f}+\mathbf{Q_p})\Psi_s^*(-\mathbf{Q_p})+A^*(\mathbf{K_f})A(\mathbf{K_f}-\mathbf{Q_p})\Psi_s^*(\mathbf{Q_p})$ equals $1+\Psi_s^*(\mathbf{Q_p})$ or $1+\Psi_s^*(-\mathbf{Q_p})$ depending on the sideband over which the sum is taken over. Since $1+\Psi_s^*(-\mathbf{Q_p})$ has no $\mathbf{K_f}$ dependence,
\begin{eqnarray}
    G_{SSB}(\mathbf{Q_p})_\text{PC}&=&\frac{N_xN_yDd_{x,y}^2}{N_\alpha}E_\text{sou}^G(\mathbf{Q_p})\sum_{SSB:\mathbf{K_f}}E_\text{chr}^G(\mathbf{K_f}, -\mathbf{Q_p})\\\notag
    &+&\frac{N_xN_yDd_{x,y}^2}{N_\alpha}E_\text{sou}^G(\mathbf{Q_p})\Psi_s^*(-\mathbf{Q_p})\sum_{SSB:\mathbf{K_f}}E_\text{chr}^G(\mathbf{K_f}, -\mathbf{Q_p})
\end{eqnarray}
Therefore, the signal power of SSB ptychography under partial coherence can be written as
\begin{equation}\label{ssb_pc_sig_sq}
    P_{\text{sig, SSB PC}}(\mathbf{Q_p})=\frac{N_xN_yDd_{x,y}^2}{N_\alpha}E_\text{sou}^G(\mathbf{Q_p})\sum_{SSB:\mathbf{K_f}}E_\text{chr}^G(\mathbf{K_f}, -\mathbf{Q_p})\text{.}
\end{equation}
\subsubsection{SSB ptychography noise power under partial coherence}
Under partial coherence the type of the noise on the diffraction patterns remains the same, Poisson noise. Partial coherence affects the interference features present in $G(\mathbf{K_f},\mathbf{Q_p})$, and therefore the expectation value in $G(\mathbf{K_f},\mathbf{Q_p})$. However, as previously discussed, the Poisson noise is incident on the detector. The assumption in the 4DSTEM dataset as well as in the case of HRTEM is that the variance across the detector is constant in the weak phase approximation. The effect of partial coherence on the expectation value of the collected diffraction patterns is therefore negligible. The square of the noise power under partial coherence is therefore described by Eq. \eqref{ssb_noise_sq}.
\subsubsection{SSB \(SNR^2\) under partial coherence}
When combining \eqref{snr_from_power}, \eqref{ssb_pc_sig_sq} and \eqref{ssb_noise_sq} this leads to the following equation for the square of the signal to noise ratio of SSB ptychography under partial coherence.
\begin{eqnarray}\label{SSB_SNR_sq_PC}
    SNR_{SSB, PC}(\mathbf{Q_p})^2&=&\frac{\Big( \frac{Dd_{x,y}^2}{N_\alpha}N_xN_yE_\text{sou}^G(\mathbf{Q_p})\sum_{SSB:\mathbf{K_f}}E_\text{chr}^G(\mathbf{K_f}, -\mathbf{Q_p})\Big)^2}{\frac{Dd_{x,y}^2}{2}N_xN_yPCTF(\mathbf{Q_p})}\notag\\
    &=&2\frac{Dd_{x,y}^2}{N_\alpha^2}N_xN_y\frac{\Big(E_\text{sou}^G(\mathbf{Q_p})\sum_{SSB:\mathbf{K_f}}E_\text{chr}^G(\mathbf{K_f}, -\mathbf{Q_p})\Big)^2}{PCTF(\mathbf{Q_p})}
\end{eqnarray}
\subsubsection{SSB DQE under partial coherence}
\begin{figure}[h]
\includegraphics[width=0.65\textwidth]{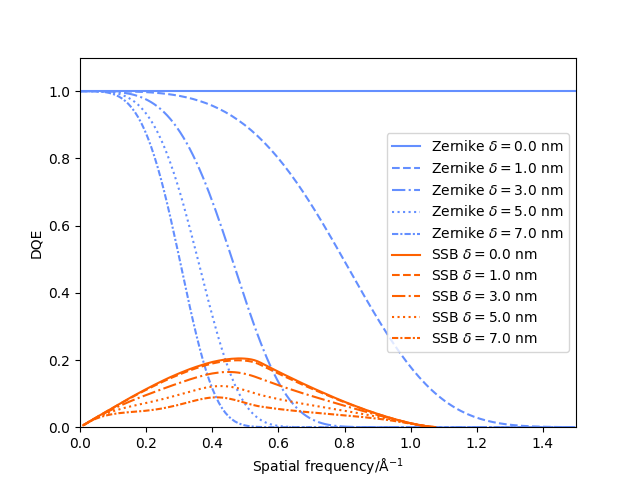}
\caption{Comparison of the DQE of SSB ptychography against Zernike phase contrast microscopy under partial temporal coherence. While even small defocus spreads of a few nanometres lead to the total loss of high spatial frequency information for Zernike microscopy, ptychography retains efficiency throughout its entire spatial frequency range. Partial temporal coherence dampens and shifts the peak in the DQE of SSB ptychography.\label{fig:dqe_zernike_ssb_pc}}
\end{figure}

An expression for the DQE of SSB ptychography under partial coherence can be obtained from combining \eqref{SSB_SNR_sq_PC} with \eqref{snr_sq_zernike} using \eqref{DQE}.
\begin{align}
    DQE_{SSB, PC}(\mathbf{Q_p})&=\frac{SNR_{SSB, PC}(\mathbf{Q_p})^2}{SNR_{Z, \text{ideal}}(\mathbf{Q_p})^2}=\frac{2\frac{Dd_{x,y}^2}{N_\alpha^2}N_xN_y\frac{\Big(E_\text{sou}^G(\mathbf{Q_p})\sum_{SSB:\mathbf{K_f}}E_\text{chr}^G(\mathbf{K_f}, -\mathbf{Q_p})\Big)^2}{PCTF(\mathbf{Q_p})}}{4N_xN_yDd_{x,y}^2}\\\notag
    &=\frac{\Big(E_\text{sou}^G(\mathbf{Q_p})\sum_{SSB:\mathbf{K_f}}E_\text{chr}^G(\mathbf{K_f}, -\mathbf{Q_p})\Big)^2}{2N_\alpha^2 PCTF(\mathbf{Q_p})}\\\notag
\end{align}
A comparison of the impact of partial temporal coherence on SSB ptychography and Zernike phase contrast microscopy is shown in Figure \ref{fig:dqe_zernike_ssb_pc}.  Here we have assumed a CTEM with a perfect phase plate to allow the effects of partial temporal coherence to be seen clearly. While HRTEM has a theoretical maximum DQE of 100\%, even small amounts of partial temporal coherence cause the total loss of high spatial frequency information. Ptychography is also affected by temporal incoherence. The peak DQE is damped while efficiency is retained at high spatial frequencies. The origin of this can be explained on the basis of the achromatic lines located in the centres of the double overlap regions which have been discussed in previous studies \citep{NELLIST199461, YANG2015232}. At low spatial frequencies the achromatic line is much broader which can be explained based on the $\vert\mathbf{Q_p}\vert^2$ term in Eq. \eqref{eq_chr_envelope}, leading to a reduced impact of partial temporal coherence. As spatial frequency increases, the double overlap region increases (see Fig. \ref{fig:ssb_do_to_overlap_illustration}) and the achromatic line becomes narrower proportional to 1/$\alpha$. The signal is located in a smaller fraction of the double overlap region, yet the noise from the entire double overlap region still contributes to the reconstruction. Therefore, the largest impact of partial temporal coherence on ptychography is seen near the maximum double overlap region size. Increasing the spatial frequency even further, the double overlap region decreases and the achromatic line narrows further. However, since it is located in the centre of the double overlap region, it represents a large fraction of the total area of the double overlap region at high spatial frequencies. Therefore, partial temporal coherence has a low impact on electron ptychography at high spatial frequencies close to 2$\alpha$.

\begin{figure}[h]
\includegraphics[width=0.65\textwidth]{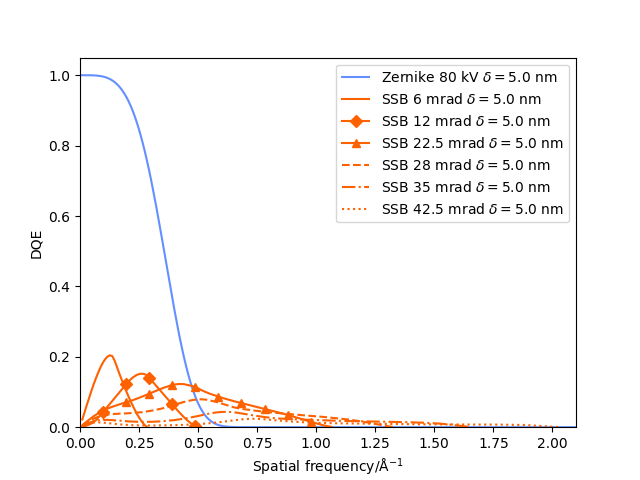}
\caption{Comparison of the DQE of SSB ptychography at different convergence angles against Zernike phase contrast microscopy with a defocus spread of 5 nm. At an assumed accelerating voltage of 80 kV this corresponds to an electron energy range of 0.4 eV if instabilities in the high voltage supply and the lens currents are neglected. The impact of partial temporal coherence on DQE increases with the convergence angle. Nevertheless, high spatial frequency information is still retained even at the high convergence angles where the impact of partial dampens the DQE peak to only a few percent. \label{fig:dqe_zernike_ssb_pc_alpha}}
\end{figure}

The impact of partial coherence on ptychography increases as the convergence angle is increased. This can be seen in Fig. \ref{fig:dqe_zernike_ssb_pc_alpha}. The 4D partial temporal coherence envelope acting on $G(\mathbf{Q_p},\mathbf{K_f})$ limits the signal in the double overlap region to the achromatic line, while the noise still affects the entire region. Increasing the convergence semi-angle $\alpha$ results in a wider double overlap region for the same spatial frequency. This means that the region around the achromatic line represents a smaller proportion of the total double overlap region for a higher convergence angle, resulting in a larger impact on the DQE. It should be noted though that even though DQE is significantly dampened, the high spatial frequency information is not entirely lost under partial coherence in ptychography unlike in the case of Zernike phase contrast microscopy for the reasons mentioned above.
\section{Discussion}
The maximum DQE of around 20.5 \% for in-focus ptychography of a weak-phase object demonstrated here does not compare well with the ideal imaging case of the Zernike phase plate.  An immediate question is whether this performance is simply related to the image retrieval method used here, which is the single side-band method.  We emphasise again that for in-focus, weak-phase ptychography, the only available information lies in the double overlap region and that this region is fully utilised in the single side-band approach.  Zernike phase imaging makes use of triple interference between the direct beam and the two diffracted beams at equal but opposite diffraction angles which both arise from a single spatial frequency in the sample.  For zero aberrations, such triple interference gives no contrast in 4D STEM, and the single side-band method relies on isolating just the interference between the direct beam and one scattered beam.  As noted previously by \cite{CDwyer2024}, this leads to a maximum of 50\% in signal compared to triple-beam interference.

To support the conclusion that an alternative reconstruction method could not improve the DQE, we note that the iCOM approach produces a very similar curve, with a slightly lower maximum but significantly worsened relative performance near the limits of transfer at high and low spatial frequencies.  The lower DQE for iCOM can be explained by the SSB approach being selective about where in the detector the signal is located, so has a somewhat enhanced noise reduction compared to iCOM especially when the double overlap regions are small.

In order to access information in the triple overlap region, a large aberration (or indeed a phase plate in the illumination aperture) is required.  High levels of defocus are employed in cryo-EM to form an approximate phase plate that transfers lower spatial frequencies.  A similar approach can be used in 4D-STEM phase imaging, and is the basis of techniques such as tilt-compensated bright field \citep{tcBFMuller, tcbf_lenak} or ptychography with large defocus \citep{Zhou2020}.  As shown in \cite{YANG2016117}, under such conditions the transfer function becomes oscillatory for $\omega < 1$, and the DQE oscillates and tends to 1.  For $\omega > 1$
, in which only the single side-band interference can contribute whether or not aberrations are present, the DQE is unchanged by aberrations.

Despite the poor DQE of ptychography for a weak phase object, high-quality images of graphene have been produced \citep{OLEARY2021113189}.  The results shown here demonstrate that for such imaging, the robustness of ptychography to chromatic aberrations outweighs the loss of DQE, and ptychography can image at such spatial frequencies more efficiently that HRTEM.  Of course, in the presence of an ideal phase plate and with a chromatic aberration corrector, these limitations for HRTEM are overcome \citep{Schwartz2019,petrov2024crossedlaserphaseplates,UteKaiserCc}.  We also note that 4D-STEM allows some inelastic scattering to contribute through the preservation of elastic contrast in the inelastic scattering.  It is not possible to use inelastically scattered electrons in conventional HRTEM imaging because of the chromatic aberration. For this reason, energy filtering is commonly used in cryo-EM, but appears to be less critical for 4D-STEM methods.

\begin{figure}[h]
\includegraphics[width=0.65\textwidth]{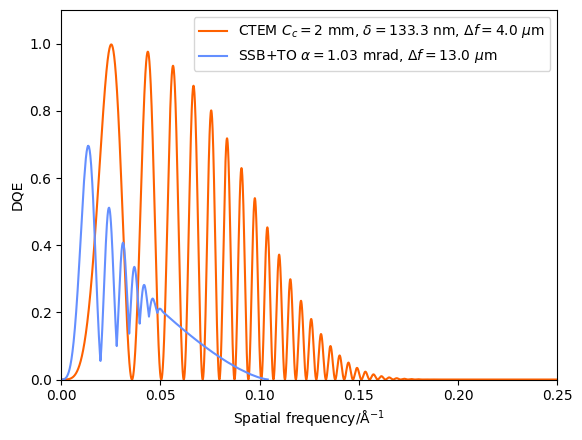}
\caption{Comparison of the DQE of previously reported conditions for imaging of biological samples at 300 kV. One can see that in the STEM geometry with a small convergence angle of 1.03 mrad \citep{Zhou2020}, the DQE tails off sharply to the 2$\alpha$ limit located at low spatial frequencies. The classical CTEM configuration on the other hand is limited by the partial coherence originating from a 20 eV energy window \citep{Laporte2025}.\label{fig:dqe_ctem_ssb_to_lowres}}
\end{figure}

When applying electron ptychography to biological samples, a small convergence angle is often used in combination with a large defocus \citep{Zhou2020}. However, when comparing the DQE of this approach to that of a CTEM \citep{Laporte2025}, one can see the sharp tail off due to the small convergence angle limiting the transfer of higher spatial frequencies, Fig. \ref{fig:dqe_ctem_ssb_to_lowres}. In \citep{Zhou2020}, the small aperture was used to strengthen the low frequencies, but larger apertures were also used in that work to pass higher frequencies.

It is important to keep in mind the severe limitations of the weak phase object itself for which these comparisons apply.  Essentially, the WPOA assumes all the signal is contained within the bright field disk and neglects diffracted beams.  This is a good approximation for a monolayer of graphene at high beam energies, but as the sample thickness grows and the atomic number of the atoms in the sample increase, the approximation breaks down rapidly and usable information can be found and extracted from higher-order scattering in 4D-STEM \citep{ma2025information4dstemisuse}. This same information may not be as simple to access from phase-plate-corrected TEM.
\section{Conclusions}
In this work we have attempted to quantify the degree to which very different imaging modes make efficient use of the electrons incident on the sample. A particular challenge was that we needed to compare a direct imaging method, namely Zernike phase contrast imaging, with ptychography which retrieves a 2D image from a much larger 4D STEM data set using an algorithm. The concept of the phase contrast transfer function is no longer effective because the strength of the contribution of a particular spatial frequency is completely controlled by the algorithm.  A more useful concept is the transfer of signal to noise, which leads readily to the idea of using the detective quantum efficiency. That ptychography appears uncompetitive against Zernike phase contrast imaging is perhaps not surprising.  Because the phase information resides in diffracted disc overlaps, it is not possible for all of the transmitted electrons to arrive in an overlap region, so some cannot contribute to the interference effects.  In principle, Zernike phase contrast imaging can use all of the incident electrons.

It is important to note that here we have used linear imaging via a weak phase object approximation. While many beam sensitive materials will consist of low atomic number species, such as organic materials, and therefore will fulfill this criterion, an obvious extension of the work is to more strongly scattering objects. Indeed, a particular strength of ptychography is the recent demonstration of overcoming the dynamical scattering problem using the multislice approach. In this case it may be hard to find an "ideal" imaging model which can form the reference for a DQE calculation.  We also note that we have not used iterative methods here. One reason is that it is impossible to calculate an analytical form of what an iterative reconstruction will produce, and so empirical methods must be used. The question then arises of whether a noise-free reconstructon can be regarded as being a ground truth, and therefore the impact of noise is hard to quantify. For the reasons explained above, it is not expected that the DQE of an iterative method applied to data recorded from a weak phase object could exceed that of SSB+TO for which all the possible interference that exists is used.

For stronger scatterers, the WPOA DQE still serves as a useful guide as to where the bulk of the signal resides in thin samples. In the extreme thin sample limit, it shows the advantage of an ideal phase-plate TEM for extracting low spatial frequencies, and the advantage of ptychography for resolving high-spatial frequencies that could not be accessed with the phase plate. Bounds for the rate at which the TEM and STEM signal change and ultimately degrade with thickness in the presence of multiple elastic and inelastic scattering still lacks a general solution.
\section*{Acknowledgments}
We acknowledge helpful discussions with Ondrej Krivanek. D.A.M. acknowledges helpful discussions with Desheng Ma and Steve Zeltmann. This work was funded by the UK Research and Innovation, Engineering and Physical Sciences Research Council. D.A.M. is supported by PARADIM under NSF Cooperative Agreement No. DMR-2039380.
\begin{appendices}
\section{Formal description of noise in $G(\mathbf{K_f},\mathbf{Q_p})$}\label{Formal_noise_G}
As stated in the main text, the intensity in the recorded 4D STEM dataset can be modelled as a random Poisson process $X_{M, I}(\mathbf{K_f})$.
\begin{equation}
    X_{M, I}(\mathbf{K_f})=E\big[X_{M, I}(\mathbf{K_f})\big]+\eta_{M, I}(\mathbf{K_f})
\end{equation}
The defining feature of a Poisson process is that the expectation value and variance are equal.
\begin{align}\label{4DSTEM_M_expectation_apdx}
    E\big[X_{M, I}(\mathbf{K_f},\mathbf{R_p})\big]&=\vert M(\mathbf{K_f}, \mathbf{R_p}))\vert^2\\
    \text{Var}(X_{M, I}(\mathbf{K_f},\mathbf{R_p}))&=\text{Var}(\eta_{M, I}(\mathbf{K_f},\mathbf{R_p}))=\vert M(\mathbf{K_f}, \mathbf{R_p}))\vert^2
\end{align}
Even though $\vert M(\mathbf{K_f}, \mathbf{R_p}))\vert^2$ varies across the 4D-STEM dataset, the intensity variations within the 2D diffraction patterns and in the entire 4D-STEM dataset are small. The same assumption was made in the main text for Zernike phase contrast microscopy. For a 4D-STEM dataset acquired on a perfect detector, the average pixel variance can be written as
\begin{equation}\label{4DSTEM_Mnoise_apdx}
    \text{Var}(X_{M, I}(\mathbf{K_f},\mathbf{R_p}))\approx\overline{\vert M(\mathbf{K_f}, \mathbf{R_p}))\vert^2}\approx\frac{Dd_{x,y}^2}{N_\alpha}
\end{equation}
where $D$ is the electron fluence in $e/$\AA$^2$, $d_x,y$ is the real space step length given in \AA\ and $N_\alpha$ is the number of pixels in the bright field disc.

To find an expression for the noise in $G(\mathbf{K_f},\mathbf{Q_p})$, the 2D DFT of $X_{M, I}(\mathbf{K_f},\mathbf{R_p})$ needs to be taken with respect to $\mathbf{R_p}$.
\begin{eqnarray}
    X_{G,I}(\mathbf{K_f},\mathbf{Q_p})&=&\sum_1^{N_x}\sum_1^{N_y}X_{M, I}(\mathbf{K_f},\mathbf{R_p})\exp\bigg(i2\pi\bigg(\frac{Q_xR_x}{N_x}+\frac{Q_yR_y}{N_y}\bigg)\bigg)\nonumber\\
    &=&\sum_1^{N_x}\sum_1^{N_y}\bigg(E\big[X_{M, I}(\mathbf{K_f},\mathbf{R_p})\big]+\eta_{M, I}(\mathbf{K_f},\mathbf{R_p})\bigg)\exp\bigg(i2\pi\bigg(\frac{Q_xR_x}{N_x}+\frac{Q_yR_y}{N_y}\bigg)\bigg)\nonumber\\
    &=&\begin{aligned}[t]
        \sum_1^{N_x}\sum_1^{N_y}E\big[X_{M, I}(\mathbf{K_f},\mathbf{R_p})\big]&\exp\bigg(i2\pi\bigg(\frac{Q_xR_x}{N_x}+\frac{Q_yR_y}{N_y}\bigg)\bigg)\nonumber\\
        &+\sum_1^{N_x}\sum_1^{N_y}\eta_{M, I}(\mathbf{K_f},\mathbf{R_p})\exp\bigg(i2\pi\bigg(\frac{Q_xR_x}{N_x}+\frac{Q_yR_y}{N_y}\bigg)\bigg)
    \end{aligned}\\
    &=&E\big[X_{G, I}(\mathbf{K_f},\mathbf{Q_p})\big]+\eta_{G, I}(\mathbf{K_f},\mathbf{Q_p})
\end{eqnarray}
Considering the complex nature of $X_{G,I}(\mathbf{K_f},\mathbf{Q_p})$, the variance of its real and imaginary components need to be evaluated separately.
\begin{eqnarray}
    \text{Var}(\text{Re}\{X_{G, I}(\mathbf{K_f},\mathbf{Q_p})\})&=&E\bigg[\big(\text{Re}\{\eta_{G, I}(\mathbf{K_f},\mathbf{Q_p})\}\big)^2\bigg]\nonumber\\
    &=&\sum_1^{N_x}\sum_1^{N_y}E\bigg[\bigg(\text{Re}\bigg\{\eta_{M, I}(\mathbf{K_f},\mathbf{R_p})\exp\bigg(i2\pi\bigg(\frac{Q_xR_x}{N_x}+\frac{Q_yR_y}{N_y}\bigg)\bigg)\bigg\}\bigg)^2\bigg]\\
    &=&\sum_1^{N_x}\sum_1^{N_y}E\bigg[\bigg(\eta_{M, I}(\mathbf{K_f},\mathbf{R_p})\cos\bigg(2\pi\bigg(\frac{Q_xR_x}{N_x}+\frac{Q_yR_y}{N_y}\bigg)\bigg)\bigg)^2\bigg]\\
    &=&\sum_1^{N_x}\sum_1^{N_y}E\bigg[\big(\eta_{M, I}(\mathbf{K_f},\mathbf{R_p})\big)^2\bigg]cos^2\bigg(2\pi\bigg(\frac{Q_xR_x}{N_x}+\frac{Q_yR_y}{N_y}\bigg)\bigg)\nonumber\\
    &=&\text{Var}(\eta_{M, I}(\mathbf{K_f},\mathbf{R_p}))\sum_1^{N_x}\sum_1^{N_y}cos^2\bigg(2\pi\bigg(\frac{Q_xR_x}{N_x}+\frac{Q_yR_y}{N_y}\bigg)\bigg)
\end{eqnarray}
Using Eq. \eqref{4DSTEM_Mnoise} and the trigonometric identity $cos^2(x)=\frac{1}{2}(1+cos(2x))$ one can evaluate the terms over the $\cos^2$ geometric series.
\begin{equation}
    \text{Var}(\text{Re}\{X_{G, I}(\mathbf{K_f},\mathbf{Q_p})\})=\frac{Dd_{x,y}^2}{2N_\alpha}N_xN_y
\end{equation}
The same process can be repeated for the imaginary component to yield
\begin{equation}
    \text{Var}(\text{Im}\{X_{G, I}(\mathbf{K_f},\mathbf{Q_p})\})=\frac{Dd_{x,y}^2}{2N_\alpha}N_xN_y
\end{equation}
\section{Equivalence of summation over double overlap regions in $G(\mathbf{K_f},\mathbf{Q_p})$}
\begin{figure}[h]
\includegraphics[width=\textwidth]{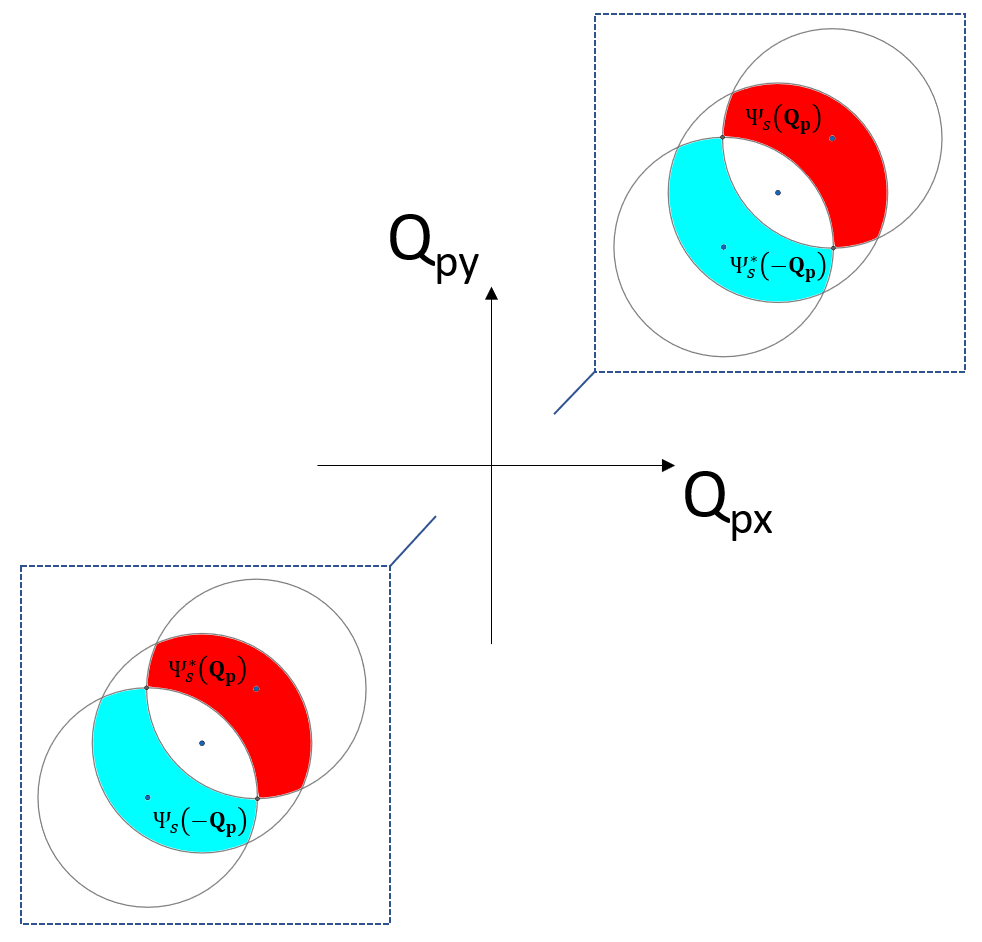}
\caption{Diagram showing the way the summation is performed in the first option of double sideband ptychography, DSB I. The red shaded region shows the area that contributes to the sum at $G_{DSB,I}(\mathbf{Q_p})$. The blue shaded region shows the area that contributes to the sum at $G_{DSB,I}(-\mathbf{Q_p})$. \label{fig:dsb_I}}
\end{figure}

\begin{figure}[h]
\includegraphics[width=\textwidth]{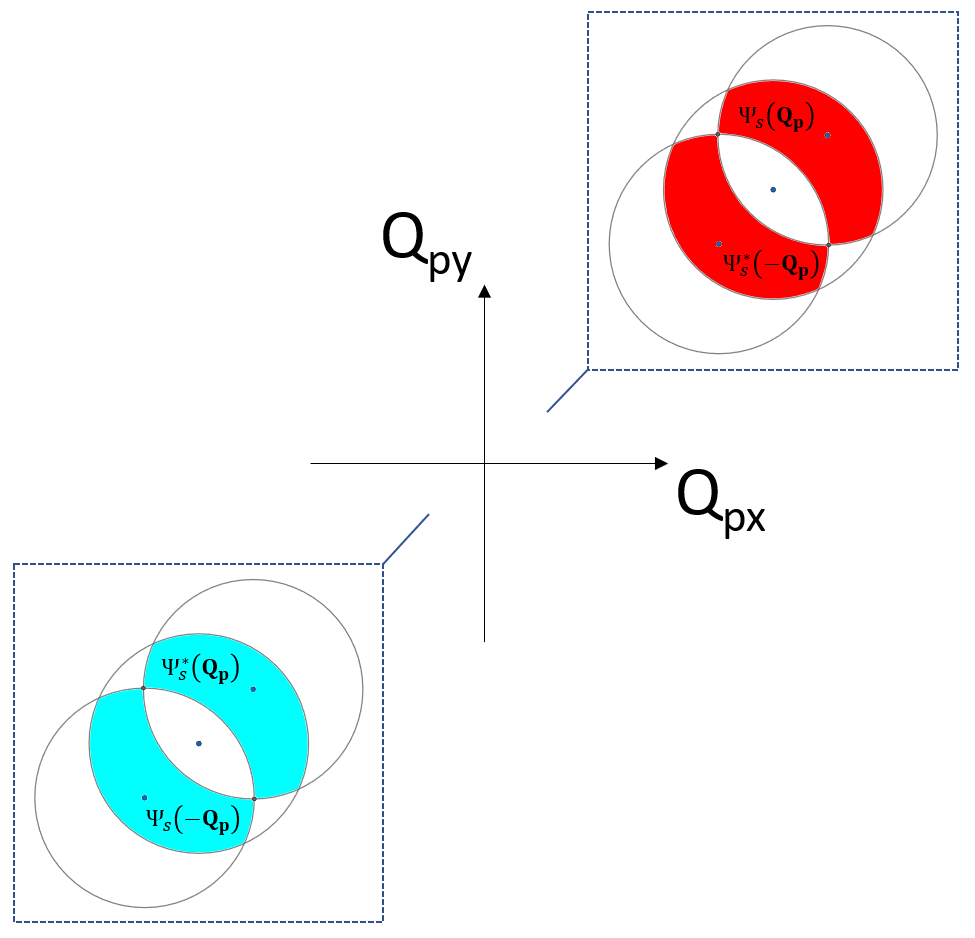}
\caption{Diagram showing the way the summation is performed in the second option of double sideband ptychography, DSB II. The red shaded region shows the area that contributes to the sum at $G_{DSB,I}(\mathbf{Q_p})$. The blue shaded region shows the area that contributes to the sum at $G_{DSB,I}(-\mathbf{Q_p})$. \label{fig:dsb_II}}
\end{figure}

$G(\mathbf{K_f},\mathbf{Q_p})$ can be described by the random process
\begin{equation}
    X_{G,I}(\mathbf{K_f},\mathbf{Q_p})=E\big[X_{G, I}(\mathbf{K_f},\mathbf{Q_p})\big]+\eta_{G, I}(\mathbf{K_f},\mathbf{Q_p})
\end{equation}
A summation can now be performed in $\mathbf{K_f}$ space. The two ways of performing the summation are shown in Fig. \ref{fig:dsb_I} and Fig. \ref{fig:dsb_II}. The first option is for a given $\mathbf{Q_p}$ to sum over one double overlap region in $\mathbf{K_f}$ space and combine it with the sum over the same pixels at $-\mathbf{Q_p}$. The second option is to sum over region both double overlap regions for a given $\mathbf{Q_p}$.
\subsection{Inner sideband at $\mathbf{Q_p}$ and conjugate of outer sideband at $\mathbf{-Q_p}$ combination before iFT}
\begin{align}
    X_{G:DSB \text{ I}, I}(\mathbf{Q_p})=\sum_{SSB,i,\mathbf{K_f}}X_{G,I}(\mathbf{K_f},\mathbf{Q_p})+\sum_{SSB,o,\mathbf{K_f}}X_{G,I}(\mathbf{K_f},\mathbf{-Q_p})^*
\end{align}
After the summation, the reconstruction is completed through an iFT
\begin{align}
    X_{\text{Image}:DSB \text{ I}, I}(\mathbf{R_p})&=\text{Im}\Bigg\{\frac{1}{N_xN_y}\sum^{N_x}_{Q_{px}=1}\sum^{N_y}_{Q_{py}=1}\bigg[\sum_{SSB,i,\mathbf{K_f}}X_{G,I}(\mathbf{K_f},Q_{px},Q_{py})\nonumber\\&+\sum_{SSB,o,\mathbf{K_f}}X_{G,I}(\mathbf{K_f},N_x-Q_{px},N_y-Q_{py})^*\bigg]\exp\bigg(i2\pi\bigg(\frac{Q_{px}R_x}{N_x}+\frac{Q_{py}R_y}{N_y}\bigg)\bigg)\Bigg\}
\end{align}
The iFT summation can now be rewritten as a sum over half as many components by combining the $\mathbf{-Q_p}$ and $\mathbf{+Q_p}$ terms. The 0 spatial frequency component is neglected.
\begin{align}
    X_{\text{Image}:DSB \text{ I}, I}(\mathbf{R_p})=\text{Im}\Bigg\{\frac{1}{N_xN_y}\sum^{\frac{N_x-1}{2}}_{Q_{px}=1}\sum^{\frac{N_y-1}{2}}_{Q_{py}=1}\Bigg[\bigg[\sum_{SSB,i,\mathbf{K_f}}X_{G,I}(\mathbf{K_f},Q_{px},Q_{py})\nonumber\\+\sum_{SSB,o,\mathbf{K_f}}X_{G,I}(\mathbf{K_f},N_x-Q_{px},N_y-Q_{py})^*\bigg]\exp\bigg(i2\pi\bigg(\frac{Q_{px}R_x}{N_x}+\frac{Q_{py}R_y}{N_y}\bigg)\bigg)\nonumber\\
    +\bigg[\sum_{SSB,i,\mathbf{K_f}}X_{G,I}(\mathbf{K_f},N_x-Q_{px},N_y-Q_{py})+\sum_{SSB,o,\mathbf{K_f}}X_{G,I}(\mathbf{K_f},Q_{px},Q_{py})^*\bigg]\exp\bigg(i2\pi\bigg(\frac{-Q_{px}R_x}{N_x}+\frac{-Q_{py}R_y}{N_y}\bigg)\bigg)\Bigg]\Bigg\}
\end{align}
The summation $\sum_{SSB,i,\mathbf{K_f}}X_{G,I}(\mathbf{K_f},Q_{px},Q_{py})$ and the summation $\sum_{SSB,o,\mathbf{K_f}}X_{G,I}(\mathbf{K_f},N_x-Q_{px},N_y-Q_{py})^*$ take place over exactly the same pixels in $\mathbf{K_f}$ space. Similarly, the summation $\sum_{SSB,i,\mathbf{K_f}}X_{G,I}(\mathbf{K_f},N_x-Q_{px},N_y-Q_{py})$ and the summation $\sum_{SSB,o,\mathbf{K_f}}X_{G,I}(\mathbf{K_f},Q_{px},Q_{py})^*$ also take place over identical regions in $\mathbf{K_f}$ space. Considering that $X_{G,I}(\mathbf{K_f},\mathbf{Q_p})=X_{G,I}(\mathbf{K_f},\mathbf{Q_p})^*$,
\begin{align}
    \sum_{SSB,i,\mathbf{K_f}}X_{G,I}(\mathbf{K_f},Q_{px},Q_{py})=\sum_{SSB,o,\mathbf{K_f}}X_{G,I}(\mathbf{K_f},N_x-Q_{px},N_y-Q_{py})^*=X_{G:SSB,i,I}(\mathbf{Q_p})\label{inner_sum_plus_qp}\\
    \sum_{SSB,i,\mathbf{K_f}}X_{G,I}(\mathbf{K_f},N_x-Q_{px},N_y-Q_{py})=\sum_{SSB,o,\mathbf{K_f}}X_{G,I}(\mathbf{K_f},Q_{px},Q_{py})^*=X_{G:SSB,i,I}(\mathbf{-Q_p})\label{inner_sum_minus_qp}
\end{align}
It is important to note that for a perfect weak phase object $\Psi_s(\mathbf{Q_p})=\Psi_s(\mathbf{-Q_p})^*$. Therefore,
\begin{equation}
    E\big[X_{G:SSB,i,I}(\mathbf{Q_p})\big]=-E\big[X_{G:SSB,i,I}(\mathbf{-Q_p})^*\big]
\end{equation}
However, the detail of the noise is not equal as it is not bound by the symmetry of the Fourier transform since the summations are not taken over the exact same pixels in $\mathbf{K_f}$ space.
\begin{equation}
    \eta_{G:SSB,i,I}(\mathbf{Q_p})\neq-\eta_{G:SSB,i,I}(\mathbf{-Q_p})^*
\end{equation}
\begin{align}
    X_{\text{Image}:DSB \text{ I}, I}(\mathbf{R_p})=\text{Im}\Bigg\{\frac{1}{N_xN_y}\sum^{\frac{N_x-1}{2}}_{Q_{px}=1}\sum^{\frac{N_y-1}{2}}_{Q_{py}=1}\Bigg[2X_{G:SSB,i,I}(\mathbf{Q_p})\exp\bigg(i2\pi\bigg(\frac{Q_{px}R_x}{N_x}+\frac{Q_{py}R_y}{N_y}\bigg)\bigg)\nonumber\\
    +2X_{G:SSB,i,I}(\mathbf{-Q_p})\bigg(\exp\bigg(i2\pi\bigg(\frac{Q_{px}R_x}{N_x}+\frac{Q_{py}R_y}{N_y}\bigg)\bigg)\bigg)^*\Bigg]\Bigg\}
\end{align}
One can now use the additive property of the imaginary part,
\begin{align}
    X_{\text{Image}:DSB \text{ I}, I}(\mathbf{R_p})=\frac{1}{N_xN_y}\sum^{\frac{N_x-1}{2}}_{Q_{px}=1}\sum^{\frac{N_y-1}{2}}_{Q_{py}=1}\Bigg[2\text{Im}\Bigg\{X_{G:SSB,i,I}(\mathbf{Q_p})\exp\bigg(i2\pi\bigg(\frac{Q_{px}R_x}{N_x}+\frac{Q_{py}R_y}{N_y}\bigg)\bigg)\Bigg\}\nonumber\\
    +2\text{Im}\Bigg\{X_{G:SSB,i,I}(\mathbf{-Q_p})\bigg(\exp\bigg(i2\pi\bigg(\frac{Q_{px}R_x}{N_x}+\frac{Q_{py}R_y}{N_y}\bigg)\bigg)\bigg)^*\Bigg\}\Bigg]\label{DSB_type_1_img}
\end{align}
\subsection{Inner sideband at $\mathbf{Q_p}$ and outer sideband at $\mathbf{Q_p}$ combination before iFT}
\begin{align}
    X_{G:DSB \text{ II}, I}(\mathbf{Q_p})=\sum_{SSB,i,\mathbf{K_f}}X_{G,I}(\mathbf{K_f},\mathbf{Q_p})-\sum_{SSB,o,\mathbf{K_f}}X_{G,I}(\mathbf{K_f},\mathbf{Q_p})
\end{align}
After the summation, the reconstruction is completed through an iFT
\begin{align}
    X_{\text{Image}:DSB \text{ II}, I}(\mathbf{R_p})=\text{Im}\Bigg\{\frac{1}{N_xN_y}\sum^{N_x}_{Q_{px}=1}\sum^{N_y}_{Q_{py}=1}\bigg[\sum_{SSB,i,\mathbf{K_f}}X_{G,I}(\mathbf{K_f},Q_{px},Q_{py})\nonumber\\-\sum_{SSB,o,\mathbf{K_f}}X_{G,I}(\mathbf{K_f},Q_{px},Q_{py})\bigg]\exp\bigg(i2\pi\bigg(\frac{Q_{px}R_x}{N_x}+\frac{Q_{py}R_y}{N_y}\bigg)\bigg)\Bigg\}
\end{align}
As before, the iFT summation can now be rewritten as a sum over half as many components through combining the $\mathbf{-Q_p}$ and $\mathbf{+Q_p}$ terms. The 0 spatial frequency component is again neglected.
\begin{align}
    X_{\text{Image}:DSB \text{ II}, I}(\mathbf{R_p})=\text{Im}\Bigg\{\frac{1}{N_xN_y}\sum^{\frac{N_x-1}{2}}_{Q_{px}=1}\sum^{\frac{N_y-1}{2}}_{Q_{py}=1}\Bigg[\bigg[\sum_{SSB,i,\mathbf{K_f}}X_{G,I}(\mathbf{K_f},Q_{px},Q_{py})\nonumber\\-\sum_{SSB,o,\mathbf{K_f}}X_{G,I}(\mathbf{K_f},Q_{px},Q_{py})\bigg]\exp\bigg(i2\pi\bigg(\frac{Q_{px}R_x}{N_x}+\frac{Q_{py}R_y}{N_y}\bigg)\bigg)\nonumber\\
    +\bigg[\sum_{SSB,i,\mathbf{K_f}}X_{G,I}(\mathbf{K_f},N_x-Q_{px},N_y-Q_{py})-\sum_{SSB,o,\mathbf{K_f}}X_{G,I}(\mathbf{K_f},N_x-Q_{px},N_y-Q_{py})\bigg]\exp\bigg(i2\pi\bigg(\frac{-Q_{px}R_x}{N_x}+\frac{-Q_{py}R_y}{N_y}\bigg)\bigg)\Bigg]\Bigg\}
\end{align}
Substituting the definitions from Eq. \eqref{inner_sum_plus_qp} and Eq. \eqref{inner_sum_minus_qp},
\begin{align}
    X_{\text{Image}:DSB \text{ II}, I}(\mathbf{R_p})=\text{Im}\Bigg\{\frac{1}{N_xN_y}\sum^{\frac{N_x-1}{2}}_{Q_{px}=1}\sum^{\frac{N_y-1}{2}}_{Q_{py}=1}\Bigg[\bigg[X_{G:SSB,i,I}(\mathbf{Q_p})-X_{G:SSB,i,I}(\mathbf{-Q_p})^*\bigg]\exp\bigg(i2\pi\bigg(\frac{Q_{px}R_x}{N_x}+\frac{Q_{py}R_y}{N_y}\bigg)\bigg)\nonumber\\
    +\bigg[X_{G:SSB,i,I}(\mathbf{-Q_p})-X_{G:SSB,i,I}(\mathbf{Q_p})^*\bigg]\bigg(\exp\bigg(i2\pi\bigg(\frac{Q_{px}R_x}{N_x}+\frac{Q_{py}R_y}{N_y}\bigg)\bigg)\bigg)^*\Bigg]\Bigg\}
\end{align}
The brackets can now be expanded and the terms reordered.
\begin{align}
    X_{\text{Image}:DSB \text{ II}, I}(\mathbf{R_p})=\text{Im}\Bigg\{\frac{1}{N_xN_y}\sum^{\frac{N_x-1}{2}}_{Q_{px}=1}\sum^{\frac{N_y-1}{2}}_{Q_{py}=1}\Bigg[X_{G:SSB,i,I}(\mathbf{Q_p})\exp\bigg(i2\pi\bigg(\frac{Q_{px}R_x}{N_x}+\frac{Q_{py}R_y}{N_y}\bigg)\bigg)\nonumber\\-X_{G:SSB,i,I}(\mathbf{Q_p})^*\bigg(\exp\bigg(i2\pi\bigg(\frac{Q_{px}R_x}{N_x}+\frac{Q_{py}R_y}{N_y}\bigg)\bigg)\bigg)^*\nonumber\\-X_{G:SSB,i,I}(\mathbf{-Q_p})^*\exp\bigg(i2\pi\bigg(\frac{Q_{px}R_x}{N_x}+\frac{Q_{py}R_y}{N_y}\bigg)\bigg)\nonumber\\
    +X_{G:SSB,i,I}(\mathbf{-Q_p})\bigg(\exp\bigg(i2\pi\bigg(\frac{Q_{px}R_x}{N_x}+\frac{Q_{py}R_y}{N_y}\bigg)\bigg)\bigg)^*\Bigg]\Bigg\}
\end{align}
Using the multiplicative property of complex conjugation,
\begin{align}
    X_{\text{Image}:DSB \text{ II}, I}(\mathbf{R_p})=\text{Im}\Bigg\{\frac{1}{N_xN_y}\sum^{\frac{N_x-1}{2}}_{Q_{px}=1}\sum^{\frac{N_y-1}{2}}_{Q_{py}=1}\Bigg[X_{G:SSB,i,I}(\mathbf{Q_p})\exp\bigg(i2\pi\bigg(\frac{Q_{px}R_x}{N_x}+\frac{Q_{py}R_y}{N_y}\bigg)\bigg)\nonumber\\-\bigg(X_{G:SSB,i,I}(\mathbf{Q_p})\exp\bigg(i2\pi\bigg(\frac{Q_{px}R_x}{N_x}+\frac{Q_{py}R_y}{N_y}\bigg)\bigg)\bigg)^*\nonumber\\-\bigg(X_{G:SSB,i,I}(\mathbf{-Q_p})\bigg(\exp\bigg(i2\pi\bigg(\frac{Q_{px}R_x}{N_x}+\frac{Q_{py}R_y}{N_y}\bigg)\bigg)\bigg)^*\bigg)^*\nonumber\\
    +X_{G:SSB,i,I}(\mathbf{-Q_p})\bigg(\exp\bigg(i2\pi\bigg(\frac{Q_{px}R_x}{N_x}+\frac{Q_{py}R_y}{N_y}\bigg)\bigg)\bigg)^*\Bigg]\Bigg\}
\end{align}
Using the additive property of the imaginary part,
\begin{align}
    X_{\text{Image}:DSB \text{ II}, I}(\mathbf{R_p})=\frac{1}{N_xN_y}\sum^{\frac{N_x-1}{2}}_{Q_{px}=1}\sum^{\frac{N_y-1}{2}}_{Q_{py}=1}\Bigg[\text{Im}\Bigg\{X_{G:SSB,i,I}(\mathbf{Q_p})\exp\bigg(i2\pi\bigg(\frac{Q_{px}R_x}{N_x}+\frac{Q_{py}R_y}{N_y}\bigg)\bigg)\Bigg\}\nonumber\\+\text{Im}\Bigg\{-\bigg(X_{G:SSB,i,I}(\mathbf{Q_p})\exp\bigg(i2\pi\bigg(\frac{Q_{px}R_x}{N_x}+\frac{Q_{py}R_y}{N_y}\bigg)\bigg)\bigg)^*\Bigg\}\nonumber\\+\text{Im}\Bigg\{-\bigg(X_{G:SSB,i,I}(\mathbf{-Q_p})\bigg(\exp\bigg(i2\pi\bigg(\frac{Q_{px}R_x}{N_x}+\frac{Q_{py}R_y}{N_y}\bigg)\bigg)\bigg)^*\bigg)^*\Bigg\}\nonumber\\
    +\text{Im}\Bigg\{X_{G:SSB,i,I}(\mathbf{-Q_p})\bigg(\exp\bigg(i2\pi\bigg(\frac{Q_{px}R_x}{N_x}+\frac{Q_{py}R_y}{N_y}\bigg)\bigg)\bigg)^*\Bigg\}\Bigg]
\end{align}
Using the fact that for any complex number $z$, $\text{Im}(z)=\text{Im}(-z^*)$ and comparing with Eq. \eqref{DSB_type_1_img},
\begin{align}
    X_{\text{Image}:DSB \text{ II}, I}(\mathbf{R_p})=\frac{1}{N_xN_y}\sum^{\frac{N_x-1}{2}}_{Q_{px}=1}\sum^{\frac{N_y-1}{2}}_{Q_{py}=1}\Bigg[2\text{Im}\Bigg\{X_{G:SSB,i,I}(\mathbf{Q_p})\exp\bigg(i2\pi\bigg(\frac{Q_{px}R_x}{N_x}+\frac{Q_{py}R_y}{N_y}\bigg)\bigg)\Bigg\}\nonumber\\+2\text{Im}\Bigg\{X_{G:SSB,i,I}(\mathbf{-Q_p})\bigg(\exp\bigg(i2\pi\bigg(\frac{Q_{px}R_x}{N_x}+\frac{Q_{py}R_y}{N_y}\bigg)\bigg)\bigg)^*\Bigg\}\Bigg]\\
    =X_{\text{Image}:DSB \text{ I}, I}(\mathbf{R_p})\nonumber
\end{align}
\section{Single sideband is all you need}\label{SSB_is_all_u_need}
A tempting way to improve the single sideband method would be to include the other sideband. To evaluate the DQE, of such a double sideband ptychography method, we again express the 4D dataset as a random process \(X_{M, I}(\mathbf{K_f},\mathbf{R_p})\) comprising of a signal component \(E\big[X_{M, I}(\mathbf{K_f},\mathbf{R_p})\big]\) and a noise component $\eta_{M, I}(\mathbf{K_f},\mathbf{R_p})$ as before in \eqref{4D_stem_random_process}.
\begin{equation}
    X_{M, I}(\mathbf{K_f},\mathbf{R_p})=E\big[X_{M, I}(\mathbf{K_f},\mathbf{R_p})\big]+\eta_{M, I}(\mathbf{K_f},\mathbf{R_p})\nonumber
\end{equation}
The 2D DFT is then applied and linearity is used.
\begin{eqnarray}
    X_{G,I}(\mathbf{K_f},\mathbf{Q_p})&=&\sum_1^{N_x}\sum_1^{N_y}X_{M, I}(\mathbf{K_f},\mathbf{R_p})\exp\bigg(i2\pi\bigg(\frac{Q_xR_x}{N_x}+\frac{Q_yR_y}{N_y}\bigg)\bigg)\nonumber\\
    &=&\sum_1^{N_x}\sum_1^{N_y}\bigg(E\big[X_{M, I}(\mathbf{K_f},\mathbf{R_p})\big]+\eta_{M, I}(\mathbf{K_f},\mathbf{R_p})\bigg)\exp\bigg(i2\pi\bigg(\frac{Q_xR_x}{N_x}+\frac{Q_yR_y}{N_y}\bigg)\bigg)\nonumber\\
    &=&\begin{aligned}[t]
        \sum_1^{N_x}\sum_1^{N_y}E\big[X_{M, I}(\mathbf{K_f},\mathbf{R_p})\big]&\exp\bigg(i2\pi\bigg(\frac{Q_xR_x}{N_x}+\frac{Q_yR_y}{N_y}\bigg)\bigg)\\
        &+\sum_1^{N_x}\sum_1^{N_y}\eta_{M, I}(\mathbf{K_f},\mathbf{R_p})\exp\bigg(i2\pi\bigg(\frac{Q_xR_x}{N_x}+\frac{Q_yR_y}{N_y}\bigg)\bigg)
    \end{aligned}\nonumber\\
    &=&E\big[X_{G, I}(\mathbf{K_f},\mathbf{Q_p})\big]+\eta_{G, I}(\mathbf{K_f},\mathbf{Q_p})\label{random_process_noise_G}
\end{eqnarray}
Since the signal in the original 4D dataset $E\big[X_{M, I}(\mathbf{K_f},\mathbf{R_p})\big]$ and the noise $\eta_{M, I}(\mathbf{K_f},\mathbf{R_p})$ are strictly real valued,
\begin{eqnarray}
    \big[X_{G,I}(\mathbf{K_f},\mathbf{-Q_p})\big]^*&=&\begin{aligned}[t]
        \sum_1^{N_x}\sum_1^{N_y}E\big[X_{M, I}(\mathbf{K_f},\mathbf{R_p})\big]^*&\exp\bigg(i2\pi\bigg(\frac{-Q_xR_x}{N_x}+\frac{-Q_yR_y}{N_y}\bigg)\bigg)^*\nonumber\\
        &+\sum_1^{N_x}\sum_1^{N_y}\eta_{M, I}(\mathbf{K_f},\mathbf{R_p})^*\exp\bigg(i2\pi\bigg(\frac{-Q_xR_x}{N_x}+\frac{-Q_yR_y}{N_y}\bigg)\bigg)^*
    \end{aligned}\\
    &=&\begin{aligned}[t]
        \sum_1^{N_x}\sum_1^{N_y}E\big[X_{M, I}(\mathbf{K_f},\mathbf{R_p})\big]&\exp\bigg(-i2\pi\bigg(\frac{-Q_xR_x}{N_x}+\frac{-Q_yR_y}{N_y}\bigg)\bigg)\nonumber\\
        &+\sum_1^{N_x}\sum_1^{N_y}\eta_{M, I}(\mathbf{K_f},\mathbf{R_p})\exp\bigg(-i2\pi\bigg(\frac{-Q_xR_x}{N_x}+\frac{-Q_yR_y}{N_y}\bigg)\bigg)
    \end{aligned}\\
    &=&\begin{aligned}[t]
        \sum_1^{N_x}\sum_1^{N_y}E\big[X_{M, I}(\mathbf{K_f},\mathbf{R_p})\big]&\exp\bigg(i2\pi\bigg(\frac{Q_xR_x}{N_x}+\frac{Q_yR_y}{N_y}\bigg)\bigg)\nonumber\\
        &+\sum_1^{N_x}\sum_1^{N_y}\eta_{M, I}(\mathbf{K_f},\mathbf{R_p})\exp\bigg(i2\pi\bigg(\frac{Q_xR_x}{N_x}+\frac{Q_yR_y}{N_y}\bigg)\bigg)
    \end{aligned}\\
    &=&E\big[X_{G, I}(\mathbf{K_f},\mathbf{Q_p})\big]+\eta_{G, I}(\mathbf{K_f},\mathbf{Q_p})=X_{G,I}(\mathbf{K_f},\mathbf{Q_p})\label{qp_neg_qp_noise_equiv}
\end{eqnarray}
This means that not only the signal components at $\mathbf{Q_p}$ and $\mathbf{-Q_p}$ are complex conjugates, but also the details in the noise.
\subsection{Double sideband ptychography signal power}
As stated in section \ref{SSB_ptych_sig_pow_sec}, the signal in $X_{G, I}(\mathbf{K_f},\mathbf{Q_p})$ can we written as
\begin{equation}
    E\big[X_{G, I}(\mathbf{K_f},\mathbf{Q_p})\big]=\frac{N_xN_yDd_{x,y}^2}{N_\alpha}\big(\vert A(\mathbf{K_f})\vert^2\delta(\mathbf{Q_p})+A(\mathbf{K_f})A^*(\mathbf{K_f}+\mathbf{Q_p})\Psi_s^*(-\mathbf{Q_p})+A^*(\mathbf{K_f})A(\mathbf{K_f}-\mathbf{Q_p})\Psi_s(\mathbf{Q_p})\big)
\end{equation}
From the above equation, it can be seen that one of the sidebands contains information about $\mathbf{Q_p}$ while the other contains information about $\mathbf{-Q_p}$. To perform the double sideband summation one now has to sum one sideband at a certain $\mathbf{Q_p}$ with the other sideband at $\mathbf{-Q_p}$ or vice versa.
\begin{eqnarray}\label{dsb_over_g_outline}
    E\big[X_{G:DSB, I}(\mathbf{K_f},\mathbf{Q_p})\big]&=&\sum_{SSB,\mathbf{-Q_p},\mathbf{K_f}}G(\mathbf{K_f},\mathbf{Q_p})+\sum_{SSB,\mathbf{+Q_p},\mathbf{K_f}}G^*(\mathbf{K_f},\mathbf{-Q_p})\nonumber\\
    &=&\begin{aligned}[t]&\frac{N_xN_yDd_{x,y}^2}{N_\alpha}\Bigg[\sum_{SSB,\mathbf{-Q_p},\mathbf{K_f}}\vert A(\mathbf{K_f})\vert^2\delta(\mathbf{Q_p})+A(\mathbf{K_f})A^*(\mathbf{K_f}+\mathbf{Q_p})\Psi_s^*(-\mathbf{Q_p})+A^*(\mathbf{K_f})A(\mathbf{K_f}-\mathbf{Q_p})\Psi_s(\mathbf{Q_p})\nonumber\\
    &+\sum_{SSB,\mathbf{+Q_p},\mathbf{K_f}}\Big[\vert A(\mathbf{K_f})\vert^2\delta(\mathbf{-Q_p})+A(\mathbf{K_f})A^*(\mathbf{K_f}-\mathbf{Q_p})\Psi_s^*(\mathbf{Q_p})+A^*(\mathbf{K_f})A(\mathbf{K_f}+\mathbf{Q_p})\Psi_s(-\mathbf{Q_p})\Big]^*\Bigg]\end{aligned}\\
    &=&\begin{aligned}[t]&\frac{N_xN_yDd_{x,y}^2}{N_\alpha}\Bigg[\sum_{SSB,\mathbf{-Q_p},\mathbf{K_f}}\vert A(\mathbf{K_f})\vert^2\delta(\mathbf{Q_p})+A(\mathbf{K_f})A^*(\mathbf{K_f}+\mathbf{Q_p})\Psi_s^*(-\mathbf{Q_p})+A^*(\mathbf{K_f})A(\mathbf{K_f}-\mathbf{Q_p})\Psi_s(\mathbf{Q_p})\nonumber\\
    &+\sum_{SSB,\mathbf{+Q_p},\mathbf{K_f}}\vert A(\mathbf{K_f})\vert^2\delta(\mathbf{
    Q_p})+A^*(\mathbf{K_f})A(\mathbf{K_f}-\mathbf{Q_p})\Psi_s(\mathbf{Q_p})+A(\mathbf{K_f})A^*(\mathbf{K_f}+\mathbf{Q_p})\Psi^*_s(-\mathbf{Q_p})\Bigg]\end{aligned}
\end{eqnarray}
where $\sum_{SSB,\mathbf{+Q_p}:\mathbf{K_f}}$ denotes the SSB summation over the double overlap region between the direct beam and the $\mathbf{+Q_p}$ beam. $\sum_{SSB,\mathbf{-Q_p}:\mathbf{K_f}}$ denotes the SSB summation over the double overlap region between the direct beam and the $\mathbf{-Q_p}$ beam. Sometimes these are referred to as the left and right sidebands. This is slightly misleading as the dexterity of the sidebands changes when $\mathbf{Q_p}$ changes. When flattening the 4D dataset, the $\mathbf{-Q_p}$-sideband is always the one closest to the centre. A more fitting description would therefore be the outer and inner sidebands. As stated above, double sideband ptychography involves combining the outer sideband at $\mathbf{+Q_p}$ with the inner sideband at $\mathbf{-Q_p}$ or vice versa.

It is critical to note that the summation of the outer sideband at $\mathbf{+Q_p}$ takes place over the exact same $\mathbf{K_f}$ values as the summation of the inner sideband at $\mathbf{-Q_p}$. The implications of this will be discussed in section \ref{dsb_noise_pow_sec}.

To find an expression for the expectation value of the output of double sideband ptychography, the summations in Eq. \eqref{dsb_over_g_outline} can be evaluated.
\begin{eqnarray}
    E\big[X_{G:DSB, I}(\mathbf{K_f},\mathbf{Q_p})\big]&=&\begin{aligned}[t]\bigg[N_xN_yDd_{x,y}^2PCTF(\mathbf{Q_p})&+N_xN_yDd_{x,y}^2PCTF(\mathbf{Q_p})\Psi_s^*(-\mathbf{Q_p})\bigg]\nonumber\\
    &+\bigg[N_xN_yDd_{x,y}^2PCTF(\mathbf{-Q_p})+N_xN_yDd_{x,y}^2PCTF(-\mathbf{Q_p})\Psi_s(\mathbf{-Q_p})\bigg]^*
    \end{aligned}\\
    &=&\begin{aligned}[t]\bigg[N_xN_yDd_{x,y}^2PCTF(\mathbf{Q_p})&+N_xN_yDd_{x,y}^2PCTF(\mathbf{Q_p})\Psi_s^*(-\mathbf{Q_p})\bigg]\nonumber\\
    &+\bigg[N_xN_yDd_{x,y}^2PCTF(\mathbf{Q_p})+N_xN_yDd_{x,y}^2PCTF(\mathbf{Q_p})\Psi^*_s(\mathbf{-Q_p})\bigg]
    \end{aligned}\\
    &=&2N_xN_yDd_{x,y}^2PCTF(\mathbf{Q_p})+2N_xN_yDd_{x,y}^2PCTF(\mathbf{Q_p})\Psi_s^*(-\mathbf{Q_p})
\end{eqnarray}

Now an expression for the signal power of double sideband ptychography can be constructed as previously in \ref{SSB_ptych_sig_pow_sec}.
\begin{equation}\label{dsb_sig_pw}
    P_\text{sig, DSB}=2N_xN_yDd_{x,y}^2 PCTF(\mathbf{Q_p})
\end{equation}
\subsection{Double sideband ptychography noise power}\label{dsb_noise_pow_sec}
As previously stated in section \ref{SSB_ptych_noise_pow_sec}, the noise in $G(\mathbf{K_f},\mathbf{Q_p})$ can be described as,
\begin{eqnarray}
    \text{Var}(Re\{X_{G,I}(\mathbf{K_f},\mathbf{Q_p})\})&=&E\big[(Re\{\eta_{G,I}(\mathbf{K_f},\mathbf{Q_p})\})^2\big]\nonumber\\&=&\frac{Dd_{x,y}^2}{2N_\alpha}N_xN_y\nonumber
\end{eqnarray}
\begin{eqnarray}
    \text{Var}(Im\{X_{G,I}(\mathbf{K_f},\mathbf{Q_p})\})&=&E\big[(Im\{\eta_{G,I}(\mathbf{K_f},\mathbf{Q_p})\})^2\big]\nonumber\\&=&\frac{Dd_{x,y}^2}{2N_\alpha}N_xN_y\nonumber
\end{eqnarray}
As mentioned before, the variance of the intensity in the imaginary part is of primary importance as the imaginary part contains the information of the scattered wave. As before, we will therefore mainly focus on the imaginary part in the remainder of this section.
\begin{equation}\label{Cumulative_DSB_variance_imag1}
    \text{Var}(Im\{X_{G:DSB, I}(\mathbf{Q_p})\})=E\Bigg[\Big(\sum_{SSB,\mathbf{-Q_p},\mathbf{K}_{\mathbf{f},i}}Im\{\eta_{G,I}(\mathbf{K}_{\mathbf{f},i},\mathbf{Q_p})\}+\sum_{SSB,\mathbf{+Q_p},\mathbf{K}_{\mathbf{f},i}}Im\{\eta_{G,I}^*(\mathbf{K}_{\mathbf{f},i},\mathbf{-Q_p})\}\Big)^2\Bigg]
\end{equation}
As described above in section \ref{SSB_ptych_noise_pow_sec}, the signal and noise of the inner sideband at spatial frequency $\mathbf{Q_p}$ are identical to the signal and noise at spatial frequency $-\mathbf{Q_p}$. This means that the noise simply doubles. Equations \eqref{Cumulative_SSB_variance_imag1} and \eqref{Cumulative_SSB_variance_imag_final} can now be combined with \eqref{Cumulative_DSB_variance_imag1} to yield an expression for the variance in the imaginary part of the double sideband ptychography reconstruction.
\begin{eqnarray}\label{Cumulative_DSB_variance_imag2}
    \text{Var}(Im\{X_{G:DSB, I}(\mathbf{Q_p})\})&=&E\Bigg[\Big(2\sum_{SSB,\mathbf{-Q_p},\mathbf{K}_{\mathbf{f},i}}Im\{\eta_{G,I}(\mathbf{K}_{\mathbf{f},i},\mathbf{Q_p})\}\Big)^2\Bigg]\nonumber\\
    &=&4E\Bigg[\Big(\sum_{SSB,\mathbf{-Q_p},\mathbf{K}_{\mathbf{f},i}}Im\{\eta_{G,I}(\mathbf{K}_{\mathbf{f},i},\mathbf{Q_p})\}\Big)^2\Bigg]\nonumber\\
    &=&4\frac{Dd_{x,y}^2}{2}N_xN_yPCTF(\mathbf{Q_p})\nonumber\\
    &=&2Dd_{x,y}^2N_xN_yPCTF(\mathbf{Q_p})
\end{eqnarray}
Of course, the same analysis can be performed for the real part.
\begin{equation}\label{Cumulative_DSB_variance_real}
    \text{Var}(Re\{X_{G:DSB, I}(\mathbf{Q_p})\})=2Dd_{x,y}^2N_xN_yPCTF(\mathbf{Q_p})
\end{equation}
As in the case of single sideband ptychography, the variance in the imaginary part of the DSB summation is in fact the square of the noise power spectrum.
\begin{eqnarray}\label{dsb_noise_sq}
    P_\text{noise, DSB}^2(\mathbf{Q_p})&=&\text{Var}(Im\{X_{G:DSB, I}(\mathbf{Q_p})\})\nonumber\\
    &=&2Dd_{x,y}^2N_xN_yPCTF(\mathbf{Q_p})
\end{eqnarray}
\subsection{Double sideband ptychography $SNR^2$}
As in the case of SSB, \eqref{dsb_sig_pw} and \eqref{dsb_noise_sq} can now be combined to find an expression for the square of the signal to noise ratio of DSB ptychography.
\begin{eqnarray}\label{dsb_snr_sq}
    \text{SNR}_{DSB}^2(\mathbf{Q_p})&=&\frac{P_\text{sig, DSB}^2(\mathbf{Q_p})}{P_\text{noise, DSB}^2(\mathbf{Q_p})}\nonumber\\&=&\frac{(2N_xN_yDd_{x,y}^2 PCTF(\mathbf{Q_p}))^2}{2Dd_{x,y}^2N_xN_yPCTF(\mathbf{Q_p})}\nonumber\\&=&2Dd_{x,y}^2N_xN_yPCTF(\mathbf{Q_p})
\end{eqnarray}
\subsection{Double sideband ptychography DQE}
\eqref{snr_z_ideal_sq} can be combined with \eqref{dsb_snr_sq} using \eqref{DQE} to find the expression for the DQE of DSB ptychography.
\begin{eqnarray}\label{DSB_DQE}
    DQE_{DSB, \text{ideal}}(\mathbf{Q_p})&=&\frac{SNR_{DSB}(\mathbf{Q_p})^2}{SNR_{Z, \text{ideal}}(\mathbf{Q_p})^2}\nonumber\\&=&\frac{2Dd_{x,y}^2N_xN_yPCTF(\mathbf{Q_p})}{4N_xN_yDd_{x,y}^2}\notag\\
    &=&\frac{PCTF(\mathbf{Q_p})}{2}
\end{eqnarray}
\section{Effect of phase correction on noise}\label{TO_phase_corr_noise}
\begin{eqnarray}
    \text{Var}(\text{Re}\{X_{G,TO\text{ sum prep},I}(\mathbf{K_f},\mathbf{Q_p})\})=E\bigg[\Big(\text{Re}\{e^{-i\text{ arg(}A^*(\mathbf{K_f})A(\mathbf{K_f}-\mathbf{Q_p})-A(\mathbf{K_f})A^*(\mathbf{K_f}+\mathbf{Q_p})\text{)}}\}\text{Re}\{\eta_{G,I}(\mathbf{K_f},\mathbf{Q_p})\}\nonumber\\
    -\text{Im}\{e^{-i\text{ arg(}A^*(\mathbf{K_f})A(\mathbf{K_f}-\mathbf{Q_p})-A(\mathbf{K_f})A^*(\mathbf{K_f}+\mathbf{Q_p})\text{)}}\}\text{Im}\{\eta_{G,I}(\mathbf{K_f},\mathbf{Q_p})\}\Big)^2\bigg]\nonumber\\
    =E\bigg[\Big(\text{Re}\{e^{-i\text{ arg(}A^*(\mathbf{K_f})A(\mathbf{K_f}-\mathbf{Q_p})-A(\mathbf{K_f})A^*(\mathbf{K_f}+\mathbf{Q_p})\text{)}}\}\text{Re}\{\eta_{G,I}(\mathbf{K_f},\mathbf{Q_p})\}\Big)^2\nonumber\\
    +\Big(\text{Im}\{e^{-i\text{ arg(}A^*(\mathbf{K_f})A(\mathbf{K_f}-\mathbf{Q_p})-A(\mathbf{K_f})A^*(\mathbf{K_f}+\mathbf{Q_p})\text{)}}\}\text{Im}\{\eta_{G,I}(\mathbf{K_f},\mathbf{Q_p})\}\Big)^2\nonumber\\-2\text{Re}\{e^{-i\text{ arg(}A^*(\mathbf{K_f})A(\mathbf{K_f}-\mathbf{Q_p})-A(\mathbf{K_f})A^*(\mathbf{K_f}+\mathbf{Q_p})\text{)}}\}\text{Re}\{\eta_{G,I}(\mathbf{K_f},\mathbf{Q_p})\}\nonumber\\\times\text{Im}\{e^{-i\text{ arg(}A^*(\mathbf{K_f})A(\mathbf{K_f}-\mathbf{Q_p})-A(\mathbf{K_f})A^*(\mathbf{K_f}+\mathbf{Q_p})\text{)}}\}\text{Im}\{\eta_{G,I}(\mathbf{K_f},\mathbf{Q_p})\}\bigg]\label{Variance_G_real_TO_sumprep_2}\\
    \text{Var}(\text{Im}\{X_{G,TO\text{ sum prep},I}(\mathbf{K_f},\mathbf{Q_p})\})=E\bigg[\Big(\text{Re}\{e^{-i\text{ arg(}A^*(\mathbf{K_f})A(\mathbf{K_f}-\mathbf{Q_p})-A(\mathbf{K_f})A^*(\mathbf{K_f}+\mathbf{Q_p})\text{)}}\}\text{Im}\{\eta_{G,I}(\mathbf{K_f},\mathbf{Q_p})\}\nonumber\\
    +\text{Im}\{e^{-i\text{ arg(}A^*(\mathbf{K_f})A(\mathbf{K_f}-\mathbf{Q_p})-A(\mathbf{K_f})A^*(\mathbf{K_f}+\mathbf{Q_p})\text{)}}\}\text{Re}\{\eta_{G,I}(\mathbf{K_f},\mathbf{Q_p})\}\Big)^2\bigg]\nonumber\\
    =E\bigg[\Big(\text{Re}\{e^{-i\text{ arg(}A^*(\mathbf{K_f})A(\mathbf{K_f}-\mathbf{Q_p})-A(\mathbf{K_f})A^*(\mathbf{K_f}+\mathbf{Q_p})\text{)}}\}\text{Im}\{\eta_{G,I}(\mathbf{K_f},\mathbf{Q_p})\}\Big)^2\nonumber\\
    +\Big(\text{Im}\{e^{-i\text{ arg(}A^*(\mathbf{K_f})A(\mathbf{K_f}-\mathbf{Q_p})-A(\mathbf{K_f})A^*(\mathbf{K_f}+\mathbf{Q_p})\text{)}}\}\text{Re}\{\eta_{G,I}(\mathbf{K_f},\mathbf{Q_p})\}\Big)^2\nonumber\\+2\text{Re}\{e^{-i\text{ arg(}A^*(\mathbf{K_f})A(\mathbf{K_f}-\mathbf{Q_p})-A(\mathbf{K_f})A^*(\mathbf{K_f}+\mathbf{Q_p})\text{)}}\}\text{Im}\{\eta_{G,I}(\mathbf{K_f},\mathbf{Q_p})\}\nonumber\\\times\text{Im}\{e^{-i\text{ arg(}A^*(\mathbf{K_f})A(\mathbf{K_f}-\mathbf{Q_p})-A(\mathbf{K_f})A^*(\mathbf{K_f}+\mathbf{Q_p})\text{)}}\}\text{Re}\{\eta_{G,I}(\mathbf{K_f},\mathbf{Q_p})\}\bigg]\label{Variance_G_imag_TO_sumprep_2}
\end{eqnarray}

Using The linarity of the expectation value operator,
\begin{eqnarray}
    \text{Var}(\text{Re}\{X_{G,TO\text{ sum prep},I}(\mathbf{K_f},\mathbf{Q_p})\})
    =E\bigg[\Big(\text{Re}\{e^{-i\text{ arg(}A^*(\mathbf{K_f})A(\mathbf{K_f}-\mathbf{Q_p})-A(\mathbf{K_f})A^*(\mathbf{K_f}+\mathbf{Q_p})\text{)}}\}\text{Re}\{\eta_{G,I}(\mathbf{K_f},\mathbf{Q_p})\}\Big)^2\bigg]\nonumber\\
    +E\bigg[\Big(\text{Im}\{e^{-i\text{ arg(}A^*(\mathbf{K_f})A(\mathbf{K_f}-\mathbf{Q_p})-A(\mathbf{K_f})A^*(\mathbf{K_f}+\mathbf{Q_p})\text{)}}\}\text{Im}\{\eta_{G,I}(\mathbf{K_f},\mathbf{Q_p})\}\Big)^2\bigg]\nonumber\\-E\bigg[2\text{Re}\{e^{-i\text{ arg(}A^*(\mathbf{K_f})A(\mathbf{K_f}-\mathbf{Q_p})-A(\mathbf{K_f})A^*(\mathbf{K_f}+\mathbf{Q_p})\text{)}}\}\text{Re}\{\eta_{G,I}(\mathbf{K_f},\mathbf{Q_p})\}\nonumber\\\times\text{Im}\{e^{-i\text{ arg(}A^*(\mathbf{K_f})A(\mathbf{K_f}-\mathbf{Q_p})-A(\mathbf{K_f})A^*(\mathbf{K_f}+\mathbf{Q_p})\text{)}}\}\text{Im}\{\eta_{G,I}(\mathbf{K_f},\mathbf{Q_p})\}\bigg]\nonumber\\
    =\Big(\text{Re}\{e^{-i\text{ arg(}A^*(\mathbf{K_f})A(\mathbf{K_f}-\mathbf{Q_p})-A(\mathbf{K_f})A^*(\mathbf{K_f}+\mathbf{Q_p})\text{)}}\Big)^2E\bigg[\Big(\text{Re}\{\eta_{G,I}(\mathbf{K_f},\mathbf{Q_p})\}\Big)^2\bigg]\nonumber\\
    +\Big(\text{Im}\{e^{-i\text{ arg(}A^*(\mathbf{K_f})A(\mathbf{K_f}-\mathbf{Q_p})-A(\mathbf{K_f})A^*(\mathbf{K_f}+\mathbf{Q_p})\text{)}}\}\Big)^2E\bigg[\Big(\text{Im}\{\eta_{G,I}(\mathbf{K_f},\mathbf{Q_p})\}\Big)^2\bigg]\nonumber\\
    -2\text{Re}\{e^{-i\text{ arg(}A^*(\mathbf{K_f})A(\mathbf{K_f}-\mathbf{Q_p})-A(\mathbf{K_f})A^*(\mathbf{K_f}+\mathbf{Q_p})\text{)}}\}\text{Im}\{e^{-i\text{ arg(}A^*(\mathbf{K_f})A(\mathbf{K_f}-\mathbf{Q_p})-A(\mathbf{K_f})A^*(\mathbf{K_f}+\mathbf{Q_p})\text{)}}\}\nonumber\\
    \times E\bigg[\Big(\text{Re}\{\eta_{G,I}(\mathbf{K_f},\mathbf{Q_p})\}\text{Im}\{\eta_{G,I}(\mathbf{K_f},\mathbf{Q_p})\}\Big)\bigg]\\
    \text{Var}(\text{Im}\{X_{G,TO\text{ sum prep},I}(\mathbf{K_f},\mathbf{Q_p})\})
    =E\bigg[\Big(\text{Re}\{e^{-i\text{ arg(}A^*(\mathbf{K_f})A(\mathbf{K_f}-\mathbf{Q_p})-A(\mathbf{K_f})A^*(\mathbf{K_f}+\mathbf{Q_p})\text{)}}\}\text{Im}\{\eta_{G,I}(\mathbf{K_f},\mathbf{Q_p})\}\Big)^2\bigg]\nonumber\\
    +E\bigg[\Big(\text{Im}\{e^{-i\text{ arg(}A^*(\mathbf{K_f})A(\mathbf{K_f}-\mathbf{Q_p})-A(\mathbf{K_f})A^*(\mathbf{K_f}+\mathbf{Q_p})\text{)}}\}\text{Re}\{\eta_{G,I}(\mathbf{K_f},\mathbf{Q_p})\}\Big)^2\bigg]\nonumber\\+E\bigg[2\text{Re}\{e^{-i\text{ arg(}A^*(\mathbf{K_f})A(\mathbf{K_f}-\mathbf{Q_p})-A(\mathbf{K_f})A^*(\mathbf{K_f}+\mathbf{Q_p})\text{)}}\}\text{Im}\{\eta_{G,I}(\mathbf{K_f},\mathbf{Q_p})\}\nonumber\\\times\text{Im}\{e^{-i\text{ arg(}A^*(\mathbf{K_f})A(\mathbf{K_f}-\mathbf{Q_p})-A(\mathbf{K_f})A^*(\mathbf{K_f}+\mathbf{Q_p})\text{)}}\}\text{Re}\{\eta_{G,I}(\mathbf{K_f},\mathbf{Q_p})\}\bigg]\nonumber\\
    =\Big(\text{Re}\{e^{-i\text{ arg(}A^*(\mathbf{K_f})A(\mathbf{K_f}-\mathbf{Q_p})-A(\mathbf{K_f})A^*(\mathbf{K_f}+\mathbf{Q_p})\text{)}}\Big)^2E\bigg[\Big(\text{Im}\{\eta_{G,I}(\mathbf{K_f},\mathbf{Q_p})\}\Big)^2\bigg]\nonumber\\
    +\Big(\text{Im}\{e^{-i\text{ arg(}A^*(\mathbf{K_f})A(\mathbf{K_f}-\mathbf{Q_p})-A(\mathbf{K_f})A^*(\mathbf{K_f}+\mathbf{Q_p})\text{)}}\}\Big)^2E\bigg[\Big(\text{Re}\{\eta_{G,I}(\mathbf{K_f},\mathbf{Q_p})\}\Big)^2\bigg]\nonumber\\
    +2\text{Re}\{e^{-i\text{ arg(}A^*(\mathbf{K_f})A(\mathbf{K_f}-\mathbf{Q_p})-A(\mathbf{K_f})A^*(\mathbf{K_f}+\mathbf{Q_p})\text{)}}\}\text{Im}\{e^{-i\text{ arg(}A^*(\mathbf{K_f})A(\mathbf{K_f}-\mathbf{Q_p})-A(\mathbf{K_f})A^*(\mathbf{K_f}+\mathbf{Q_p})\text{)}}\}\nonumber\\
    \times E\bigg[\Big(\text{Im}\{\eta_{G,I}(\mathbf{K_f},\mathbf{Q_p})\}\text{Re}\{\eta_{G,I}(\mathbf{K_f},\mathbf{Q_p})\}\Big)\bigg]
\end{eqnarray}
Considering that the real and imaginary parts of $\eta_{G, I}(\mathbf{K_f},\mathbf{Q_p})$ are independent,
\begin{equation}\label{G_real_imag_noise_indep}
    E[\text{Re}\{\eta_{G,I}(\mathbf{K_f},\mathbf{Q_p})\}\text{Im}\{\eta_{G,I}(\mathbf{K_f},\mathbf{Q_p})\}]=E[\text{Re}\{\eta_{G,I}(\mathbf{K_f},\mathbf{Q_p})\}]E[\text{Im}\{\eta_{G,I}(\mathbf{K_f},\mathbf{Q_p})\}]=0
\end{equation}
Using Eq. \eqref{G_real_imag_noise_indep},
\begin{align}
    \text{Var}(\text{Re}\{X_{G,TO\text{ sum prep},I}(\mathbf{K_f},\mathbf{Q_p})\})=\Big(\text{Re}\{e^{-i\text{ arg(}A^*(\mathbf{K_f})A(\mathbf{K_f}-\mathbf{Q_p})-A(\mathbf{K_f})A^*(\mathbf{K_f}+\mathbf{Q_p})\text{)}}\Big)^2E\bigg[\Big(\text{Re}\{\eta_{G,I}(\mathbf{K_f},\mathbf{Q_p})\}\Big)^2\bigg]\nonumber\\
    +\Big(\text{Im}\{e^{-i\text{ arg(}A^*(\mathbf{K_f})A(\mathbf{K_f}-\mathbf{Q_p})-A(\mathbf{K_f})A^*(\mathbf{K_f}+\mathbf{Q_p})\text{)}}\}\Big)^2E\bigg[\Big(\text{Im}\{\eta_{G,I}(\mathbf{K_f},\mathbf{Q_p})\}\Big)^2\bigg]\\
    \text{Var}(\text{Im}\{X_{G,TO\text{ sum prep},I}(\mathbf{K_f},\mathbf{Q_p})\})=\Big(\text{Re}\{e^{-i\text{ arg(}A^*(\mathbf{K_f})A(\mathbf{K_f}-\mathbf{Q_p})-A(\mathbf{K_f})A^*(\mathbf{K_f}+\mathbf{Q_p})\text{)}}\Big)^2E\bigg[\Big(\text{Im}\{\eta_{G,I}(\mathbf{K_f},\mathbf{Q_p})\}\Big)^2\bigg]\nonumber\\
    +\Big(\text{Im}\{e^{-i\text{ arg(}A^*(\mathbf{K_f})A(\mathbf{K_f}-\mathbf{Q_p})-A(\mathbf{K_f})A^*(\mathbf{K_f}+\mathbf{Q_p})\text{)}}\}\Big)^2E\bigg[\Big(\text{Re}\{\eta_{G,I}(\mathbf{K_f},\mathbf{Q_p})\}\Big)^2\bigg]
\end{align}
Using Eq. \eqref{Variance_G_real} and Eq. \eqref{Variance_G_imag}
\begin{align}
    \text{Var}(\text{Re}\{X_{G,TO\text{ sum prep},I}(\mathbf{K_f},\mathbf{Q_p})\})=\frac{Dd_{x,y}^2}{2N_\alpha}N_xN_y\nonumber\\
    \times\Big(\Big(\text{Im}\{e^{-i\text{ arg(}A^*(\mathbf{K_f})A(\mathbf{K_f}-\mathbf{Q_p})-A(\mathbf{K_f})A^*(\mathbf{K_f}+\mathbf{Q_p})\text{)}}\}\Big)^2+\Big(\text{Re}\{e^{-i\text{ arg(}A^*(\mathbf{K_f})A(\mathbf{K_f}-\mathbf{Q_p})-A(\mathbf{K_f})A^*(\mathbf{K_f}+\mathbf{Q_p})\text{)}}\}\Big)^2\Big)\nonumber\\
    =\frac{Dd_{x,y}^2}{2N_\alpha}N_xN_y\\
    \text{Var}(\text{Im}\{X_{G,TO\text{ sum prep},I}(\mathbf{K_f},\mathbf{Q_p})\})=\frac{Dd_{x,y}^2}{2N_\alpha}N_xN_y\nonumber\\
    \times\Big(\Big(\text{Im}\{e^{-i\text{ arg(}A^*(\mathbf{K_f})A(\mathbf{K_f}-\mathbf{Q_p})-A(\mathbf{K_f})A^*(\mathbf{K_f}+\mathbf{Q_p})\text{)}}\}\Big)^2+\Big(\text{Re}\{e^{-i\text{ arg(}A^*(\mathbf{K_f})A(\mathbf{K_f}-\mathbf{Q_p})-A(\mathbf{K_f})A^*(\mathbf{K_f}+\mathbf{Q_p})\text{)}}\}\Big)^2\Big)\nonumber\\
    =\frac{Dd_{x,y}^2}{2N_\alpha}N_xN_y
\end{align}
\section{Symmetry of $X_{G,TO\text{ sum prep},I}(\mathbf{K_f},\mathbf{Q_p})$}\label{app_symm_xgto_sumprep}
As stated in Eq. \eqref{4D_STEM_TO_sumprep_random_proc}, the modified 4D-STEM dataset, prepared for triple overlap region summation under defocus can be described as
\begin{align}
    X_{G,TO\text{ sum prep},I}(\mathbf{K_f},\mathbf{Q_p})=\frac{N_xN_yDd_{x,y}^2}{N_\alpha}&\Big(\vert A(\mathbf{K_f})\vert^2\delta(\mathbf{Q_p})+\Big\vert A^*(\mathbf{K_f})A(\mathbf{K_f}-\mathbf{Q_p})-A(\mathbf{K_f})A^*(\mathbf{K_f}+\mathbf{Q_p})\Big\vert\Psi_s(\mathbf{Q_p})\Big)\nonumber\\&+e^{-i\text{ arg(}A^*(\mathbf{K_f})A(\mathbf{K_f}-\mathbf{Q_p})-A(\mathbf{K_f})A^*(\mathbf{K_f}+\mathbf{Q_p})\text{)}}\eta_{G, I}(\mathbf{K_f},\mathbf{Q_p})
\end{align}
To aid readability, we introduce the complex number 
\begin{equation}
    B(\mathbf{K_f},\mathbf{Q_p})=A^*(\mathbf{K_f})A(\mathbf{K_f}-\mathbf{Q_p})-A(\mathbf{K_f})A^*(\mathbf{K_f}+\mathbf{Q_p})
\end{equation}
\begin{align}
    X_{G,TO\text{ sum prep},I}(\mathbf{K_f},\mathbf{Q_p})=\frac{N_xN_yDd_{x,y}^2}{N_\alpha}\Big(\big\vert A(\mathbf{K_f})\vert^2\delta(\mathbf{Q_p})+\big\vert B(\mathbf{K_f},\mathbf{Q_p})\big\vert\Psi_s(\mathbf{Q_p})\Big)+e^{-i\text{ arg(}B(\mathbf{K_f},\mathbf{Q_p})\text{)}}\eta_{G, I}(\mathbf{K_f},\mathbf{Q_p})\label{x_g_to_sumprep_simplified}
\end{align}
Using the linearity of complex conjugation, the symmetry of $B(\mathbf{K_f},\mathbf{Q_p})$ in $\mathbf{Q_p}$ space can be determined.
\begin{align}
    B(\mathbf{K_f},\mathbf{-Q_p})&=A^*(\mathbf{K_f})A(\mathbf{K_f}+\mathbf{Q_p})-A(\mathbf{K_f})A^*(\mathbf{K_f}-\mathbf{Q_p})\nonumber\\&=-\big(A(\mathbf{K_f})A^*(\mathbf{K_f}-\mathbf{Q_p})-A^*(\mathbf{K_f})A(\mathbf{K_f}+\mathbf{Q_p})\big)\nonumber\\
    &=-\big(\big(A^*(\mathbf{K_f})A(\mathbf{K_f}-\mathbf{Q_p})-A(\mathbf{K_f})A^*(\mathbf{K_f}+\mathbf{Q_p})\big)^*\big)\nonumber\\
    &=-B^*(\mathbf{K_f},\mathbf{Q_p})\label{B_Qp_symmetry}
\end{align}
Eq. \eqref{B_Qp_symmetry} can now be used to determine the symmetry of $e^{-i\text{ arg(}B(\mathbf{Q_p})\text{)}}$.
\begin{align}
    e^{-i\text{ arg(}B(\mathbf{K_f},\mathbf{-Q_p})\text{)}}=e^{-i\text{ arg(}-B^*(\mathbf{K_f},\mathbf{Q_p})\text{)}}=e^{-i(\pi-\text{ arg(}B(\mathbf{K_f},\mathbf{Q_p})\text{)})}=e^{-i\pi+i\text{ arg(}B(\mathbf{K_f},\mathbf{Q_p})\text{)}}\nonumber\\
    =-\big(e^{-i\text{ arg(}B(\mathbf{K_f},\mathbf{Q_p})\text{)}}\big)^*\label{exp_B_Qp_symmetry}
\end{align}
Eq. \eqref{exp_B_Qp_symmetry} can now be used to find an expression for $X_{G,TO\text{ sum prep},I}(\mathbf{K_f},\mathbf{-Q_p})$.
\begin{align}
    X_{G,TO\text{ sum prep},I}(\mathbf{K_f},\mathbf{-Q_p})=\frac{N_xN_yDd_{x,y}^2}{N_\alpha}\Big(\big\vert A(\mathbf{K_f})\vert^2\delta(-\mathbf{Q_p})+\big\vert B(\mathbf{K_f},-\mathbf{Q_p})\big\vert\Psi_s(-\mathbf{Q_p})\Big)\nonumber\\+e^{-i\text{ arg(}B(\mathbf{K_f},-\mathbf{Q_p})\text{)}}\eta_{G, I}(\mathbf{K_f},-\mathbf{Q_p})\nonumber\\
    =\frac{N_xN_yDd_{x,y}^2}{N_\alpha}\Big(\big\vert A(\mathbf{K_f})\vert^2\delta(\mathbf{Q_p})-\big\vert B(\mathbf{K_f},\mathbf{Q_p})\big\vert\Psi_s^*(\mathbf{Q_p})\Big)-\big(e^{-i\text{ arg(}B(\mathbf{K_f},\mathbf{Q_p})\text{)}}\big)^*\eta_{G, I}^*(\mathbf{K_f},\mathbf{Q_p})\nonumber\\
    =\frac{N_xN_yDd_{x,y}^2}{N_\alpha}\big\vert A(\mathbf{K_f})\vert^2\delta(\mathbf{Q_p})-\bigg[\frac{N_xN_yDd_{x,y}^2}{N_\alpha}\big\vert B(\mathbf{K_f},\mathbf{Q_p})\big\vert\Psi_s(\mathbf{Q_p})+e^{-i\text{ arg(}B(\mathbf{K_f},\mathbf{Q_p})\text{)}}\eta_{G, I}(\mathbf{K_f},\mathbf{Q_p})\bigg]^*\label{x_g_to_minus_sumprep_simplified}
\end{align}
Using the fact that for any complex number $z$, $\text{Im}(z)=\text{Im}(-z^*)$ and comparing with Eq. \eqref{x_g_to_sumprep_simplified},
\begin{align}
    \text{Im}\{X_{G,TO\text{ sum prep},I}(\mathbf{K_f},\mathbf{-Q_p})\}=\text{Im}\bigg\{\frac{N_xN_yDd_{x,y}^2}{N_\alpha}\big\vert A(\mathbf{K_f})\vert^2\delta(\mathbf{Q_p})\bigg\}\nonumber\\+\text{Im}\bigg\{-\bigg[\frac{N_xN_yDd_{x,y}^2}{N_\alpha}\big\vert B(\mathbf{K_f}, \mathbf{Q_p})\big\vert\Psi_s(\mathbf{Q_p})+e^{-i\text{ arg(}B(\mathbf{Q_p})\text{)}}\eta_{G, I}(\mathbf{K_f},\mathbf{Q_p})\bigg]^*\bigg\}\nonumber\\
    =\text{Im}\bigg\{\frac{N_xN_yDd_{x,y}^2}{N_\alpha}\big\vert B(\mathbf{K_f},\mathbf{Q_p})\big\vert\Psi_s(\mathbf{Q_p})+e^{-i\text{ arg(}B(\mathbf{Q_p})\text{)}}\eta_{G, I}(\mathbf{K_f},\mathbf{Q_p})\bigg\}\nonumber\\
    =\text{Im}\{X_{G,TO\text{ sum prep},I}(\mathbf{K_f},\mathbf{Q_p})\}
\end{align}
\section{Equivalence of summation over triple overlap regions in $G_{SSB,TO \text{ sum prep}, I}(\mathbf{K_f},\mathbf{Q_p})$}
To aid readability in this section, we introduce a complex quantity $X_{F,I}(\mathbf{K_f},\mathbf{Q_p})$,
\begin{equation}\label{F_helper_var}
    X_{F,I}(\mathbf{K_f},\mathbf{Q_p})=\frac{N_xN_yDd_{x,y}^2}{N_\alpha}\big\vert B(\mathbf{K_f},\mathbf{Q_p})\big\vert\Psi_s(\mathbf{Q_p})+e^{-i\text{ arg(}B(\mathbf{K_f},\mathbf{Q_p})\text{)}}\eta_{G, I}(\mathbf{K_f},\mathbf{Q_p})
\end{equation}
By substituting Eq. \eqref{F_helper_var} into Eq. \eqref{x_g_to_sumprep_simplified} and into Eq. \eqref{x_g_to_minus_sumprep_simplified}, simplified expressions of the modified dataset prior to SSB and triple overlap summation can be found.
\begin{align}
    X_{G,TO\text{ sum prep},I}(\mathbf{K_f},\mathbf{Q_p})=\frac{N_xN_yDd_{x,y}^2}{N_\alpha}\big\vert A(\mathbf{K_f})\vert^2\delta(\mathbf{Q_p})+X_{F,I}(\mathbf{K_f},\mathbf{Q_p})\\
    X_{G,TO\text{ sum prep},I}(\mathbf{K_f},-\mathbf{Q_p})=\frac{N_xN_yDd_{x,y}^2}{N_\alpha}\big\vert A(\mathbf{K_f})\vert^2\delta(\mathbf{Q_p})-X_{F,I}^*(\mathbf{K_f},\mathbf{Q_p})
\end{align}
It is important to note that there is no modification in $\mathbf{K_f}$ space. That means if a summations are performed, they would also be equivalent by linearity as long as they are performed over the same pixel indices. For example the pixel indices are the same for a summation over the inner triple overlap region at $\mathbf{Q_p}$ and the outer triple overlap region at $-\mathbf{Q_p}$ and vice versa. Using these facts the sums over the 4 separate triple overlap regions can now be related to each other
\begin{align}
    X_{G:TO,i,I}(\mathbf{Q_p})&=\sum_{TO,i,\mathbf{K_f}}\frac{N_xN_yDd_{x,y}^2}{N_\alpha}\big\vert A(\mathbf{K_f})\vert^2\delta(\mathbf{Q_p})+X_{F,I}(\mathbf{K_f},\mathbf{Q_p})\nonumber\\
    &=N_{TO/2,I}(\mathbf{Q_p})+X_{F:TO,i,I}(\mathbf{Q_p})\\
    X_{G:TO,o,I}(\mathbf{Q_p})&=\sum_{TO,o,\mathbf{K_f}}\frac{N_xN_yDd_{x,y}^2}{N_\alpha}\big\vert A(\mathbf{K_f})\vert^2\delta(\mathbf{Q_p})+X_{F,I}(\mathbf{K_f},\mathbf{Q_p})\nonumber\\
    &=N_{TO/2,I}(\mathbf{Q_p})+X_{F:TO,o,I}(\mathbf{Q_p})\\
    X_{G:TO,i,I}(-\mathbf{Q_p})&=\sum_{TO,i,\mathbf{K_f}}\frac{N_xN_yDd_{x,y}^2}{N_\alpha}\big\vert A(\mathbf{K_f})\vert^2\delta(\mathbf{Q_p})+X_{F,I}(\mathbf{K_f},-\mathbf{Q_p})\nonumber\\
    &=\sum_{TO,o,\mathbf{K_f}}\frac{N_xN_yDd_{x,y}^2}{N_\alpha}\big\vert A(\mathbf{K_f})\vert^2\delta(\mathbf{Q_p})-X_{F,I}^*(\mathbf{K_f},\mathbf{Q_p})\nonumber\\&=N_{TO/2,I}(\mathbf{Q_p})-X_{F:TO,o,I}^*(\mathbf{Q_p})\\
    X_{G:TO,o,I}(-\mathbf{Q_p})&=\sum_{TO,o,\mathbf{K_f}}\frac{N_xN_yDd_{x,y}^2}{N_\alpha}\big\vert A(\mathbf{K_f})\vert^2\delta(\mathbf{Q_p})+X_{F,I}(\mathbf{K_f},-\mathbf{Q_p})\nonumber\\&=\sum_{TO,i,\mathbf{K_f}}\frac{N_xN_yDd_{x,y}^2}{N_\alpha}\big\vert A(\mathbf{K_f})\vert^2\delta(\mathbf{Q_p})-X_{F,I}^*(\mathbf{K_f},\mathbf{Q_p})\nonumber\\
    &=N_{TO/2,I}(\mathbf{Q_p})-X_{F:TO,i,I}^*(\mathbf{Q_p})
\end{align}
where $N_{TO/2,I}(\mathbf{Q_p})$ is the number of pixels in either the inner or outer half of the triple overlap region scaled by the constant $\frac{N_xN_yDd_{x,y}^2}{N_\alpha}$.
\subsection{TO I}
The first option for the summation of a single sideband and the triple overlap region can be written as,
\begin{align}
    X_{G:SSB,TO\text{ I},I}(\mathbf{Q_p})&=X_{G:SSB,I}(\mathbf{Q_p})+X_{G:TO,i,I}(\mathbf{Q_p})-(X_{G:TO,o,I}(-\mathbf{Q_p}))^*\nonumber\\
    &=X_{G:SSB,I}(\mathbf{Q_p})+N_{TO/2,I}(\mathbf{Q_p})+X_{F:TO,i,I}(\mathbf{Q_p})-(N_{TO/2,I}(\mathbf{Q_p})-X_{F:TO,i,I}^*(\mathbf{Q_p}))^*\nonumber\\
    &=X_{G:SSB,I}(\mathbf{Q_p})+2X_{F:TO,i,I}(\mathbf{Q_p})\\
    X_{G:SSB,TO\text{ I},I}(-\mathbf{Q_p})&=X_{G:SSB,I}(-\mathbf{Q_p})+X_{G:TO,i,I}(-\mathbf{Q_p})-(X_{G:TO,o,I}(\mathbf{Q_p}))^*\nonumber\\
    &=X_{G:SSB,I}(-\mathbf{Q_p})+N_{TO/2,I}(\mathbf{Q_p})-X_{F:TO,o,I}^*(\mathbf{Q_p})-(N_{TO/2,I}(\mathbf{Q_p})+X_{F:TO,o,I}(\mathbf{Q_p}))^*\nonumber\\
    &=X_{G:SSB,I}(-\mathbf{Q_p})-2X_{F:TO,o,I}^*(\mathbf{Q_p})
\end{align}
The image formed from this summation is defined as
\begin{align}
    X_{\text{Image}:SSB,TO \text{ I}, I}(\mathbf{R_p})=\text{Im}\Bigg\{\frac{1}{N_xN_y}\sum^{N_x}_{Q_{px}=1}\sum^{N_y}_{Q_{py}=1}\bigg[X_{G:SSB,TO\text{ I},I}(Q_{px},Q_{py})\bigg]\exp\bigg(i2\pi\bigg(\frac{Q_{px}R_x}{N_x}+\frac{Q_{py}R_y}{N_y}\bigg)\bigg)\Bigg\}\nonumber\\
    =\text{Im}\Bigg\{\frac{1}{N_xN_y}\sum^{N_x}_{Q_{px}=1}\sum^{N_y}_{Q_{py}=1}\bigg[X_{G:SSB,I}(Q_{px},Q_{py})+2X_{F:TO,i,I}(Q_{px},Q_{py})\bigg]\exp\bigg(i2\pi\bigg(\frac{Q_{px}R_x}{N_x}+\frac{Q_{py}R_y}{N_y}\bigg)\bigg)\Bigg\}
\end{align}
The iFT summation can now be rewritten as a sum over half as many components by combining the $\mathbf{-Q_p}$ and $\mathbf{+Q_p}$ terms. The 0 spatial frequency component is neglected.
\begin{align}
    X_{\text{Image}:SSB, TO \text{ I}, I}(\mathbf{R_p})=\text{Im}\Bigg\{\frac{1}{N_xN_y}\sum^{\frac{N_x-1}{2}}_{Q_{px}=1}\sum^{\frac{N_y-1}{2}}_{Q_{py}=1}\Bigg[X_{G:SSB,TO\text{ I},I}(Q_{px},Q_{py})\exp\bigg(i2\pi\bigg(\frac{Q_{px}R_x}{N_x}+\frac{Q_{py}R_y}{N_y}\bigg)\bigg)\nonumber\\+X_{G:SSB,TO\text{ I},I}(N_x-Q_{px},N_y-Q_{py})\exp\bigg(i2\pi\bigg(\frac{-Q_{px}R_x}{N_x}+\frac{-Q_{py}R_y}{N_y}\bigg)\bigg)\Bigg]\nonumber\\=\text{Im}\Bigg\{\frac{1}{N_xN_y}\sum^{\frac{N_x-1}{2}}_{Q_{px}=1}\sum^{\frac{N_y-1}{2}}_{Q_{py}=1}\Bigg[\bigg[X_{G:SSB,I}(Q_{px},Q_{py})+2X_{F:TO,i,I}(Q_{px},Q_{py})\bigg]\exp\bigg(i2\pi\bigg(\frac{Q_{px}R_x}{N_x}+\frac{Q_{py}R_y}{N_y}\bigg)\bigg)\nonumber\\
    +\bigg[X_{G:SSB,I}(N_x-Q_{px},N_y-Q_{py})-2X_{F:TO,o,I}^*(Q_{px},Q_{py})\bigg]\exp\bigg(i2\pi\bigg(\frac{-Q_{px}R_x}{N_x}+\frac{-Q_{py}R_y}{N_y}\bigg)\bigg)\Bigg]\Bigg\}\nonumber\\
    =\text{Im}\Bigg\{\frac{1}{N_xN_y}\sum^{\frac{N_x-1}{2}}_{Q_{px}=1}\sum^{\frac{N_y-1}{2}}_{Q_{py}=1}\Bigg[X_{G:SSB,I}(Q_{px},Q_{py})\exp\bigg(i2\pi\bigg(\frac{Q_{px}R_x}{N_x}+\frac{Q_{py}R_y}{N_y}\bigg)\bigg)\nonumber\\+2X_{F:TO,i,I}(Q_{px},Q_{py})\exp\bigg(i2\pi\bigg(\frac{Q_{px}R_x}{N_x}+\frac{Q_{py}R_y}{N_y}\bigg)\bigg)\nonumber\\
    +X_{G:SSB,I}(N_x-Q_{px},N_y-Q_{py})\bigg(\exp\bigg(i2\pi\bigg(\frac{Q_{px}R_x}{N_x}+\frac{Q_{py}R_y}{N_y}\bigg)\bigg)\bigg)^*\nonumber\\-2X_{F:TO,o,I}^*(Q_{px},Q_{py})\bigg(\exp\bigg(i2\pi\bigg(\frac{Q_{px}R_x}{N_x}+\frac{Q_{py}R_y}{N_y}\bigg)\bigg)\bigg)^*\Bigg]\Bigg\}\nonumber\\
    =\frac{1}{N_xN_y}\sum^{\frac{N_x-1}{2}}_{Q_{px}=1}\sum^{\frac{N_y-1}{2}}_{Q_{py}=1}\Bigg[\text{Im}\bigg\{X_{G:SSB,I}(Q_{px},Q_{py})\exp\bigg(i2\pi\bigg(\frac{Q_{px}R_x}{N_x}+\frac{Q_{py}R_y}{N_y}\bigg)\bigg)\bigg\}\nonumber\\+2\text{Im}\bigg\{X_{F:TO,i,I}(Q_{px},Q_{py})\exp\bigg(i2\pi\bigg(\frac{Q_{px}R_x}{N_x}+\frac{Q_{py}R_y}{N_y}\bigg)\bigg)\Bigg\}\nonumber\\
    +\text{Im}\bigg\{X_{G:SSB,I}(N_x-Q_{px},N_y-Q_{py})\bigg(\exp\bigg(i2\pi\bigg(\frac{Q_{px}R_x}{N_x}+\frac{Q_{py}R_y}{N_y}\bigg)\bigg)\bigg)^*\bigg\}\nonumber\\+2\text{Im}\bigg\{X_{F:TO,o,I}(Q_{px},Q_{py})\exp\bigg(i2\pi\bigg(\frac{Q_{px}R_x}{N_x}+\frac{Q_{py}R_y}{N_y}\bigg)\bigg)\bigg\}\Bigg]
\end{align}
\begin{figure}[h]
\includegraphics[width=\textwidth]{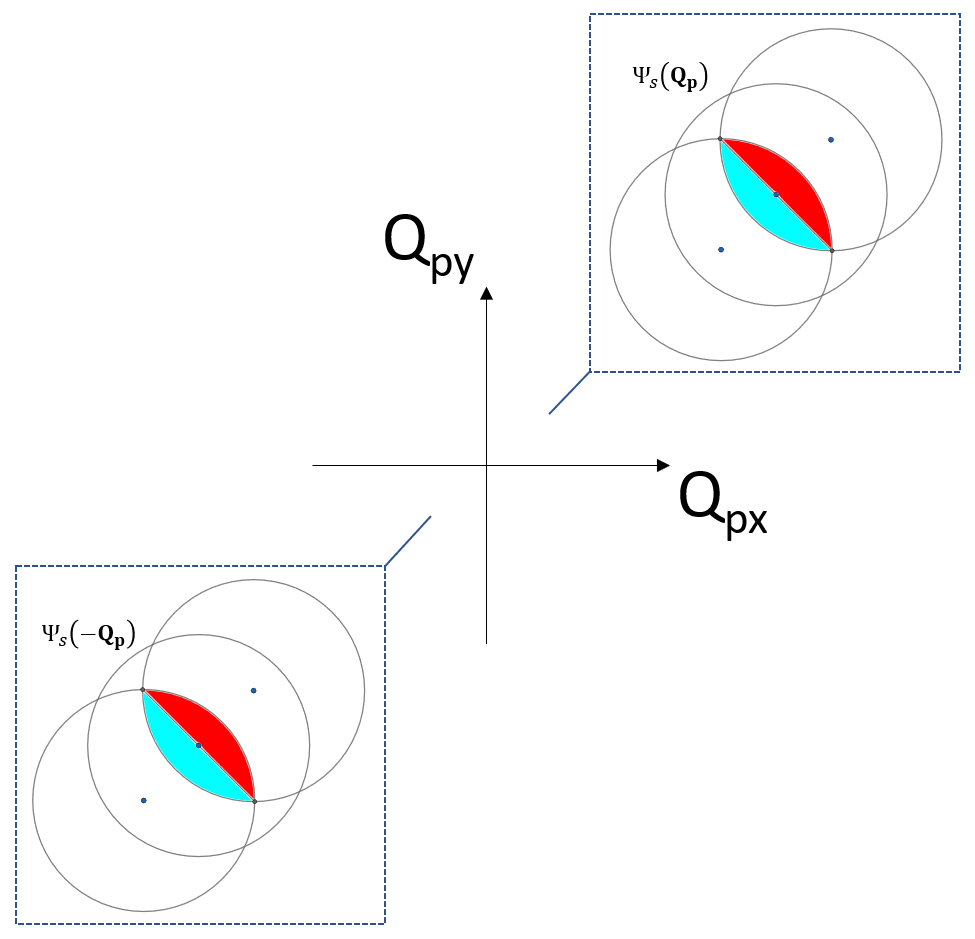}
\caption{Diagram showing the way the summation is performed in the first option of single sideband and triple overlap ptychography. The red shaded region shows the area that contributes to the sum at $\mathbf{Q_p}$. The blue shaded region shows the area that contributes to the sum at $-\mathbf{Q_p}$. \label{fig:to_I}}
\end{figure}
\clearpage
\subsection{TO II}
\begin{figure}[h]
\includegraphics[width=\textwidth]{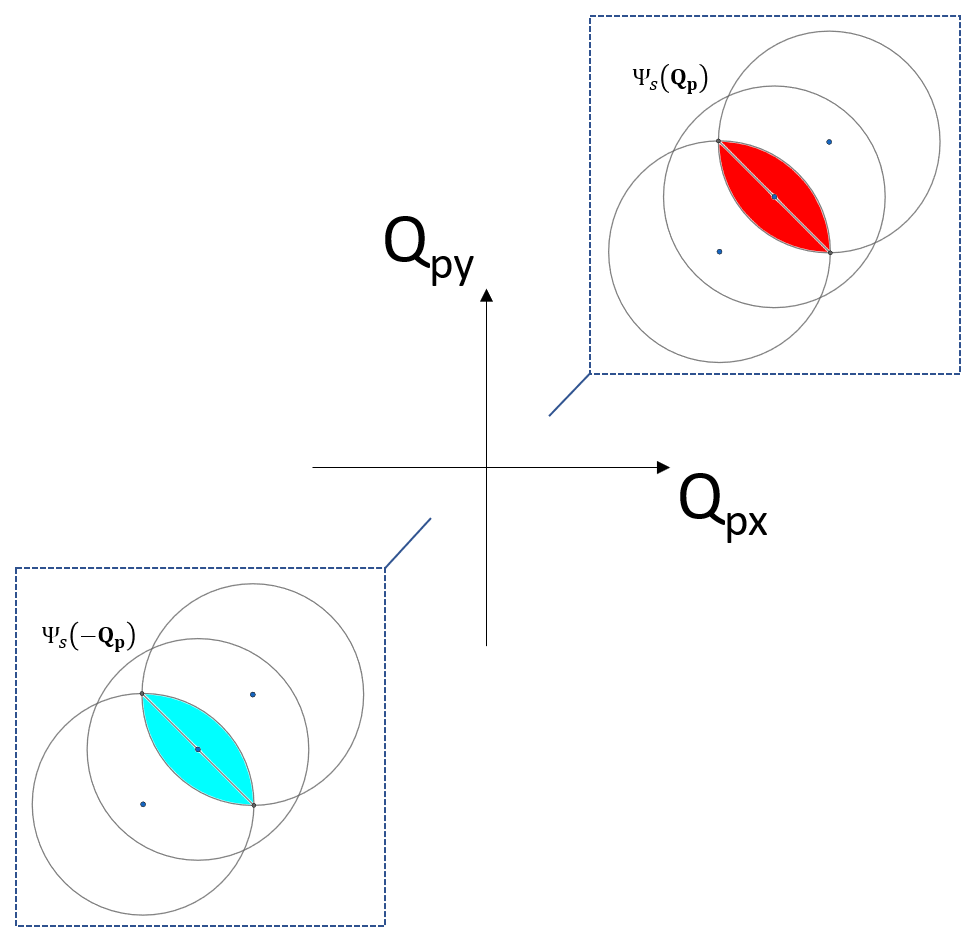}
\caption{Diagram showing the way the summation is performed in the second option of single sideband and triple overlap ptychography. The red shaded region shows the area that contributes to the sum at $\mathbf{Q_p}$. The blue shaded region shows the area that contributes to the sum at $-\mathbf{Q_p}$. \label{fig:to_II}}
\end{figure}
The second option for the summation of a single sideband and the triple overlap region can be written as,
\begin{equation}
    X_{G:SSB,TO\text{ II},I}(\mathbf{Q_p})=X_{G:SSB,I}(\mathbf{Q_p})+X_{G:TO,i,I}(\mathbf{Q_p})+X_{G:TO,o,I}(\mathbf{Q_p})
\end{equation}
\begin{align}
    X_{G:SSB,TO\text{ II},I}(\mathbf{Q_p})&=X_{G:SSB,I}(\mathbf{Q_p})+X_{G:TO,i,I}(\mathbf{Q_p})+X_{G:TO,o,I}(\mathbf{Q_p})\nonumber\\
    &=X_{G:SSB,I}(\mathbf{Q_p})+N_{TO/2,I}(\mathbf{Q_p})+X_{F:TO,i,I}(\mathbf{Q_p})+N_{TO/2,I}(\mathbf{Q_p})+X_{F:TO,o,I}(\mathbf{Q_p})\nonumber\\
    &=X_{G:SSB,I}(\mathbf{Q_p})+2N_{TO/2,I}(\mathbf{Q_p})+X_{F:TO,i,I}(\mathbf{Q_p})+X_{F:TO,o,I}(\mathbf{Q_p})\\
    X_{G:SSB,TO\text{ II},I}(-\mathbf{Q_p})&=X_{G:SSB,I}(-\mathbf{Q_p})+X_{G:TO,i,I}(-\mathbf{Q_p})+X_{G:TO,o,I}(-\mathbf{Q_p})\nonumber\\
    &=X_{G:SSB,I}(-\mathbf{Q_p})+N_{TO/2,I}(\mathbf{Q_p})-X_{F:TO,o,I}^*(\mathbf{Q_p})+N_{TO/2,I}(\mathbf{Q_p})-X_{F:TO,i,I}^*(\mathbf{Q_p})\nonumber\\
    &=X_{G:SSB,I}(-\mathbf{Q_p})+2N_{TO/2,I}(\mathbf{Q_p})-X_{F:TO,o,I}^*(\mathbf{Q_p})-X_{F:TO,i,I}^*(\mathbf{Q_p})
\end{align}
The image formed from this summation is defined as
\begin{align}
    X_{\text{Image}:SSB,TO \text{ I}, I}(\mathbf{R_p})=\text{Im}\Bigg\{\frac{1}{N_xN_y}\sum^{N_x}_{Q_{px}=1}\sum^{N_y}_{Q_{py}=1}\bigg[X_{G:SSB,TO\text{ II},I}(Q_{px},Q_{py})\bigg]\exp\bigg(i2\pi\bigg(\frac{Q_{px}R_x}{N_x}+\frac{Q_{py}R_y}{N_y}\bigg)\bigg)\Bigg\}\nonumber\\
    =\text{Im}\Bigg\{\frac{1}{N_xN_y}\sum^{N_x}_{Q_{px}=1}\sum^{N_y}_{Q_{py}=1}\bigg[X_{G:SSB,I}(Q_{px},Q_{py})+2N_{TO/2,I}(Q_{px},Q_{py})\nonumber\\+X_{F:TO,i,I}(Q_{px},Q_{py})+X_{F:TO,o,I}(Q_{px},Q_{py})\bigg]\exp\bigg(i2\pi\bigg(\frac{Q_{px}R_x}{N_x}+\frac{Q_{py}R_y}{N_y}\bigg)\bigg)\Bigg\}
\end{align}
The iFT summation can now be rewritten as a sum over half as many components by combining the $\mathbf{-Q_p}$ and $\mathbf{+Q_p}$ terms. The 0 spatial frequency component is neglected.
\begin{align}
    X_{\text{Image}:SSB, TO \text{ II}, I}(\mathbf{R_p})=\text{Im}\Bigg\{\frac{1}{N_xN_y}\sum^{\frac{N_x-1}{2}}_{Q_{px}=1}\sum^{\frac{N_y-1}{2}}_{Q_{py}=1}\Bigg[X_{G:SSB,TO\text{ II},I}(Q_{px},Q_{py})\exp\bigg(i2\pi\bigg(\frac{-Q_{px}R_x}{N_x}+\frac{-Q_{py}R_y}{N_y}\bigg)\bigg)\nonumber\\+X_{G:SSB,TO\text{ II},I}(N_x-Q_{px},N_y-Q_{py})\exp\bigg(i2\pi\bigg(\frac{-Q_{px}R_x}{N_x}+\frac{-Q_{py}R_y}{N_y}\bigg)\bigg)\Bigg]\nonumber\\=\text{Im}\Bigg\{\frac{1}{N_xN_y}\sum^{\frac{N_x-1}{2}}_{Q_{px}=1}\sum^{\frac{N_y-1}{2}}_{Q_{py}=1}\Bigg[\bigg[X_{G:SSB,I}(Q_{px},Q_{py})+2N_{TO/2,I}(Q_{px},Q_{py})+X_{F:TO,i,I}(Q_{px},Q_{py})\nonumber\\+X_{F:TO,o,I}(Q_{px},Q_{py})\bigg]\exp\bigg(i2\pi\bigg(\frac{Q_{px}R_x}{N_x}+\frac{Q_{py}R_y}{N_y}\bigg)\bigg)\nonumber\\
    +\bigg[X_{G:SSB,I}(N_x-Q_{px},N_y-Q_{py})+2N_{TO/2,I}(Q_{px},Q_{py})-X_{F:TO,o,I}^*(Q_{px},Q_{py})\nonumber\\-X_{F:TO,i,I}^*(Q_{px},Q_{py})\bigg]\exp\bigg(i2\pi\bigg(\frac{-Q_{px}R_x}{N_x}+\frac{-Q_{py}R_y}{N_y}\bigg)\bigg)\Bigg]\Bigg\}\nonumber\\
    =\text{Im}\Bigg\{\frac{1}{N_xN_y}\sum^{\frac{N_x-1}{2}}_{Q_{px}=1}\sum^{\frac{N_y-1}{2}}_{Q_{py}=1}\Bigg[X_{G:SSB,I}(Q_{px},Q_{py})\exp\bigg(i2\pi\bigg(\frac{Q_{px}R_x}{N_x}+\frac{Q_{py}R_y}{N_y}\bigg)\bigg)\nonumber\\+2N_{TO/2,I}(Q_{px},Q_{py})\exp\bigg(i2\pi\bigg(\frac{Q_{px}R_x}{N_x}+\frac{Q_{py}R_y}{N_y}\bigg)\bigg)+X_{F:TO,i,I}(Q_{px},Q_{py})\exp\bigg(i2\pi\bigg(\frac{Q_{px}R_x}{N_x}+\frac{Q_{py}R_y}{N_y}\bigg)\bigg)\nonumber\\+X_{F:TO,o,I}(Q_{px},Q_{py})\exp\bigg(i2\pi\bigg(\frac{Q_{px}R_x}{N_x}+\frac{Q_{py}R_y}{N_y}\bigg)\bigg)\nonumber\\
    +X_{G:SSB,I}(N_x-Q_{px},N_y-Q_{py})\bigg(\exp\bigg(i2\pi\bigg(\frac{Q_{px}R_x}{N_x}+\frac{Q_{py}R_y}{N_y}\bigg)\bigg)\bigg)^*\nonumber\\+2N_{TO/2,I}(Q_{px},Q_{py})\bigg(\exp\bigg(i2\pi\bigg(\frac{Q_{px}R_x}{N_x}+\frac{Q_{py}R_y}{N_y}\bigg)\bigg)\bigg)^*-X_{F:TO,o,I}^*(Q_{px},Q_{py})\bigg(\exp\bigg(i2\pi\bigg(\frac{Q_{px}R_x}{N_x}+\frac{Q_{py}R_y}{N_y}\bigg)\bigg)\bigg)^*\nonumber\\-X_{F:TO,i,I}^*(Q_{px},Q_{py})\bigg(\exp\bigg(i2\pi\bigg(\frac{Q_{px}R_x}{N_x}+\frac{Q_{py}R_y}{N_y}\bigg)\bigg)\bigg)^*\Bigg]\Bigg\}\nonumber\\
    =\frac{1}{N_xN_y}\sum^{\frac{N_x-1}{2}}_{Q_{px}=1}\sum^{\frac{N_y-1}{2}}_{Q_{py}=1}\Bigg[\text{Im}\bigg\{X_{G:SSB,I}(Q_{px},Q_{py})\exp\bigg(i2\pi\bigg(\frac{Q_{px}R_x}{N_x}+\frac{Q_{py}R_y}{N_y}\bigg)\bigg)\bigg\}\nonumber\\+2\text{Im}\bigg\{X_{F:TO,i,I}(Q_{px},Q_{py})\exp\bigg(i2\pi\bigg(\frac{Q_{px}R_x}{N_x}+\frac{Q_{py}R_y}{N_y}\bigg)\bigg)\Bigg\}\nonumber\\
    +\text{Im}\bigg\{X_{G:SSB,I}(N_x-Q_{px},N_y-Q_{py})\bigg(\exp\bigg(i2\pi\bigg(\frac{Q_{px}R_x}{N_x}+\frac{Q_{py}R_y}{N_y}\bigg)\bigg)\bigg)^*\bigg\}\nonumber\\+2\text{Im}\bigg\{X_{F:TO,o,I}(Q_{px},Q_{py})\exp\bigg(i2\pi\bigg(\frac{Q_{px}R_x}{N_x}+\frac{Q_{py}R_y}{N_y}\bigg)\bigg)\bigg\}\Bigg]\nonumber\\
    =X_{\text{Image}:SSB, TO \text{ I}, I}(\mathbf{R_p})
\end{align}
\end{appendices}

\bibliographystyle{unsrtnat}
\bibliography{reference}

\begin{thebibliography}{34}
\providecommand{\natexlab}[1]{#1}
\providecommand{\url}[1]{\texttt{#1}}
\expandafter\ifx\csname urlstyle\endcsname\relax
  \providecommand{\doi}[1]{doi: #1}\else
  \providecommand{\doi}{doi: \begingroup \urlstyle{rm}\Url}\fi

\bibitem[Henderson et~al.(1990)Henderson, Baldwin, Ceska, Zemlin, Beckmann, and Downing]{HENDERSON1990899}
R.~Henderson, J.M. Baldwin, T.A. Ceska, F.~Zemlin, E.~Beckmann, and K.H. Downing.
\newblock Model for the structure of bacteriorhodopsin based on high-resolution electron cryo-microscopy.
\newblock \emph{Journal of Molecular Biology}, 213\penalty0 (4):\penalty0 899--929, 1990.
\newblock ISSN 0022-2836.
\newblock \doi{https://doi.org/10.1016/S0022-2836(05)80271-2}.
\newblock URL \url{https://www.sciencedirect.com/science/article/pii/S0022283605802712}.

\bibitem[Penczek et~al.(1992)Penczek, Radermacher, and Frank]{PENCZEK199233}
Pawel Penczek, Michael Radermacher, and Joachim Frank.
\newblock Three-dimensional reconstruction of single particles embedded in ice.
\newblock \emph{Ultramicroscopy}, 40\penalty0 (1):\penalty0 33--53, 1992.
\newblock ISSN 0304-3991.
\newblock \doi{https://doi.org/10.1016/0304-3991(92)90233-A}.
\newblock URL \url{https://www.sciencedirect.com/science/article/pii/030439919290233A}.

\bibitem[Dubochet et~al.(1988)Dubochet, Adrian, Chang, Homo, Lepault, McDowall, and Schultz]{Dubochet_Adrian_Chang_Homo_Lepault_McDowall_Schultz_1988}
Jacques Dubochet, Marc Adrian, Jiin-Ju Chang, Jean-Claude Homo, Jean Lepault, Alasdair~W. McDowall, and Patrick Schultz.
\newblock Cryo-electron microscopy of vitrified specimens.
\newblock \emph{Quarterly Reviews of Biophysics}, 21\penalty0 (2):\penalty0 129–228, 1988.
\newblock \doi{10.1017/S0033583500004297}.

\bibitem[Nakane et~al.(2020)Nakane, Kotecha, Sente, McMullan, Masiulis, Brown, Grigoras, Malinauskaite, Malinauskas, Miehling, Ucha{\'{n}}ski, Yu, Karia, Pechnikova, de~Jong, Keizer, Bischoff, McCormack, Tiemeijer, Hardwick, Chirgadze, Murshudov, Aricescu, and Scheres]{Nakane2020}
Takanori Nakane, Abhay Kotecha, Andrija Sente, Greg McMullan, Simonas Masiulis, Patricia M. G.~E. Brown, Ioana~T. Grigoras, Lina Malinauskaite, Tomas Malinauskas, Jonas Miehling, Tomasz Ucha{\'{n}}ski, Lingbo Yu, Dimple Karia, Evgeniya~V. Pechnikova, Erwin de~Jong, Jeroen Keizer, Maarten Bischoff, Jamie McCormack, Peter Tiemeijer, Steven~W. Hardwick, Dimitri~Y. Chirgadze, Garib Murshudov, A.~Radu Aricescu, and Sjors H.~W. Scheres.
\newblock Single-particle cryo-em at atomic resolution.
\newblock \emph{Nature}, 587\penalty0 (7832):\penalty0 152--156, Nov 2020.
\newblock ISSN 1476-4687.
\newblock \doi{10.1038/s41586-020-2829-0}.
\newblock URL \url{https://doi.org/10.1038/s41586-020-2829-0}.

\bibitem[Yip et~al.(2020)Yip, Fischer, Paknia, Chari, and Stark]{Yip2020}
Ka~Man Yip, Niels Fischer, Elham Paknia, Ashwin Chari, and Holger Stark.
\newblock Atomic-resolution protein structure determination by cryo-em.
\newblock \emph{Nature}, 587\penalty0 (7832):\penalty0 157--161, Nov 2020.
\newblock ISSN 1476-4687.
\newblock \doi{10.1038/s41586-020-2833-4}.
\newblock URL \url{https://doi.org/10.1038/s41586-020-2833-4}.

\bibitem[Zhou et~al.(2020)Zhou, Song, Kim, Pei, Huang, Boyce, Mendon{\c{c}}a, Clare, Siebert, Allen, Liberti, Stuart, Pan, Nellist, Zhang, Kirkland, and Wang]{Zhou2020}
Liqi Zhou, Jingdong Song, Judy~S. Kim, Xudong Pei, Chen Huang, Mark Boyce, Luiza Mendon{\c{c}}a, Daniel Clare, Alistair Siebert, Christopher~S. Allen, Emanuela Liberti, David Stuart, Xiaoqing Pan, Peter~D. Nellist, Peijun Zhang, Angus~I. Kirkland, and Peng Wang.
\newblock Low-dose phase retrieval of biological specimens using cryo-electron ptychography.
\newblock \emph{Nature Communications}, 11\penalty0 (1):\penalty0 2773, Jun 2020.
\newblock ISSN 2041-1723.
\newblock \doi{10.1038/s41467-020-16391-6}.
\newblock URL \url{https://doi.org/10.1038/s41467-020-16391-6}.

\bibitem[Seki et~al.(2018)Seki, Ikuhara, and Shibata]{SEKI2018118}
Takehito Seki, Yuichi Ikuhara, and Naoya Shibata.
\newblock Theoretical framework of statistical noise in scanning transmission electron microscopy.
\newblock \emph{Ultramicroscopy}, 193:\penalty0 118--125, 2018.
\newblock ISSN 0304-3991.
\newblock \doi{https://doi.org/10.1016/j.ultramic.2018.06.014}.
\newblock URL \url{https://www.sciencedirect.com/science/article/pii/S0304399118300603}.

\bibitem[Dwyer and Paganin(2024)]{CDwyer2024}
Christian Dwyer and David~M. Paganin.
\newblock Quantum and classical fisher information in four-dimensional scanning transmission electron microscopy.
\newblock \emph{Phys. Rev. B}, 110:\penalty0 024110, Jul 2024.
\newblock \doi{10.1103/PhysRevB.110.024110}.
\newblock URL \url{https://link.aps.org/doi/10.1103/PhysRevB.110.024110}.

\bibitem[Yu et~al.(2024)Yu, Spoth, Colletta, Nguyen, Zeltmann, Zhang, Paraan, Kopylov, Dubbeldam, Serwas, Siems, Muller, and Kourkoutis]{tcBFMuller}
Yue Yu, Katherine~A. Spoth, Michael Colletta, Kayla~X. Nguyen, Steven~E. Zeltmann, Xiyue~S. Zhang, Mohammadreza Paraan, Mykhailo Kopylov, Charlie Dubbeldam, Daniel Serwas, Hannah Siems, David~A. Muller, and Lena~F. Kourkoutis.
\newblock Dose-efficient cryo-electron microscopy for thick samples using tilt-corrected scanning transmission electron microscopy, demonstrated on cells and single particles.
\newblock \emph{bioRxiv}, 2024.
\newblock \doi{10.1101/2024.04.22.590491}.
\newblock URL \url{https://www.biorxiv.org/content/early/2024/04/22/2024.04.22.590491}.

\bibitem[Bissonnette et~al.(1997)Bissonnette, Cunningham, Jaffray, Fenster, and Munro]{JPBissonnetteCunnigham}
Jean-Pierre Bissonnette, I.~A. Cunningham, David~A. Jaffray, A.~Fenster, and P.~Munro.
\newblock A quantum accounting and detective quantum efficiency analysis for video-based portal imaging.
\newblock \emph{Medical Physics}, 24\penalty0 (6):\penalty0 815--826, 1997.
\newblock \doi{https://doi.org/10.1118/1.598009}.

\bibitem[Siewerdsen and Jaffray(2000)]{ConebeamCT}
Jeffrey~H. Siewerdsen and David~A. Jaffray.
\newblock {Cone-beam CT with a flat-panel imager: noise considerations for fully 3D computed tomography}.
\newblock In James T.~Dobbins III and John~M. Boone, editors, \emph{Medical Imaging 2000: Physics of Medical Imaging}, volume 3977, pages 408 -- 416. International Society for Optics and Photonics, SPIE, 2000.
\newblock \doi{10.1117/12.384515}.
\newblock URL \url{https://doi.org/10.1117/12.384515}.

\bibitem[Tward and Siewerdsen(2008)]{SiewardsenConebeam3D}
Daniel~J. Tward and Jeffrey~H. Siewerdsen.
\newblock Cascaded systems analysis of the 3d noise transfer characteristics of flat-panel cone-beam ct.
\newblock \emph{Medical Physics}, 35\penalty0 (12):\penalty0 5510--5529, 2008.
\newblock \doi{https://doi.org/10.1118/1.3002414}.
\newblock URL \url{https://aapm.onlinelibrary.wiley.com/doi/abs/10.1118/1.3002414}.

\bibitem[Jones(1959)]{jones1959advances}
R.~C. Jones.
\newblock \emph{Advances in Electronics and Electron Physics {XI} (Academic Press}.
\newblock Inc., New York, 1959.

\bibitem[Cowley(1981)]{Cowley1981}
J.~M. Cowley.
\newblock \emph{Diffraction Physics}.
\newblock North-Holland, Amsterdam, 2nd edition, 1981.

\bibitem[Zernike(1942)]{ZERNIKE1942686}
F.~Zernike.
\newblock Phase contrast, a new method for the microscopic observation of transparent objects.
\newblock \emph{Physica}, 9\penalty0 (7):\penalty0 686--698, 1942.
\newblock ISSN 0031-8914.
\newblock \doi{https://doi.org/10.1016/S0031-8914(42)80035-X}.
\newblock URL \url{https://www.sciencedirect.com/science/article/pii/S003189144280035X}.

\bibitem[Scherzer(1949)]{ScherzerDefocus}
O.~Scherzer.
\newblock The theoretical resolution limit of the electron microscope.
\newblock \emph{Journal of Applied Physics}, 20\penalty0 (1):\penalty0 20--29, 01 1949.
\newblock ISSN 0021-8979.
\newblock \doi{10.1063/1.1698233}.
\newblock URL \url{https://doi.org/10.1063/1.1698233}.

\bibitem[Kirkland(2020)]{Kirkland2020}
Earl~J. Kirkland.
\newblock \emph{Some Image Approximations}, pages 37--80.
\newblock Springer International Publishing, Cham, 2020.
\newblock ISBN 978-3-030-33260-0.
\newblock \doi{10.1007/978-3-030-33260-0_3}.
\newblock URL \url{https://doi.org/10.1007/978-3-030-33260-0_3}.

\bibitem[Ibanez and Verbeeck(2024)]{ibanez2024}
Francisco~Vega Ibanez and Jo~Verbeeck.
\newblock Retrieval of phase information from low-dose electron microscopy experiments: are we at the limit yet?, 2024.
\newblock URL \url{https://arxiv.org/abs/2408.10590}.

\bibitem[Wiener(1964)]{wiener1964time}
N.~Wiener.
\newblock \emph{Time Series}.
\newblock M.I.T. Press, 1964.
\newblock URL \url{https://books.google.co.uk/books?id=wNQxngAACAAJ}.

\bibitem[Khintchine(1934)]{Khintchine1934}
A.~Khintchine.
\newblock Korrelationstheorie der station{\"a}ren stochastischen prozesse.
\newblock \emph{Mathematische Annalen}, 109\penalty0 (1):\penalty0 604--615, Dec 1934.
\newblock ISSN 1432-1807.
\newblock \doi{10.1007/BF01449156}.
\newblock URL \url{https://doi.org/10.1007/BF01449156}.

\bibitem[Rodenburg et~al.(1993)Rodenburg, McCallum, and Nellist]{RODENBURG1993304}
J.M. Rodenburg, B.C. McCallum, and P.D. Nellist.
\newblock Experimental tests on double-resolution coherent imaging via stem.
\newblock \emph{Ultramicroscopy}, 48\penalty0 (3):\penalty0 304--314, 1993.
\newblock ISSN 0304-3991.
\newblock \doi{https://doi.org/10.1016/0304-3991(93)90105-7}.
\newblock URL \url{https://www.sciencedirect.com/science/article/pii/0304399193901057}.

\bibitem[Pennycook et~al.(2015)Pennycook, Lupini, Yang, Murfitt, Jones, and Nellist]{PENNYCOOK2015160}
Timothy~J. Pennycook, Andrew~R. Lupini, Hao Yang, Matthew~F. Murfitt, Lewys Jones, and Peter~D. Nellist.
\newblock Efficient phase contrast imaging in stem using a pixelated detector. part 1: Experimental demonstration at atomic resolution.
\newblock \emph{Ultramicroscopy}, 151:\penalty0 160--167, 2015.
\newblock ISSN 0304-3991.
\newblock \doi{https://doi.org/10.1016/j.ultramic.2014.09.013}.
\newblock URL \url{https://www.sciencedirect.com/science/article/pii/S0304399114001934}.

\bibitem[Yang et~al.(2015)Yang, Pennycook, and Nellist]{YANG2015232}
Hao Yang, Timothy~J. Pennycook, and Peter~D. Nellist.
\newblock Efficient phase contrast imaging in stem using a pixelated detector. part ii: Optimisation of imaging conditions.
\newblock \emph{Ultramicroscopy}, 151:\penalty0 232--239, 2015.
\newblock ISSN 0304-3991.
\newblock \doi{https://doi.org/10.1016/j.ultramic.2014.10.013}.
\newblock URL \url{https://www.sciencedirect.com/science/article/pii/S0304399114002058}.

\bibitem[Lazić et~al.(2016)Lazić, Bosch, and Lazar]{LAZIC2016265}
Ivan Lazić, Eric~G.T. Bosch, and Sorin Lazar.
\newblock Phase contrast stem for thin samples: Integrated differential phase contrast.
\newblock \emph{Ultramicroscopy}, 160:\penalty0 265--280, 2016.
\newblock ISSN 0304-3991.
\newblock \doi{https://doi.org/10.1016/j.ultramic.2015.10.011}.
\newblock URL \url{https://www.sciencedirect.com/science/article/pii/S0304399115300449}.

\bibitem[Rodenburg and Bates(1992)]{rodenburg1992theory}
J.~M. Rodenburg and R.~H.~T. Bates.
\newblock The theory of super-resolution electron microscopy via wigner-distribution deconvolution.
\newblock \emph{Philosophical Transactions of the Royal Society of London. Series A: Physical and Engineering Sciences}, 339\penalty0 (1655), 1992.

\bibitem[Nellist and Rodenburg(1994)]{NELLIST199461}
P.D. Nellist and J.M. Rodenburg.
\newblock Beyond the conventional information limit: the relevant coherence function.
\newblock \emph{Ultramicroscopy}, 54\penalty0 (1):\penalty0 61--74, 1994.
\newblock ISSN 0304-3991.
\newblock \doi{https://doi.org/10.1016/0304-3991(94)90092-2}.
\newblock URL \url{https://www.sciencedirect.com/science/article/pii/0304399194900922}.

\bibitem[Yu et~al.(2021)Yu, Spoth, Muller, and Kourkoutis]{tcbf_lenak}
Yue Yu, Katherine Spoth, David Muller, and Lena Kourkoutis.
\newblock Dose-efficient tcbf-stem imaging with real-space information beyond the scan sampling limit.
\newblock \emph{Microscopy and Microanalysis}, 27\penalty0 (S1):\penalty0 758--760, 08 2021.
\newblock ISSN 1431-9276.
\newblock \doi{10.1017/S1431927621003032}.

\bibitem[Yang et~al.(2016)Yang, Ercius, Nellist, and Ophus]{YANG2016117}
Hao Yang, Peter Ercius, Peter~D. Nellist, and Colin Ophus.
\newblock Enhanced phase contrast transfer using ptychography combined with a pre-specimen phase plate in a scanning transmission electron microscope.
\newblock \emph{Ultramicroscopy}, 171:\penalty0 117--125, 2016.
\newblock ISSN 0304-3991.
\newblock \doi{https://doi.org/10.1016/j.ultramic.2016.09.002}.
\newblock URL \url{https://www.sciencedirect.com/science/article/pii/S0304399116301966}.

\bibitem[O’Leary et~al.(2021)O’Leary, Martinez, Liberti, Humphry, Kirkland, and Nellist]{OLEARY2021113189}
Colum~M. O’Leary, Gerardo~T. Martinez, Emanuela Liberti, Martin~J. Humphry, Angus~I. Kirkland, and Peter~D. Nellist.
\newblock Contrast transfer and noise considerations in focused-probe electron ptychography.
\newblock \emph{Ultramicroscopy}, 221:\penalty0 113189, 2021.
\newblock ISSN 0304-3991.
\newblock \doi{https://doi.org/10.1016/j.ultramic.2020.113189}.

\bibitem[Schwartz et~al.(2019)Schwartz, Axelrod, Campbell, Turnbaugh, Glaeser, and M{\"u}ller]{Schwartz2019}
Osip Schwartz, Jeremy~J. Axelrod, Sara~L. Campbell, Carter Turnbaugh, Robert~M. Glaeser, and Holger M{\"u}ller.
\newblock Laser phase plate for transmission electron microscopy.
\newblock \emph{Nature Methods}, 16\penalty0 (10):\penalty0 1016--1020, Oct 2019.
\newblock ISSN 1548-7105.
\newblock \doi{10.1038/s41592-019-0552-2}.

\bibitem[Petrov et~al.(2024)Petrov, Zhang, Axelrod, and Müller]{petrov2024crossedlaserphaseplates}
Petar~N. Petrov, Jessie~T. Zhang, Jeremy~J. Axelrod, and Holger Müller.
\newblock Crossed laser phase plates for transmission electron microscopy, 2024.
\newblock URL \url{https://arxiv.org/abs/2410.11328}.

\bibitem[Linck et~al.(2016)Linck, Hartel, Uhlemann, Kahl, M\"uller, Zach, Haider, Niestadt, Bischoff, Biskupek, Lee, Lehnert, B\"orrnert, Rose, and Kaiser]{UteKaiserCc}
Martin Linck, Peter Hartel, Stephan Uhlemann, Frank Kahl, Heiko M\"uller, Joachim Zach, Max. Haider, Marcel Niestadt, Maarten Bischoff, Johannes Biskupek, Zhongbo Lee, Tibor Lehnert, Felix B\"orrnert, Harald Rose, and Ute Kaiser.
\newblock Chromatic aberration correction for atomic resolution tem imaging from 20 to 80 kv.
\newblock \emph{Phys. Rev. Lett.}, 117:\penalty0 076101, Aug 2016.
\newblock \doi{10.1103/PhysRevLett.117.076101}.
\newblock URL \url{https://link.aps.org/doi/10.1103/PhysRevLett.117.076101}.

\bibitem[Laporte et~al.(2025)Laporte, Jochmans, Bardiot, Desmarets, Debski-Antoniak, Mizzon, Abdelnabi, Leyssen, Chiu, Zhang, Nomura, Boland, Ohto, Stahl, Wuyts, De~Jonghe, Stevaert, van Hemert, Bontes, Wanningen, Groenewold, Zegar, Owczarek, Joshi, Koukni, Arzel, Klaassen, Vanherck, Vandecaetsbeek, Cremers, Donckers, Francken, Van~Buyten, Rymenants, Schepers, Pyrc, Hilgenfeld, Dubuisson, Bosch, Van~Kuppeveld, Eydoux, Decroly, Canard, Naesens, Weynand, Snijder, Belouzard, Shimizu, Bartenschlager, Hurdiss, Marchand, Chaltin, and Neyts]{Laporte2025}
Manon Laporte, Dirk Jochmans, Doroth{\'e}e Bardiot, Lowiese Desmarets, Oliver~J. Debski-Antoniak, Giulia Mizzon, Rana Abdelnabi, Pieter Leyssen, Winston Chiu, Zhikuan Zhang, Norimichi Nomura, Sandro Boland, Umeharu Ohto, Yannick Stahl, Jurgen Wuyts, Steven De~Jonghe, Annelies Stevaert, Martijn~J. van Hemert, Brenda~W. Bontes, Patrick Wanningen, G.~J.~Mirjam Groenewold, Aneta Zegar, Katarzyna Owczarek, Sanjata Joshi, Mohamed Koukni, Philippe Arzel, Hugo Klaassen, Jean-Christophe Vanherck, Ilse Vandecaetsbeek, Niels Cremers, Kim Donckers, Thibault Francken, Tina Van~Buyten, Jasper Rymenants, Joost Schepers, Krzysztof Pyrc, Rolf Hilgenfeld, Jean Dubuisson, Berend-Jan Bosch, Frank Van~Kuppeveld, Cecilia Eydoux, Etienne Decroly, Bruno Canard, Lieve Naesens, Birgit Weynand, Eric~J. Snijder, Sandrine Belouzard, Toshiyuki Shimizu, Ralf Bartenschlager, Daniel~L. Hurdiss, Arnaud Marchand, Patrick Chaltin, and Johan Neyts.
\newblock A coronavirus assembly inhibitor that targets the viral membrane protein.
\newblock \emph{Nature}, 640\penalty0 (8058):\penalty0 514--523, Apr 2025.
\newblock ISSN 1476-4687.
\newblock \doi{10.1038/s41586-025-08773-x}.
\newblock URL \url{https://doi.org/10.1038/s41586-025-08773-x}.

\bibitem[Ma et~al.(2025)Ma, Li, Muller, and Zeltmann]{ma2025information4dstemisuse}
Desheng Ma, Guanxing Li, David~A Muller, and Steven~E Zeltmann.
\newblock Information in 4d-stem: Where it is, and how to use it, 2025.
\newblock URL \url{https://arxiv.org/abs/2507.21034}.

\end{thebibliography}
\end{document}